\documentclass[aps,prd,floatfix,footinbib,longbibliography,twocolumn,superscriptaddress]{revtex4-2}

\usepackage{amsmath}
\usepackage{amsfonts}
\usepackage{amssymb}
\usepackage{graphicx}
\usepackage{color}
\usepackage{accents}
\usepackage[colorlinks=true, citecolor=blue]{hyperref}
\usepackage[all]{hypcap}
\usepackage{subfigure}
\usepackage[normalem]{ulem}
\usepackage{float}
\usepackage{dsfont}
\usepackage{soul}

\usepackage[pagewise,left]{lineno}

\newcommand{\ket}[1]{\left|#1\right\rangle}
\newcommand{\bra}[1]{\left\langle #1\right|}

\usepackage{todonotes}

\newcommand{\bea}{\begin{eqnarray}}
\newcommand{\eea}{\end{eqnarray}}
\newcommand{\be}{\begin{equation}}
\newcommand{\ee}{\end{equation}}
\newcommand{\ba}{\begin{align}}
\newcommand{\ea}{\end{align}}

\makeatletter 
\newsavebox{\@brx}
\newcommand{\llangle}[1][]{\savebox{\@brx}{\(\m@th{#1\langle}\)}%
  \mathopen{\copy\@brx\kern-0.5\wd\@brx\usebox{\@brx}}}
\newcommand{\rrangle}[1][]{\savebox{\@brx}{\(\m@th{#1\rangle}\)}%
  \mathclose{\copy\@brx\kern-0.5\wd\@brx\usebox{\@brx}}}
\makeatother

\newlength{\dhatheight} 

\newenvironment{definition}[1][Definition]{\begin{trivlist}
\item[\hskip \labelsep {\bfseries #1}]}{\end{trivlist}}

\newcommand{\qed}{\nobreak \ifvmode \relax \else
      \ifdim\lastskip<1.5em \hskip-\lastskip
      \hskip1.5em plus0em minus0.5em \fi \nobreak
      \vrule height0.75em width0.5em depth0.25em\fi}

\usepackage{fancyhdr}
\begin{document}

\title{Causal State Updates in Real Scalar Quantum Field Theory}
\author{I. Jubb\href{https://orcid.org/0000-0001-7339-2058}{\includegraphics[scale=0.066]{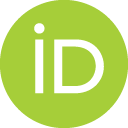}}}
\affiliation{Dublin Institute for Advanced Studies, School of Theoretical Physics, 10 Burlington Rd, Dublin 4, Ireland.}

\date{\today}
\begin{abstract}
In relativistic Quantum Field Theory (QFT) ideal measurements of certain observables are physically impossible without violating causality. This prompts two questions: i) can a given observable be ideally measured in QFT, and ii) if not, in what sense can it be measured? Here we formulate a necessary and sufficient condition that any measurement, and more generally any state update (quantum operation), must satisfy to respect causality in real scalar QFT. We argue that for unitary `kicks' and operations involving 1-parameter families of Kraus operators, e.g. Gaussian measurements, the only causal observables are smeared fields and the identity --- the basic observables in real scalar QFT. We provide examples with more complicated operators such as products of smeared fields, and show that the associated state updates are acausal, and hence impossible. Despite this, one can still recover expectation values of such operators, and we show how to do this using only causal measurements of smeared fields.
\end{abstract}
\maketitle

\section{Introduction}\label{sec:Introduction}

While quantum theory is mathematically and philosophically disparate from General Relativity, it is nonetheless understood that it too must obey the universal speed limit of causal influence. The relativistic setting of Quantum Field Theory (QFT) has this speed limit hard-coded into the spacetime commutation relations, i.e. any pair of spacelike operators commute~\footnote{Here we work in the Heisenberg picture where operators carry the dynamics, and hence it makes sense to talk about operators at points, or more accurately in regions, of spacetime.}. Most discussions of causality in QFT end here (e.g.~\cite{Hellwig_Kraus}), as local operations on the state cannot affect expectation values of observables at spacelike points, that is, points in space and time that are causally disconnected. 

\begin{figure}
    \centering
    \includegraphics[scale=0.7]{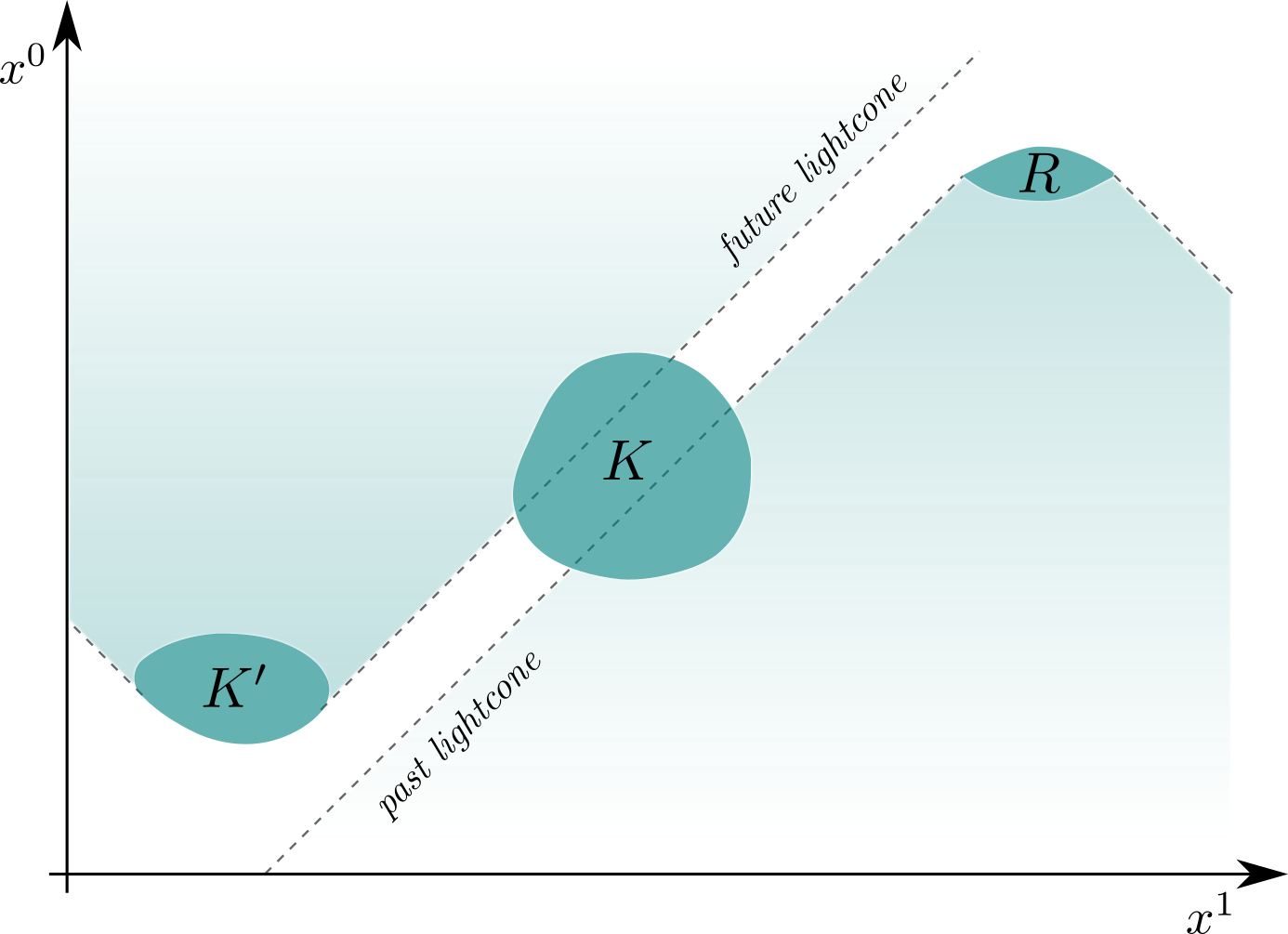}
    \caption{Spacetime diagram with time and space coordinates $x^0$ and $x^1$ respectively. The compact (closed and bounded) subset $K'$ is spacelike, or causally disconnected from the region $R$, as can be seen from the lightcones (dashed lines). $K'$/$R$ is also partly to the past/future of the compact subset $K$. Any local operations occurring at points in time and space within $K$ should not be able to transmit the effects of any local operations occurring in $K'$ to measurements in $R$. While the diagrams in this paper are for spacetime dimension $d=2$, they are only illustrative, as the discussions apply more generally to all $d\geq 2$.}
    \label{fig:protocol}
\end{figure}

This is not the end of the story for causality in QFT, however. In 1993 Sorkin pointed out that local operations must satisfy a further causality condition regarding their properties under composition~\cite{Sorkin:1993gg}. It is not enough to say that a local operation, contained in spatial extent and duration in some portion of spacetime $K$, cannot affect measurements occurring at points spacelike to $K$. To respect causality it must also not transmit the effects of some other local operation, contained in $K'$, to a region spacelike to $K'$ (Fig.~\ref{fig:protocol}). In other words, \emph{it cannot enable other local operations to violate causality}.

This puts an additional, but physically justified, constraint on the allowed quantum operations, or state updates in QFT. Surprisingly, some standard state updates in Non-Relativistic Quantum Mechanics (NRQM) and Quantum Information (QI) fail this causality condition when applied to the relativistic setting of QFT, e.g. ideal measurements of certain observables, including projectors onto wave-packet states~\cite{Sorkin:1993gg,Benincasa_2014,borsten2021impossible} and Wilson loops in gauge theory~\cite{Wilson_loops}. To avoid any causality violations, such ideal measurements in QFT must be \emph{impossible} to implement experimentally, by \emph{any} measurement apparatus~\footnote{Note we are not saying that the analogous state updates in NRQM are impossible, just that this is the case in relativistic QFT.}. Related questions of causality in QFT have also been studied using Unruh-DeWitt detectors~\cite{Tjoa_2019,Mart_n_Mart_nez_2021,de_Ram_n_2021,perche2021antiparticle,pologomez2021detectorbased}, and are of broader relevance to QI~\cite{PhysRevA.64.052309,Popescu_1994}.

The fact that not all self-adjoint operators in QFT can be measured in the standard sense of quantum theory prompts two questions: \textit{which operators are measurable in QFT, and to what extent?} For example, if an ideal measurement is not possible, then perhaps something less sharp is. One route to answering this is to construct specific measurement models which do not superluminally signal, e.g. using local probes~\cite{Mart_n_Mart_nez_2015} or probe fields~\cite{Bostelmann_2021}, although this transforms the question into what measurements are possible on probes. Alternatively, one can remain agnostic to the details of the measurement apparatus, and ask more generally which state updates are possible with respect to this additional causality constraint~\cite{borsten2021impossible}.

In this paper we precisely characterise the class of state updates that are causal (Section~\ref{sec:Setup}), and we provide several simple examples of causal and acausal maps using local \emph{unitary kicks} (Section~\ref{sec:Unitary kicks}) and \emph{Gaussian measurements} (Section~\ref{sec:Measurements}), a less sharp alternative to ideal measurements. Furthermore, the acausal examples presented here, unlike those presented in~\cite{Sorkin:1993gg,Benincasa_2014}, will be local update maps, thus eliminating the worry that the acausality of a given map is entirely due to its non-locality. Specifically, in~\cite{Sorkin:1993gg,Benincasa_2014} they considered ideal measurements of a projector of the form $P=\ket{\Psi}\bra{\Psi}$, for some spatially compact wave-packet state $\ket{\Psi}$. While the shape of the wave-packet is local (in the sense that it is of finite spatial extent), the projector $P$ is a non-local operator, in the precise sense that it is not localisable in any sub-region of spacetime\footnote{This follows as $P=\ket{\Psi}\bra{\Psi}$ is a rank 1 operator, and so $P$ cannot
be localisable in any spacetime region, as all localisable projectors must be of infinite rank --- a common feature of type III von Neumann algebras (see~\cite{fewster2019algebraic} for example).}. 

Surprisingly, our results suggest that the only causality respecting observables (those for which the corresponding measurement is described by a causal update map) are the smeared fields and the identity --- the basic observables of real scalar QFT. This also seems to be the case for unitary kicks and operations described by a 1-parameter family of Kraus operators. Conversely, to update the state according to the measurement of, or unitary kick with, some other more complicated observable, e.g. the product of two smeared fields, it appears one must violate causality, and thus such operations must be physically impossible. Our calculations also suggest that ideal measurements of smeared fields are acausal, which motivates our focus on the less sharp Gaussian measurements.

It is important to note that this conclusion, that only measurement updates for smeared fields are possible with respect to causality, does not preclude the recovery of correlation functions and other expectation values of products of smeared fields, and in Section~\ref{sec:Extracting expectation values from measurements of smeared fields} we describe how this can be done with causal measurements of smeared fields alone. Alternatively, expectation values can also be recovered through some other measurement prescription, e.g. using probes~\cite{Fewster2020,fewster2019generally,Bostelmann_2021,ruep2021weakly}.

In Section~\ref{sec:Interactions} we extend our results to interacting QFT, and show that in the case of a compact self-interaction smeared fields can still be measured in a causal manner. In~\ref{sec:Discussion} we briefly comment on their relevance to continuous measurement models~\cite{Brun_2000,Jacobs_2006}, and discuss the potential philosophical implications to the ontology of QFT. Lastly, in Section~\ref{sec:Conclusion} we summarise our results.

In what follows some definitions and results will be generalisable to complex scalar and fermionic QFT, since they rely only on certain basic concepts in Algebraic (A)QFT~\cite{fewster2019algebraic}, namely that there is a net of subalgebras of observables associated to regions of spacetime satisfying certain properties. We will be careful to highlight at which points such generalisations are possible.

\section{Setup}\label{sec:Setup}

\subsection{Spacetime Geometry}\label{sec:Spacetime Geometry}

\begin{figure}
    \centering
    \includegraphics[scale=0.7]{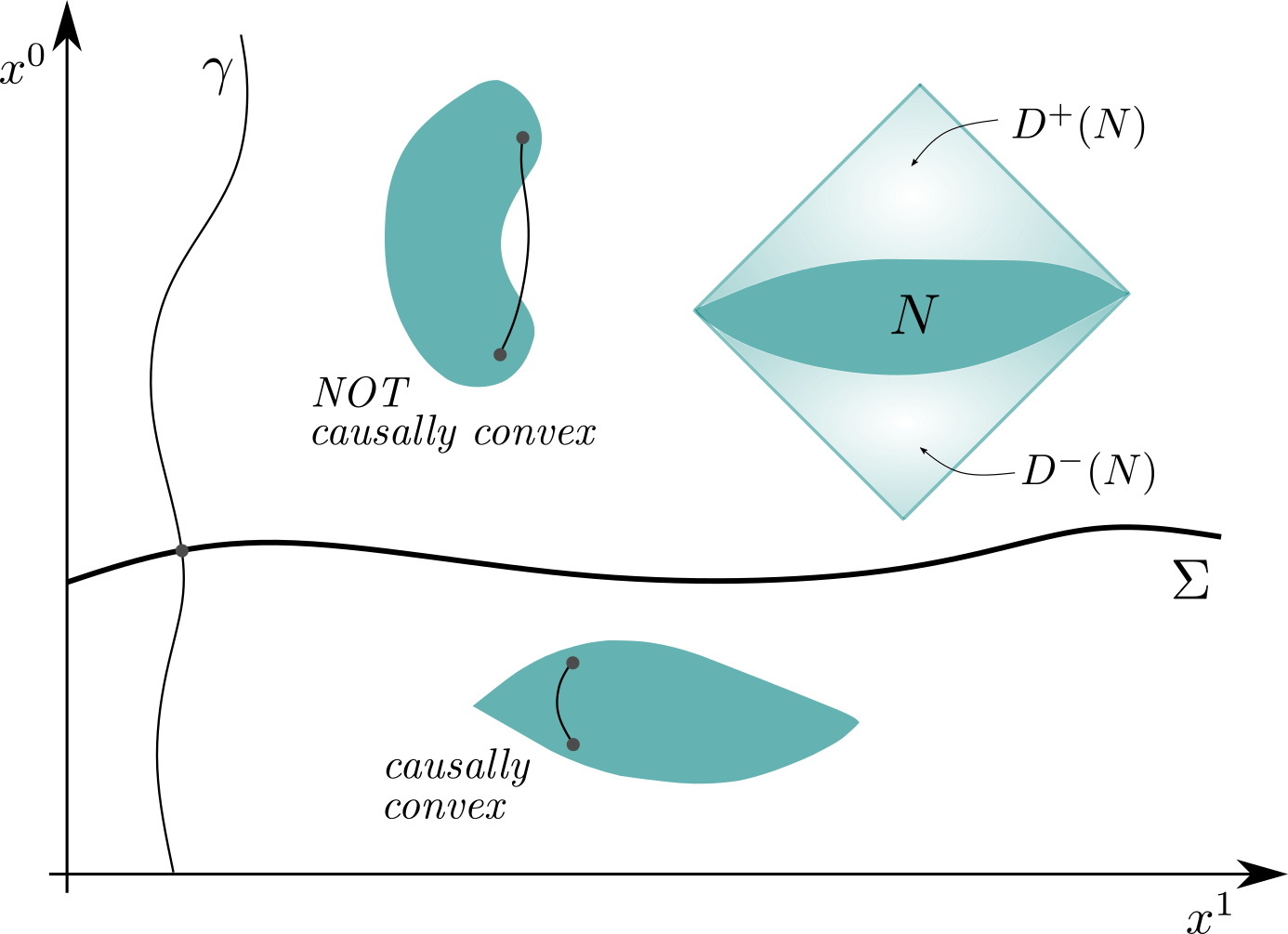}
    \caption{Spacetime diagram with a Cauchy surface $\Sigma$. All inextendible timelike curves, e.g. $\gamma$, cross $\Sigma$ exactly once. Also illustrated are examples of subsets that are (not) causally convex. Finally, for the subset $N$ we have shown its future and past domains of dependence (which both include $N$).}
    \label{fig:spacetime_geometry}
\end{figure}

Here we consider some potentially curved spacetime $M$ with a Lorentzian metric. $M$ must be \emph{globally hyperbolic}, meaning that it contains a \emph{Cauchy surface} $\Sigma\subset M$. Recall that a spatial surface $\Sigma$ is a Cauchy surface if all inextendible timelike curves, i.e. all slower than light trajectories with no future or past endpoints, intersect $\Sigma$ exactly once. This, and the other concepts below, are illustrated in Fig.~\ref{fig:spacetime_geometry}. See~\cite{Wald} for more details. 

The causal future/past of some subset of spacetime $N\subset M$ is denoted by $J^{\pm}(N)$, e.g. $J^+ (K' )$ and $J^-(R)$ in Fig.~\ref{fig:protocol}. A subset $N\subseteq M$ is \emph{causally convex} if any causal curve, i.e. any timelike (slower than light) or lightlike curve, with endpoints in $N$ is itself contained in $N$.

In the following we reserve the word \emph{region} for any open causally convex subset $R\subseteq M$ which, if treated as a spacetime in its own right, is globally hyperbolic.

For a subset $N$, the \emph{domain of dependence} is given by $D(N)=D^+(N)\cup D^-(N)$, where $D^{\pm}(N)$ denotes the future/past domain of dependence, and consists of all spacetime points $x\in M$ for which all past/future inextendible causal curves from $x$ pass through $N$.

The \emph{causal complement} of a subset $N$ is denoted by $N^{\perp} = M\setminus (J^+(N)\cup J^-(N))$, and consists of all points spacelike to, or causally disconnected from, $N$.

\begin{figure}
    \centering
    \includegraphics[scale=0.6]{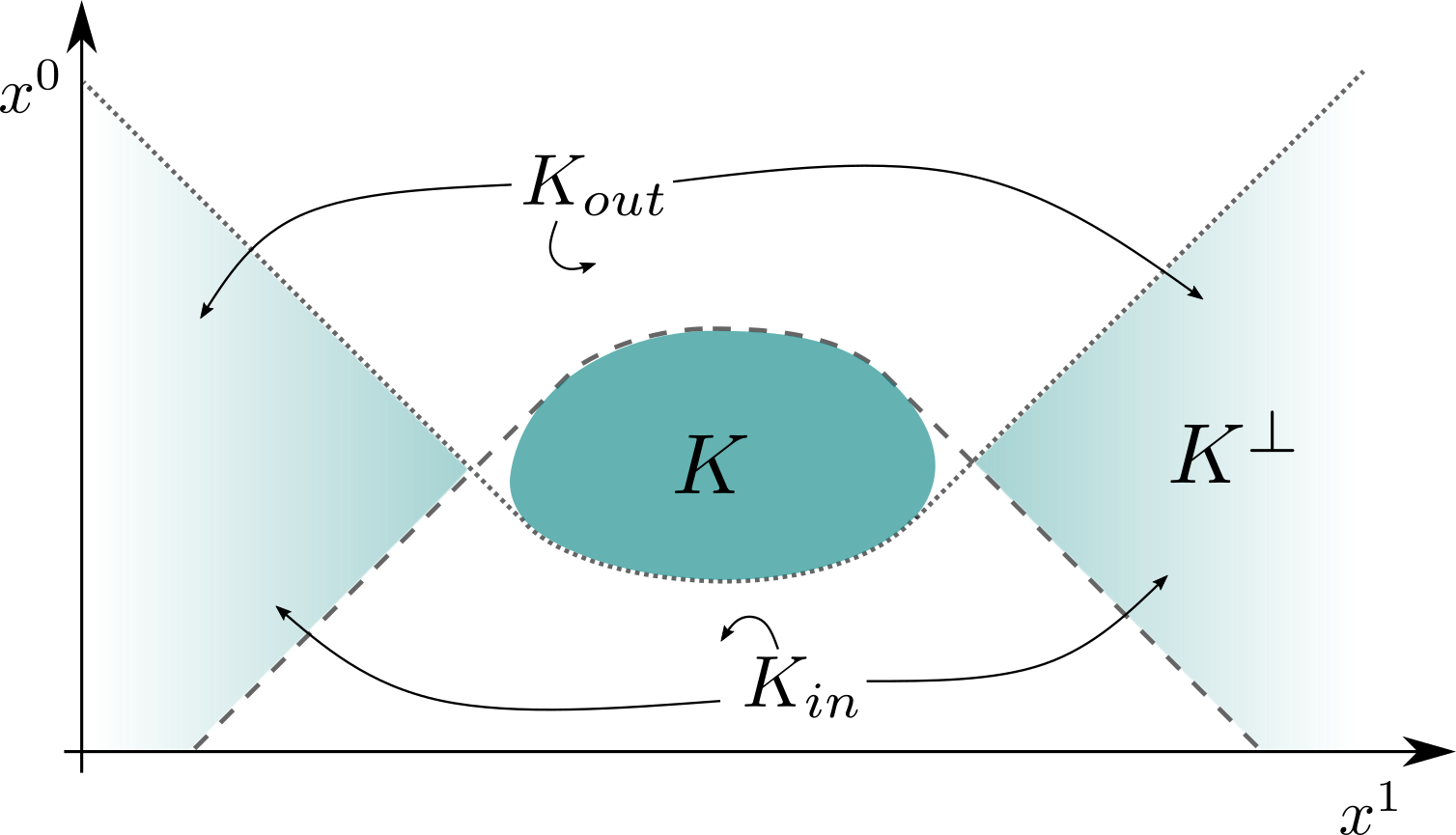}
    \caption{Spacetime diagram of a compact subset $K\subset M$, and the corresponding in/out-region $K_{in}$/$K_{out}$ (all points below/above the dotted/dashed line). The causal complement, $K^{\perp}$, consisting of all spacelike points to $K$, has also been illustrated (shaded with gradient).}
    \label{fig:K_in_out_perp}
\end{figure}

To a compact (closed and bounded) subset $K$ we associate an in-region and an out-region, consisting of points not to the future and past of $K$ respectively (see Fig.~\ref{fig:K_in_out_perp}). We denote these regions (so called because they are open, causally convex, and constitute globally hyperbolic spacetimes in their own right) as $K_{in} = M\setminus J^+(K)$ and $K_{out} = M\setminus J^-(K)$ respectively. Note these regions intersect at points spacelike to $K$, i.e. in $K^{\perp}$.

For any function over spacetime, $f : \mathcal{M}\mapsto \mathbb{R}$ (or valued in $\mathbb{C}$), we denote its support as $\text{supp}f = \lbrace x\in M : f(x)\neq 0 \rbrace$, and we say $f$ is compactly supported if $\overline{\text{supp}f}$ is compact, i.e. $\text{supp}f$ has compact closure (where we use $\overline{S}$ to denote the closure of a set $S$).

\subsection{QFT}\label{sec:QFT}

\subsubsection{Smeared field operators}

\begin{figure}[b]
    \centering
    \includegraphics[scale=0.7]{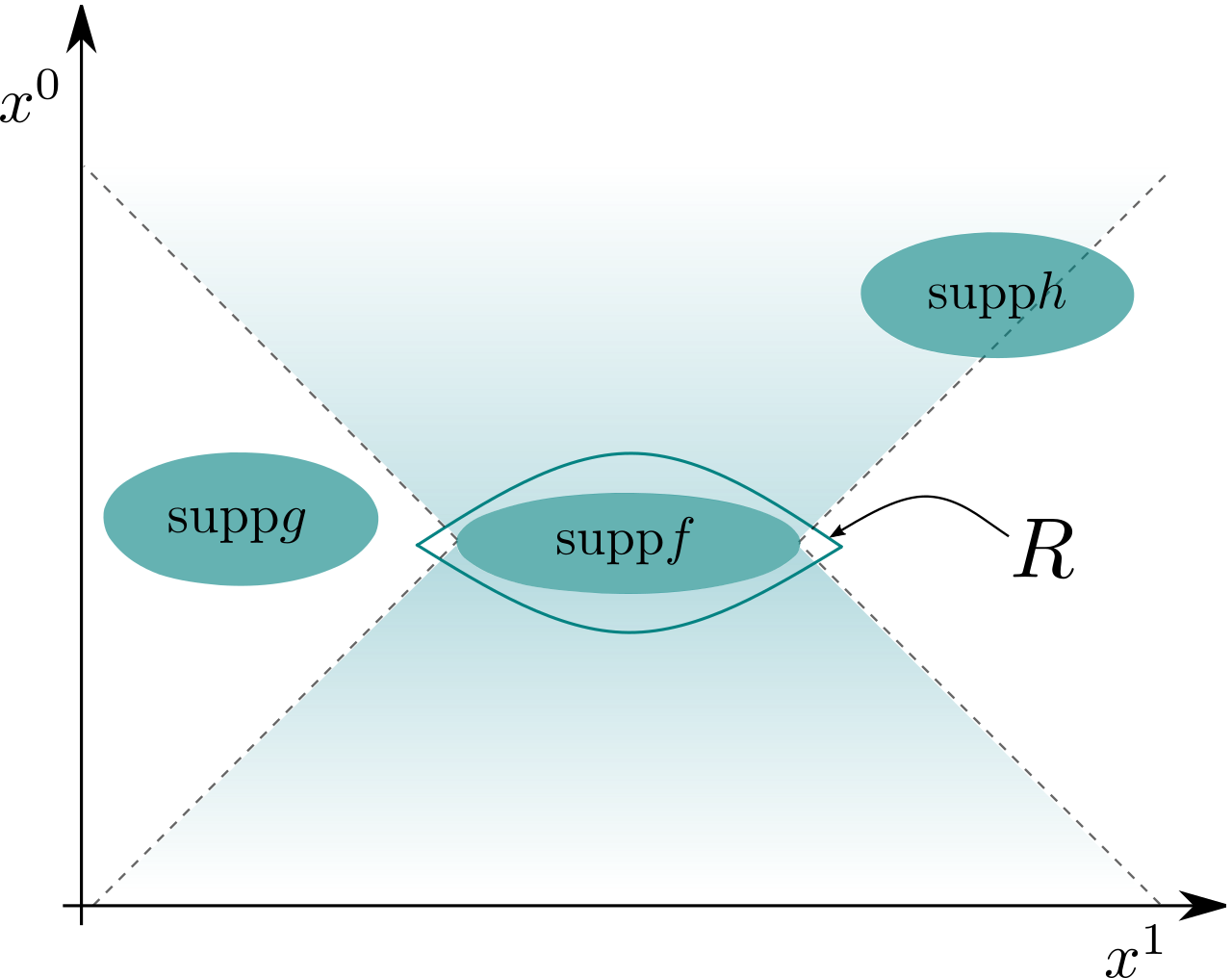}
    \caption{Spacetime diagram illustrating the supports of three test, or smearing, functions $f$, $g$, and $h$. The smeared field operator $\phi(f)$ is constructed by integrating $\phi(x)$ with $f$ over $\text{supp}f$ (contained in the region $R$), and similarly for the smeared fields $\phi(g)$ and $\phi(h)$. $\text{supp}f$ is spacelike to $\text{supp}g$, but not to $\text{supp}h$. Therefore, $\phi(f)$ and $\phi(g)$ commute, while $\phi(f)$ and $\phi(h)$ may not.}
    \label{fig:support_of_functions}
\end{figure}

Consider free real scalar QFT in $M$, with the field operator $\phi(x)$ acting on the bosonic Fock space in the usual way. We are working in the Heisenberg picture where the fields carry the dynamics. Technically speaking, the field `operator' $\phi(x)$ is really an operator-valued distribution, and hence we must integrate it against a test function $f$ to form a proper operator on the Fock space. Recall that test functions must be smooth and compactly supported. The result of this integration, or smearing, with $f$ gives the \emph{smeared field operator}
\begin{equation}\label{eq:smeared_field_in_terms_of_op_valued_dist}
\phi(f) = \int_M dx \, f(x) \phi(x) \; ,
\end{equation}
where $dx$ denotes the spacetime volume element. Note we have used the symbol ``$\phi$'' again for the smeared field. Any ambiguity between the smeared field $\phi(f)$ and the operator-valued distribution $\phi(x)$ can be resolved by inspecting whether the argument is a test function or a spacetime point respectively. If $\text{supp}f\subseteq R$ for some spacetime region $R$, the operator $\phi(f)$ is said to be \emph{localisable} in $R$. This is shown in Fig.~\ref{fig:support_of_functions}. Similarly to the position operator in NRQM, $\phi(f)$ is self-adjoint (for real-valued $f$) and unbounded.

The identity, $\mathds{1}$, together with the set of all smeared fields $\phi(f)$ for all test functions $f$, form the \emph{generators} of the QFT operator algebra $\mathfrak{A}$. That is, any operator in $\mathfrak{A}$ is some complex algebraic combination of the identity and the smeared fields. As an analogy, in a lattice of qubits the identity and the Pauli matrices local to each site generate the entire algebra of operators in the same way. We can also generate the subalgebra $\mathfrak{A}(R)\subseteq \mathfrak{A}$ associated to some region $R$ by only considering algebraic combinations of smeared fields supported in $R$.

Note that $a\phi(f)+b\phi(g) = \phi(a f + b g)$ for any test functions $f$ and $g$, and any $a,b\in\mathbb{C}$. The dynamics of the theory --- that $\phi(x)$ satisfies the wave equation $(\Box + m^2)\phi = 0$ --- imply that $\phi( (\Box + m^2)f ) = 0$ for any test function $f$. This can be seen using~\eqref{eq:smeared_field_in_terms_of_op_valued_dist} and integration by parts. Alternatively, $\phi( f ) = \phi(g)$ whenever $f-g = (\Box + m^2 ) h$ for some compactly supported $h$. In this case we say that $f$ and $g$ are equivalent.

Given some $f$, it is always possible to find an equivalent $g$ supported in a region $R$ that contains $\text{supp}f$ in its domain of dependence, that is, $D(R)\supseteq \text{supp}f$. An example of this is shown in Fig.~\ref{fig:solution}, and a procedure for doing this is described in Section~\ref{sec:Interactions}. Since $\text{supp}g$ can be different from $\text{supp}f$, and even disjoint, this means that $\phi(f)$ is localisable in different, possibly disjoint regions.

See~\cite{fewster2019algebraic} for an introduction to AQFT. It should be noted that in AQFT one usually starts with an abstract algebra of observables, such as the algebra of smeared fields, and \emph{then} represents that algebra as operators on some Hilbert space. Here we have implicitly assumed such a representation, and hence we work entirely at the level of operators on a Hilbert space.

\subsubsection{Covariant commutation relations}

The causal structure of the spacetime is encoded via the Covariant Commutation Relations (CCR's) for smeared fields:
\begin{equation}\label{eq:field_CCR}
[\phi(f) , \phi(g)] = i \Delta(f,g) \mathds{1} \; ,
\end{equation} 
where 
\begin{equation}
\Delta(f,g) = \int_{M\times M} dx dy \, f(x) \Delta(x,y)g(y) \; ,
\end{equation}
is the \emph{smeared} Pauli-Jordan function (smeared with $f$ and $g$), and $\Delta(x,y)=G_R(x,y) - G_A(x,y)$ is the usual Pauli-Jordan function, i.e.~the difference between the retarded and advanced Green functions of the classical field theory. That is, $(\Box + m^2)G_{R/A}(x,y) = \delta(x,y)$ and $G_{R/A}(x,y)=0$ whenever $x$ is not to the future/past of $y$. Note we use the notation ``$\Delta(\cdot , \cdot)$'' for both the smeared and standard Pauli-Jordan functions. Any ambiguity can again be resolved by inspecting whether the arguments are functions or spacetime points respectively.

\begin{figure}
    \centering
    \includegraphics[scale=0.7]{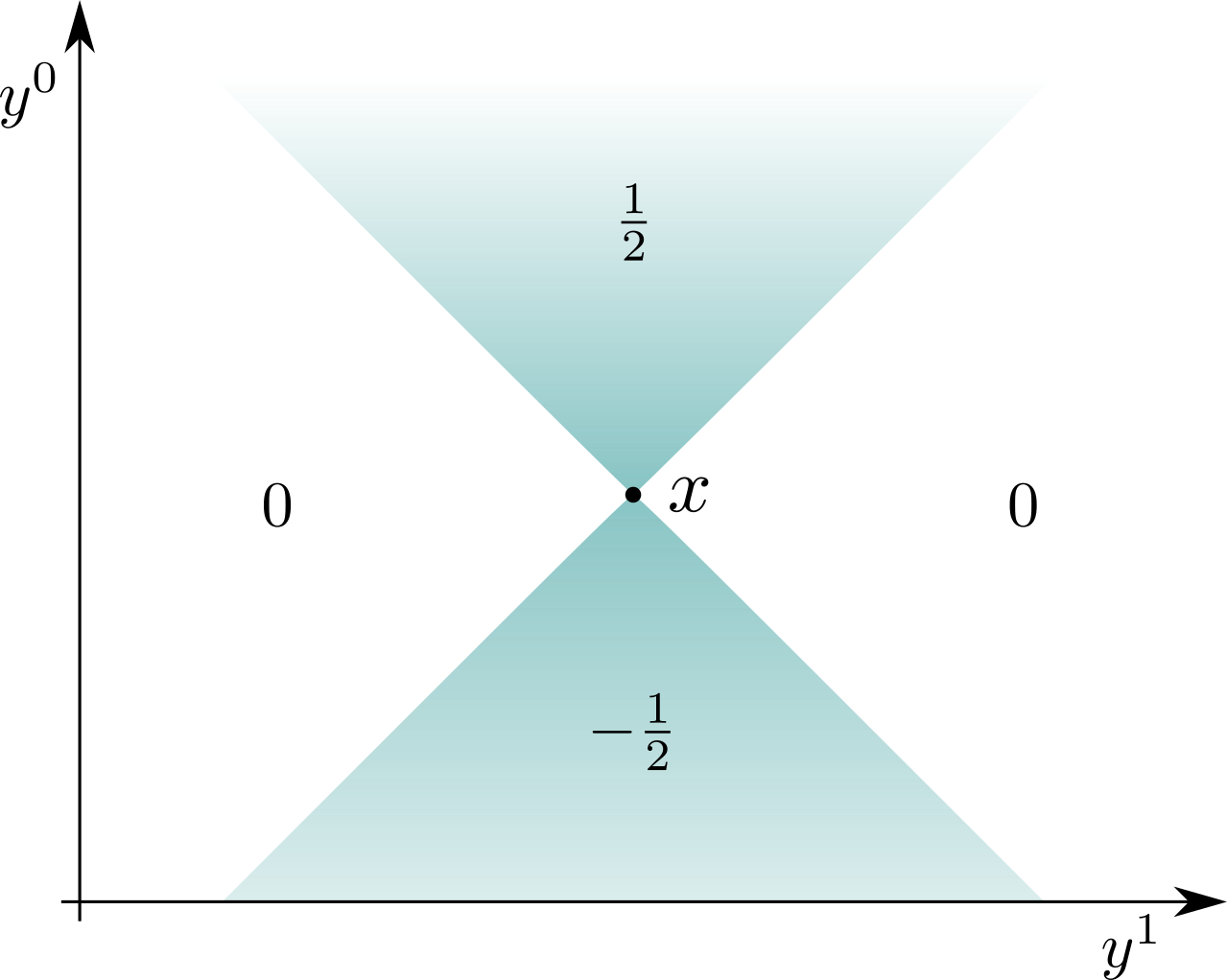}
    \caption{Plot of $\Delta(x,y)$ for a massless scalar field in 1+1 Minkowski spacetime. The spacetime point $x$ has been fixed and $\Delta(x,y)$ has been plotted as a function of the spacetime point $y$, i.e. as a function of the time and space coordinates $y^0$ and $y^1$ respectively. Note that, for $x$ and $y$ spacelike, $\Delta(x,y)=0$. The massless 1+1 case is especially simple in that $\Delta(x,y)$ is constant inside the lightcone. This is not the case for non-zero mass, or in higher dimensional Minkowski (or other curved) spacetimes. What is true in any spacetime, however, is that $\Delta(x,y)=0$ for $x$ and $y$ spacelike.}
    \label{fig:delta_massless_2d_mink}
\end{figure}

Some readers may be more used to expressing the spacetime commutation relations as
\begin{equation}
[\phi(x),\phi(y)] = i \Delta(x,y) \mathds{1} \; ,
\end{equation}
in terms of the operator-valued distribution $\phi(x)$. Indeed, the CCR's in~\eqref{eq:field_CCR} follow from these relations by integrating over the spacetime points $x$ and $y$, weighted by the smearing functions $f(x)$ and $g(y)$. To have a concrete picture in mind, we plot the functional form of $\Delta(x,y)$ in Fig.~\ref{fig:delta_massless_2d_mink} for the simple case of the massless theory in 1+1 Minkowski spacetime. To visualise $\Delta(f,g)$ for this example, one can imagine integrating $\Delta(x,y)$ against two functions $f(x)$ and $g(y)$.

In Fig.~\ref{fig:support_of_functions}, $f$ and $g$ have spacelike supports. In this case $\Delta(f,g)=0$, and hence $\phi(f)$ and $\phi(g)$ commute. For a test function $h$ that does not overlap with $f$, but is also not spacelike to $f$ (see Fig.~\ref{fig:support_of_functions}), $\Delta(f,h)$ may not vanish, and hence $\phi(f)$ and $\phi(h)$ may not commute. Therefore, the fact that $\phi(f)$ and $\phi(h)$ are localisable in disjoint regions does not imply they commute. 

\subsubsection{General properties}

The above properties of the smeared fields imply the \emph{Einstein causality} property, namely that spacelike subalgebras commute, i.e. $[ \mathfrak{A}(R) , \mathfrak{A}(R')] = 0$ for any spacelike regions $R$ and $R'$. Additionally, we have the \emph{isotony} property: $\mathfrak{A}(R)\subseteq \mathfrak{A}(R')$ whenever $R\subseteq R'$. We also have the useful \emph{time-slice} property: $\mathfrak{A}(R) = \mathfrak{A}(R')$ whenever $R\subseteq R'$ and $R$ contains a Cauchy surface for $R'$. These properties are usually assumed at the algebraic level in AQFT, before any representation of the algebra on a Hilbert space is given. Importantly, they also apply more generally to complex scalar and fermionic QFT, but only to the \emph{physical} subalgebras in each case, namely the even degree combinations of the fields which are invariant under any unobservable gauge transformations. For this reason these general properties are often taken as a starting point for constructing physical QFT's.

Given some subalgebra $\mathfrak{B}\subseteq \mathfrak{A}$, we denote the commutant as $\mathfrak{B}^{\perp}$, i.e. the set of all operators that commute with everything in $\mathfrak{B}$. We will assume the \emph{Haag property}~\cite{Haag:1992hx} (proved for scalar fields in~\cite{araki}): for any compact subset $K\subset M$, and every region $R\supset K$, then $\mathfrak{A}(K^{\perp})^{\perp}\subseteq\mathfrak{A}(R)$. That is, the subalgebras $\mathfrak{A}(R)$, for all regions $R$ that contain $K$, contain all operators that commute with those spacelike to $K$. This property is sometimes weakened to only apply to any connected compact $K$, though we do not do this here.

\subsection{Causality conditions on update maps}\label{sec:Causality conditions on update maps}

Given some state, or density matrix, $\rho$, and some self-adjoint operator $X\in\mathfrak{A}$, its expectation value is given by $\text{tr}(\rho X)$\footnote{In what follows we will implicitly restrict to states $\rho$ for which such expectation values are well defined, specifically quasifree states (described in the AQFT framework in~\cite{fewster2019algebraic}) and any states that can be constructed from these via the action of elements in $\mathfrak{A}$.}.

Any quantum operation is described by a completely-positive (CP) update map, $\tilde{\mathcal{E}}(\cdot)$, on the state: $\rho\mapsto\rho' = \tilde{\mathcal{E}}(\rho)$. Under expectation values we can instead consider the dual update map on the operators: $\text{tr}(\rho' X) = \text{tr}(\tilde{\mathcal{E}}(\rho) X) = \text{tr}(\rho \mathcal{E}(X))$. In what follows we will mostly be concerned with update maps, $\mathcal{E}(\cdot)$, acting on the operators instead of the state.

Note that under a composition of two maps on the state, e.g. $\rho \mapsto \tilde{\mathcal{E}}'(\tilde{\mathcal{E}}(\rho))$, the composition on the operators is order-reversed, e.g. $X\mapsto \mathcal{E}(\mathcal{E}'(X))$.

Our focus will usually be on trace-preserving maps, such that $\mathcal{E}(\mathds{1})=\mathds{1}$. In the case of an ideal measurement this amounts to the non-selective case where no outcome is conditioned on. Recall that for any compact self-adjoint operator, $X$, with projectors $E_n$ onto the eigenspaces associated to distinct eigenvalues $x_n$, the update map for an ideal measurement of $X$ is given by
\begin{equation}\label{eq:ideal_measurement}
\mathcal{E}_X^0(Y) = \sum_n E_n Y E_n \; ,
\end{equation}
for any operator $Y\in\mathfrak{A}$. Note that $\mathcal{E}_X^0(\mathds{1})=\mathds{1}$ since the projectors square to themselves and resolve the identity. Furthermore, if $X\in\mathfrak{A}(R)$, i.e. it is localisable in a region $R\subset M$, and $Y\in\mathfrak{A}(R')$ where $R'$ is spacelike to $R$, then $[X,Y]=0$ and $[E_n , Y]=0$. Therefore, $\mathcal{E}_X^0(Y)=Y$.

This property of an update map, that it acts trivially on operators that are spacelike to some subset of spacetime, can be concisely stated as
\begin{definition}[Definition (local)]
An update map $\mathcal{E}(\cdot)$ is \emph{local} to a compact subset $K$ if
\begin{equation}
\mathcal{E}(\cdot)\big|_{\mathfrak{A}(K^{\perp})} = 1 \; .
\end{equation}
\end{definition}

That is, $\mathcal{E}(\cdot)$ acts trivially on operators spacelike to its associated subset $K$. This ensures that expectation values of any $Y\in\mathfrak{A}(K^{\perp})$ are the same in the updated state as the original state. 

Furthermore, if we impose that the expected value of any operator $Y\in\mathfrak{A}(K^{\perp})$ is unchanged under the update $Y\mapsto \mathcal{E}(Y)$, in \emph{any} state $\rho$, we arrive at the above locality condition on $\mathcal{E}(\cdot)$. To see this let $Y' = \mathcal{E}(Y)-Y$. For any pure state $\ket{\psi}$ we then have $\bra{\psi}Y'\ket{\psi}=0$ by assumption. If, for any orthonormal states $\ket{1},\ket{2}$, we pick $\ket{\psi} = a \ket{1} + b \ket{2}$ and $\ket{\varphi} = a \ket{1} + i b \ket{2}$, for any $a,b\in\mathbb{R}$, the fact that $\bra{\psi}Y'\ket{\psi}=\bra{\varphi}Y'\ket{\varphi}=0$ implies that $\bra{1}Y' \ket{2}=0$. Since this is true for any orthonormal states, we have that $Y'=0$, in the sense that, as an operator, its matrix elements vanish. This then implies the operator equation $\mathcal{E}(Y)=Y$.

So long as $\mathcal{E}(\cdot)$ is constructed through functions of an operator $X\in\mathfrak{A}(R)$, for some region $R\subset K$, then the map $\mathcal{E}(\cdot)$ is local to $K$. In the above example of an ideal measurement of $X$, the projectors $E_n$ are functions of $X$, specifically indicator functions, and hence they commute with all operators spacelike to $X$, and thus $\mathcal{E}(\cdot)$ acts trivially on such operators.

As discussed above, many local update maps fail a further causal constraint regarding compositions with other local maps. To make this precise we make the following
\begin{definition}[Definition (causal w.r.t.)]
An update map $\mathcal{E}(\cdot)$, local to a compact subset $K$, is \emph{causal with respect to} a map $\mathcal{E}'(\cdot)$, local to some compact $K'\subset K_{in}$, if 
\begin{equation}\label{eq:causal_wrt_conditions}
\mathcal{E}'(\mathcal{E}(\cdot))\big|_{\mathfrak{A}({K'^{\perp}\cap K_{out}})} = \mathcal{E}(\cdot) \; .
\end{equation}
\end{definition}

\begin{figure}
    \centering
    \includegraphics[scale=0.6]{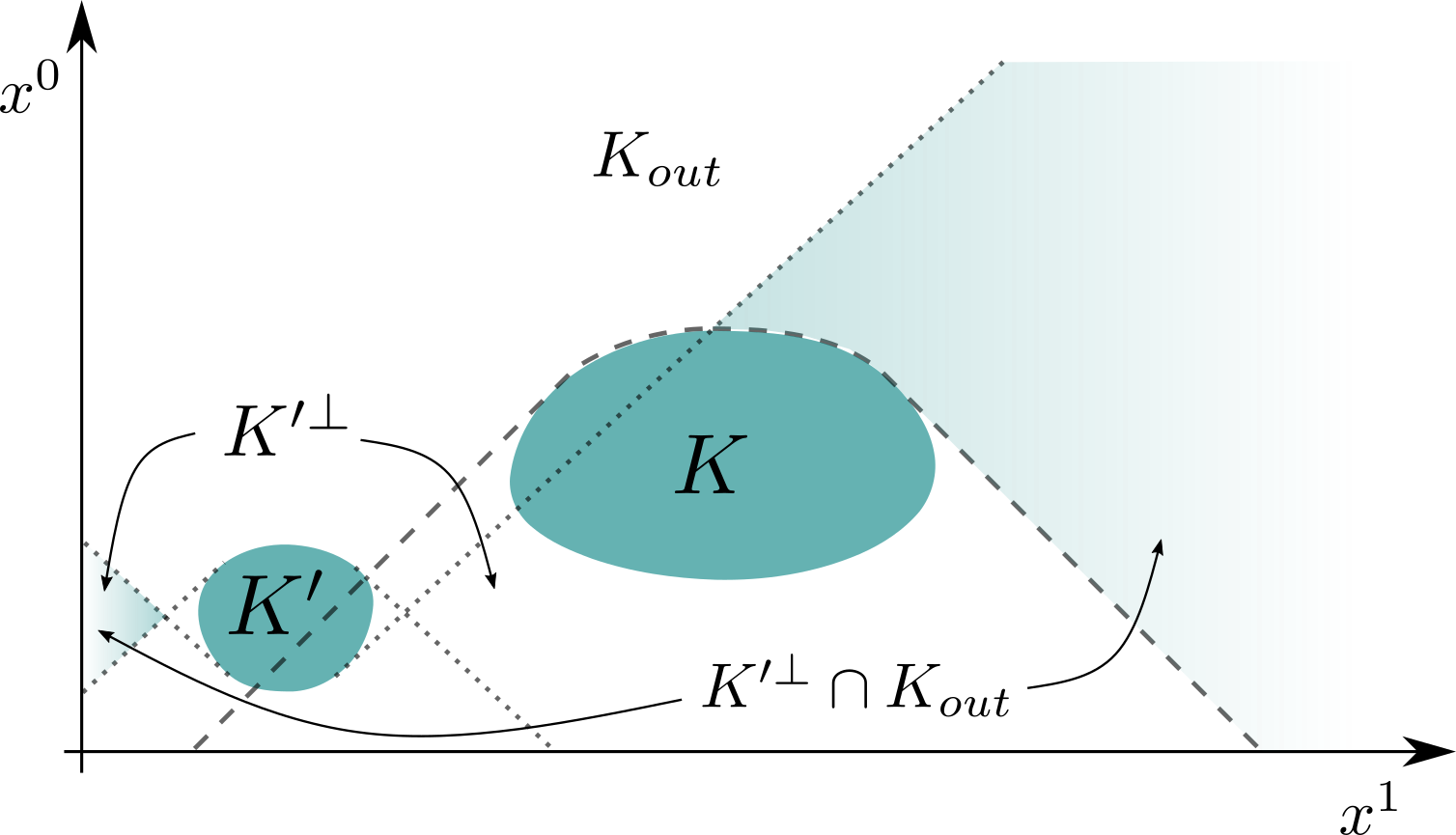}
    \caption{Spacetime diagram of a compact subset $K$, its corresponding out-region $K_{out}$ (all points above the dashed line), and a compact subset $K'\subset K_{in}$ (note that $K_{in}$ is not shown, but it should be clear that $K'$ is not in the future of $K$). The dotted lines illustrate the causal complement $K'^{\perp}$, and the areas shaded with a gradient show the intersection $K'^{\perp}\cap K_{out}$ used in the definition of the term \emph{causal w.r.t.} This definition encodes the fact that any expectation values measured in $K'^{\perp}\cap K_{out}$, and hence measured in a region spacelike to $K'$, should only depend on the map $\mathcal{E}(\cdot )$ local to $K$ and not on $\mathcal{E}'(\cdot )$ local to $K'$.}
    \label{fig:K_prime_intersect_K_out}
\end{figure}

In other words, $\mathcal{E}'(\cdot)$ drops out when the pair of maps act on operators localisable in $K_{out}$ and spacelike to $K'$. This implies that $\mathcal{E}'(\cdot)$ drops out of any expectation values of operators $Y\in\mathfrak{A}(K'^{\perp}\cap K_{out})$, i.e. $\text{tr}(\rho \mathcal{E}'(\mathcal{E}(Y))) = \text{tr}(\rho \mathcal{E}(Y))$. See Fig.~\ref{fig:K_prime_intersect_K_out} for an illustration of the intersection $K'^{\perp}\cap K_{out}$ used in the definition. Similarly to above, if we impose this condition on expectation values for all states we find that it must be true at the operator level, that is, we arrive at~\eqref{eq:causal_wrt_conditions}.

In the following we remove the dependence of the map $\mathcal{E}'(\cdot)$.
\begin{definition}[Definition (strongly causal)]
An update map $\mathcal{E}(\cdot)$, local to a compact subset $K$, is \emph{strongly causal} if it is causal w.r.t. all maps $\mathcal{E}'(\cdot)$ local to all compact $K'\subset K_{in}$.
\end{definition}

We use the term ``strongly causal'' (and apologise for doing so, given the standard meaning in Lorentzian geometry) because this property may seem too strong at a first glance. For instance, it could be too much to ask of a map to be causal w.r.t. all local maps, especially if those local maps are themselves not causal w.r.t. some other maps. With this in mind we make the following weaker definition.
\begin{definition}[Definition (weakly causal)]
An update map $\mathcal{E}(\cdot)$, local to a compact subset $K$, is \emph{weakly causal} if it is causal w.r.t. all strongly causal maps $\mathcal{E}'(\cdot)$ local to all compact $K'\subset K_{in}$.
\end{definition}

Any strongly causal map is causal w.r.t. all local maps, and so is clearly causal w.r.t. the subset of local maps which are strongly causal themselves. That is, any strongly causal map is also weakly causal; hence why the latter condition is weaker.

It seems physically reasonable to think that strong causality is as strong as it gets for update maps, since, on the contrary, it seems physically \emph{unreasonable} to demand that a map $\mathcal{E}(\cdot)$ is causal w.r.t. maps that are not even local (as well as all local maps). Strong causality being the strongest condition then implies that weak causality is the weakest condition, as to define a weaker condition on a map $\mathcal{E}(\cdot)$ requires a smaller set of local maps (smaller than the set of strongly causal maps) with which $\mathcal{E}(\cdot)$ must be causal w.r.t. In this way strong and weak causality seem to determine natural upper and lower limits of what one can expect from causality respecting maps under composition.

In Section~\ref{sec:Kicking causality conditions into shape} we will sketch an argument as to why strong and weak causality are in fact the same, and hence we will simply refer to maps as \emph{causal} if they satisfy strong/weak causality. Furthermore, we will also argue that the causal maps are precisely those that have the physically intuitive \emph{past-support non-increasing} (PSNI) property, where
\begin{definition}[Definition (PSNI)]
An update map $\mathcal{E}(\cdot)$, local to a compact subset $K$, is \emph{past-support non-increasing} (PSNI) if it satisfies
\begin{equation}\label{eq:psni_condition}
\mathcal{E}(\mathfrak{A}(R_+)) \subseteq \mathfrak{A}(R_-) \; ,
\end{equation}
for all regions $R_+\subseteq K_{out}$ and $R_-\subseteq K_{in}$ with $\overline{R}_+ \subset D(R_-)$.
\end{definition}

\begin{figure}
    \centering
    \includegraphics[scale=0.7]{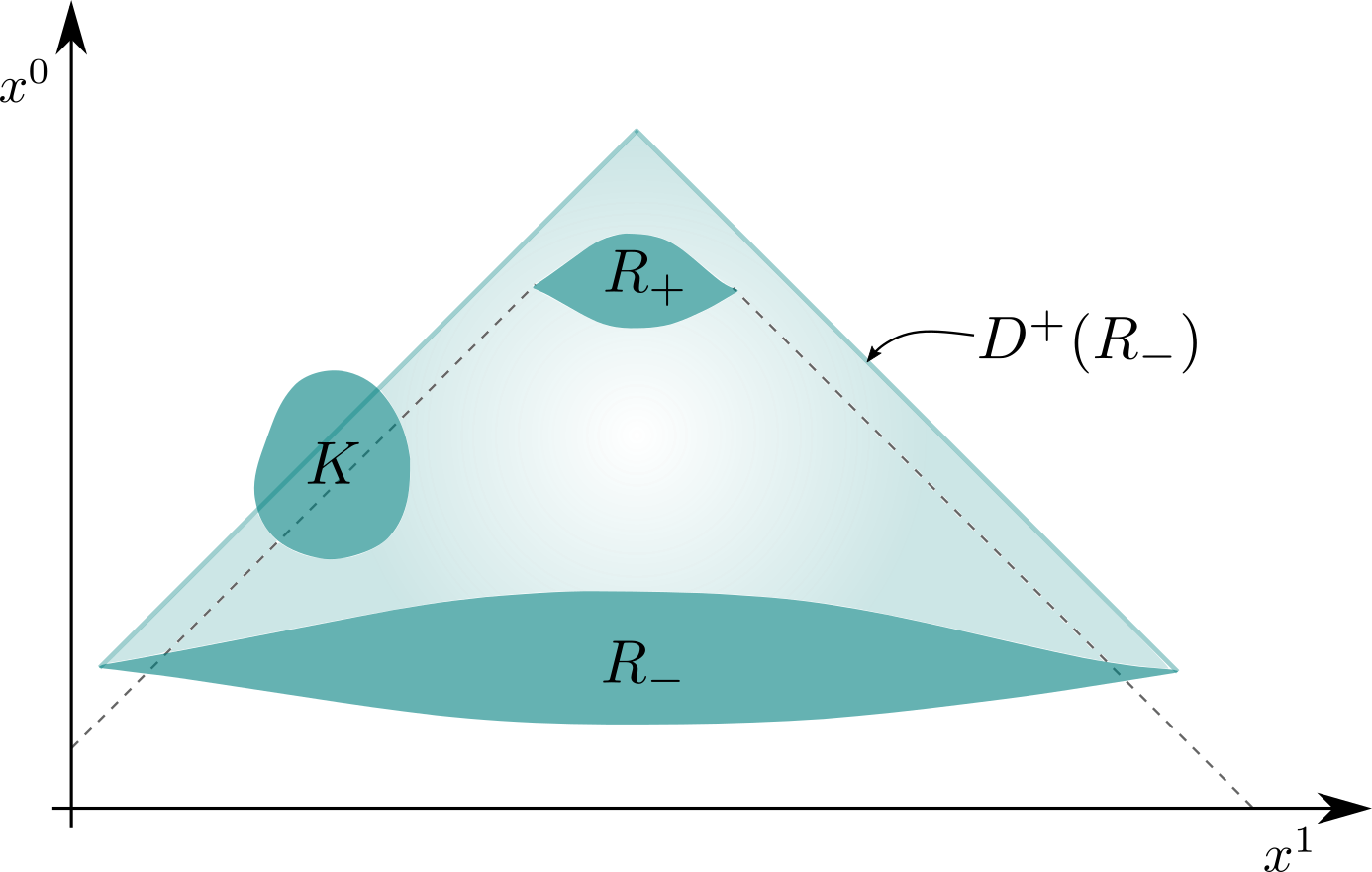}
    \caption{Illustration of the \emph{past-support non-increasing} (PSNI) property for an update map $\mathcal{E}(\cdot)$ local to some compact $K$. For any region $R_+$ in the out-region for $K$, and any region $R_-$ in the in-region for $K$, such that the closure of $R_+$ is contained in the domain of dependence of $R_-$ (as can be seen in the figure), then~\eqref{eq:psni_condition} must be satisfied for $\mathcal{E}(\cdot)$ to be PSNI. Heuristically, $\mathcal{E}(\cdot)$ cannot `push' operators outside their past lightcone.}
    \label{fig:psni}
\end{figure} 

This is physically intuitive as for any operator $X$, localisable in some region $R_+$ within the out-region for $\mathcal{E}(\cdot)$, if we localise $\mathcal{E}(X)$ in the in-region for $\mathcal{E}(\cdot)$, then the PSNI property says that its support is not pushed outside the past lightcone of $R_+$. Conversely, imagine for some $R_+\subseteq K_{out}$ there exists some $R_-\subseteq K_{in}$ with $\overline{R}_+\subset D(R_-)$ for which $\mathcal{E}(\mathfrak{A}(R_+)) \nsubseteq \mathfrak{A}(R_-)$. Any valid localisation region containing a Cauchy surface for $R_-$, that is, any $R'_-\subseteq K_{in}$ with $D(R_-)\subseteq D(R'_-)$ and $\mathcal{E}(\mathfrak{A}(R_+)) \subseteq \mathfrak{A}(R'_-)$, must be strictly larger in spatial extent than $R_-$, in the sense that $D(R_-)\subset D(R'_-)$. This is required since we need $\mathfrak{A}(R_-)\subset \mathfrak{A}(R'_-)$ to have any hope of localising $\mathcal{E}(\mathfrak{A}(R_+))$ in $R'_-$. Given that $\overline{R}_+\subset D(R_-)$, $R_-$ necessarily extends outside the past lightcone of $\overline{R}_+$, and hence any such $R'_-$ must do as well, e.g Fig.~\ref{fig:psni}. This PSNI property is almost exactly that given in~\cite{Fewster2020,Bostelmann_2021} in the case of a scattering map for an interaction of a quantum field with another probe quantum field.

One can consider the $n$-map generalisation of~\eqref{eq:causal_wrt_conditions}, where a given map drops out if it is spacelike to the operator that the composition acts on. For a sequence of $n$ PSNI maps $\mathcal{E}_1(\cdot), ... , \mathcal{E}_n(\cdot)$, local to $K_1 , ... , K_n$ respectively, where $K_r \subset (K_s)_{in}$ whenever $r<s$, this $n$-map generalisation of~\eqref{eq:causal_wrt_conditions} is satisfied. This can be shown by starting from the inner-most map and working outwards. Applying any map can only change the operator it acts on in a way that keeps its support in the past lightcone of $X$, and hence, at any stage of the composition, the application of a map spacelike to $X$ will be trivial. Given this $n$-map property follows from the PSNI property, which (we will show) follows from the $2$-map causality conditions above, we see that further $n$-map causality conditions for $n>2$ are redundant.

It should also be noted that the above definitions (local, strongly/weakly causal, and PSNI) apply more generally to maps on the physical subalgebra of a complex scalar or fermionic QFT. That being said, the argument in Section~\ref{sec:Kicking causality conditions into shape} for the equivalence of the different causality conditions does not generalise so straightforwardly, and is left for future work.

\subsection{Analogous picture in lattice systems}\label{sec:Analogous picture in finite systems}

The above locality and causality conditions on update maps are even more transparent in a lattice, or multipartite, system.

Consider a lattice of $N$ sites. Recall that an operator $X$ is local to site $n$ if it can be expressed as $X = \mathds{1}_1 \otimes ... \otimes \mathds{1}_{n-1}\otimes X_n \otimes ... \otimes \mathds{1}_N$, i.e. it is only non-trivial (not the identity) on site $n$. Similarly, an update map, $\mathcal{E}(\cdot)$ is local to some subset of sites if, when expressed in terms of operators, it is only non-trivial (not the identity) on those sites.

Furthermore, $\mathcal{E}(\cdot)$ is PSNI if, for any operator $X$ local to some subset of sites, $\mathcal{E}(X)$ is also local to the same subset of sites; otherwise the support of $X$ has been increased.

Update maps that increase support can be used to (subluminally) signal between parts of a multipartite system~\cite{Popescu_1994,PhysRevA.64.052309,borsten2021impossible}, and are routinely considered in QI. Causality is not violated in these cases because any experimental realisations of the update maps take at least the light-travel time between the sites to complete. 

That is, in NRQM if an update map is support increasing, or signalling, we do not need to rule it out as physically unrealisable by any experiment. It is only physically unrealisable on time-scales shorter than the relevant light-travel time. On the other hand, in QFT certain update maps must be ruled out completely. Essentially, the relativistic setting necessitates the specification of the spacetime regions in which any quantum operations take place. The causal relations between the specified regions then, potentially, adds additional constraints (e.g.~\eqref{eq:causal_wrt_conditions}) on the physically allowed update maps.

\section{Unitary kicks}\label{sec:Unitary kicks}

\subsection{Smeared field kicks}\label{sec:Smeared field kicks}

Let us consider one of the simplest update maps --- a local unitary kick. Specifically, for some self-adjoint operator $A$, localisable in some compact $K$, we consider the map
\begin{equation}\label{eq:unitary_kick_map_defintion}
B \mapsto \mathcal{U}_{A}(B) = e^{i A}B e^{-i A} \; ,
\end{equation}
for any operator $B\in\mathfrak{A}$. Clearly, if $B$ is localisable in a region spacelike to $K$, then $[A,B]=0$ and hence $\mathcal{U}_{A}(B)= B$. Thus, the map $\mathcal{U}_{A}(\cdot)$ is local to $K$ for any $A$ in $K$. It is also clear that $\mathcal{U}_{A}(\mathds{1})=\mathds{1}$. First we consider the simplest case of smeared field kick, i.e. $A=\phi(f)$ for $f$ supported in $K$.

For any region $R$, the subalgebra $\mathfrak{A}(R)$ is generated by algebraic combinations of smeared fields in $R$, i.e. by smeared fields $\phi(g)$ for test functions $g$ supported in $R$.

Alternatively, one can construct any $B\in\mathfrak{A}(R)$ through suitable derivatives of linear combinations of the unitary Weyl generators $e^{i \phi(g)}$, where again $g$ is any test function supported in $R$~\footnote{Note that these two approaches are only equivalent given a suitable representation of the algebra. We meet this requirement through our use of the usual bosonic Fock space and our implicit assumption of an appropriate ground state on which the Fock space is built.}. For instance, to recover the smeared field $\phi(g)$ we can consider the 1-parameter family of Weyl generators $e^{i t\phi( g)} = e^{i \phi( t.g )}$, where $t\in\mathbb{R}$. We can then write $\phi(g) = -i \partial_t ( e^{i t\phi( g)} )\big|_{t=0}$. Similarly, for the square we have $\phi(g)^2 = -\partial^2_t ( e^{i t\phi( g)} )\big|_{t=0}$. To recover a product of two smeared fields, $\phi(g) \phi(h)$ say, we instead consider the 2-parameter family of Weyl generators $e^{i\phi(t. g + s. h)} = e^{i t\phi(g)+i s \phi(h)}$, where $t,s\in\mathbb{R}$. Using the Baker-Campbell-Hausdorff (BCH) formula~\cite{Baker,Campbell,Hausdorff,Casas_2009} and the CCR's in~\eqref{eq:field_CCR} one can verify that
\begin{equation}
e^{i t\phi(g)+i s \phi(h)} = e^{\frac{i}{2}\Delta(g,h)ts}e^{i t\phi(g)} e^{i s \phi(h)} \; ,
\end{equation}
and hence
\begin{align}
-\partial_t & \partial_s ( e^{-\frac{i}{2}\Delta(g,h)ts} e^{i t\phi(t . g+s.h)} )\big|_{t=0,s=0} 
\nonumber
\\
& =  -\partial_t\partial_s ( e^{i t\phi(g)} e^{i s \phi(h)} )\big|_{t=0,s=0} 
\nonumber
\\
& = \left( -i \partial_t ( e^{i t\phi( g)} )\big|_{t=0} \right)\left( -i \partial_t ( e^{i s\phi( h)} )\big|_{s=0} \right)
\nonumber
\\
& = \phi(g) \phi(h) \; ,
\end{align}
as desired. In a similar fashion one can recover any $B\in\mathfrak{A}(R)$, i.e. any algebraic combination of smeared fields, through appropriate complex sums and derivatives of Weyl generators.

We will say that any complex linear combination of Weyl generators, e.g. $\sum_{i=1}^n c_i e^{i\phi(g_i )}$ for $c_i\in\mathbb{C}$, is localisable in a given region $R$ if it contains the supports of all the functions $g_i$. This ensures that localisation regions are unchanged when taking derivates of the Weyl generators to recover sums and products of smeared fields.


Since $\mathcal{U}_{A}(\cdot)$ is linear we can also recover $\mathcal{U}_{A}(B)$, for any $B\in\mathfrak{A}(R)$, through suitable complex sums and derivatives of terms of the form $\mathcal{U}_{A}(e^{i t\phi(g)})$. Therefore, we need only determine the action of $\mathcal{U}_{A}(\cdot)$ on a general Weyl generator $e^{i t \phi(g)}$.

Using the Baker–Campbell–Hausdorff (BCH) formula one can verify that
\begin{align}\label{eq:smeared_field_kick_on_weyl}
\mathcal{U}_{\phi(f)}(e^{i t\phi(g)}) & = e^{i \phi(f)}e^{i t\phi(g)} e^{-i \phi(f)}
\nonumber
\\
& = e^{i t\Delta(g,f)}e^{it\phi(g)} \; .
\end{align}
Since the RHS is proportional to the original Weyl generator $e^{i t\phi(g)}$, which is localisable in $R$, we see that $\mathcal{U}_{\phi(f)}(e^{i t\phi(g)})$ is also localisable in $R$. From this we see that if $B$ is any complex sum of Weyl generators in $\mathfrak{A}(R)$, then $\mathcal{U}_{\phi(f)}(B) \in\mathfrak{A}(R)$, and hence $\mathcal{U}_{\phi(f)}(\mathfrak{A}(R))\subseteq \mathfrak{A}(R)$. This then implies that, for any $\mathcal{E}'(\cdot)$ local to some compact $K'\subset K_{in}$,
\begin{equation}
\mathcal{E}'( \mathcal{U}_{\phi(f)}( B ) ) = \mathcal{U}_{\phi(f)}( B ) \; ,
\end{equation}
for any $B$ localisable in $K'^{\perp}\cap K_{out}$. This follows as $\mathcal{U}_{\phi(f)}( \cdot )$ does not change the localisation of $B$, thus $ \mathcal{U}_{\phi(f)}( B )$ is also localisable in $K'^{\perp}\cap K_{out}\subseteq K'^{\perp}$, and hence also localisable in $K'^{\perp}$. Since $\mathcal{E}'( \cdot )$ is local to $K'$, $\mathcal{E}'( \cdot )$ then acts trivially on $\mathcal{U}_{\phi(f)}( B )$.

This argument holds for all maps local to any compact $K'\subset K_{in}$, and hence $\mathcal{U}_{\phi(f)}( \cdot )$ is \emph{strongly causal}. Since this is the strongest (physically reasonable) causality condition we can impose on a given map, there is no reason (at least at the level of the theory) to think that $\mathcal{U}_{\phi(f)}( \cdot )$ is not physically realisable in experiments. Of course, this may not be surprising to many readers given the simple form of the map.

\subsection{Kicking causality conditions into shape}\label{sec:Kicking causality conditions into shape}

We will now use these strongly causal smeared field kicks to show that any weakly causal map satisfies the PSNI property. Following this we will show that the PSNI property implies strong causality. Since strongly causal maps are also weakly causal, this implies that weak causality is in fact equivalent to strong causality, and to the PSNI property. Thus, for real scalar QFT, causal maps are precisely those that are PSNI.

\subsubsection{\emph{Weak causality} $\Rightarrow$ \emph{PSNI} for real scalar QFT}

Recall that a weakly causal map $\mathcal{E}(\cdot)$ is causal w.r.t. all strongly causal maps. This means any weakly causal map, local to some compact $K$, must be causal w.r.t. all smeared field kicks $\mathcal{U}_{\phi(f)}(\cdot)$ with $f$ compactly supported in $K_{in}$. That is,
\begin{equation}\label{eq:causal_wrt_smeared_field_kicks}
\mathcal{U}_{\phi(f)} ( \mathcal{E} (\cdot)) \big|_{\mathfrak{A}(K'^{\perp}\cap K_{out})}=\mathcal{E} (\cdot)  \; ,
\end{equation}
for all $f$ supported in any compact $K'\subset K_{in}$. In particular, this implies that $\mathcal{E}(\cdot)$ is causal w.r.t. the 1-parameter family of smeared field kicks $\mathcal{U}_{\phi(\lambda . f)} (\cdot) = \mathcal{U}_{\lambda\phi(f)} (\cdot)$ for $\lambda\in\mathbb{R}$. Substituting $\mathcal{U}_{\lambda\phi(f)} (\cdot)$ into~\eqref{eq:causal_wrt_smeared_field_kicks} we see that the RHS does not depend on $\lambda$, and hence derivatives w.r.t. $\lambda$ must kill both sides. Using this as a condition on the LHS gives, for any $B\in\mathfrak{A}(K'^{\perp}\cap K_{out})$,
\begin{align}
0 & = i\partial_{\lambda} \left(  \mathcal{U}_{\lambda\phi(f_1)} ( \mathcal{E} (B)) \right)\big|_{\lambda = 0}
\nonumber
\\
 & = 
i\partial_{\lambda} \left( e^{i\lambda \phi(f_1)} \mathcal{E} (B)) e^{-i\lambda \phi(f_1)} \right)\big|_{\lambda = 0}
\nonumber
\\
& = [\mathcal{E}(B),\phi(f)] \; .
\end{align}
The vanishing of this commutator is also sufficient for~\eqref{eq:causal_wrt_smeared_field_kicks}, as it implies that $\mathcal{E}(B)$ commutes with $e^{i\phi(f)}$, and hence $\mathcal{U}_{\phi(f)} (\mathcal{E}(B) ) = \mathcal{U}_{\phi(f)} (\mathds{1}) \mathcal{E}(B) = \mathcal{E}(B)$.

In short, for any weakly causal map $\mathcal{E}(\cdot)$ (local to compact $K$), any $f$ supported in any compact $K'\subset K_{in}$, and any $B$ localisable in $K'^{\perp}\cap K_{out}$, we have $[\mathcal{E}(B),\phi(f)]=0$. For any region $R\subset K'$, any operator $A\in\mathfrak{A}(R)$ is some algebraic combination of smeared fields in $K'$, and hence $[\mathcal{E}(B),A]=0$. Since this is true for all $A$ and $B$, localisable in their respective regions, we get $[\mathcal{E}(\mathfrak{A}(K'^{\perp}\cap K_{out})),\mathfrak{A}(R)]=0$. In other words,
\begin{equation}\label{eq:psni_commutant}
\mathcal{E}(\mathfrak{A}(K'^{\perp}\cap K_{out})) \subseteq \mathfrak{A}(R)^{\perp} \; ,
\end{equation}
for any region $R\subset K'$, and any compact $K'\subset K_{in}$. To help visualise these subsets one can imagine Fig.~\ref{fig:K_prime_intersect_K_out} but with the addition of a region $R\subset K'$.

\begin{figure}
    \centering
    \includegraphics[scale=0.7]{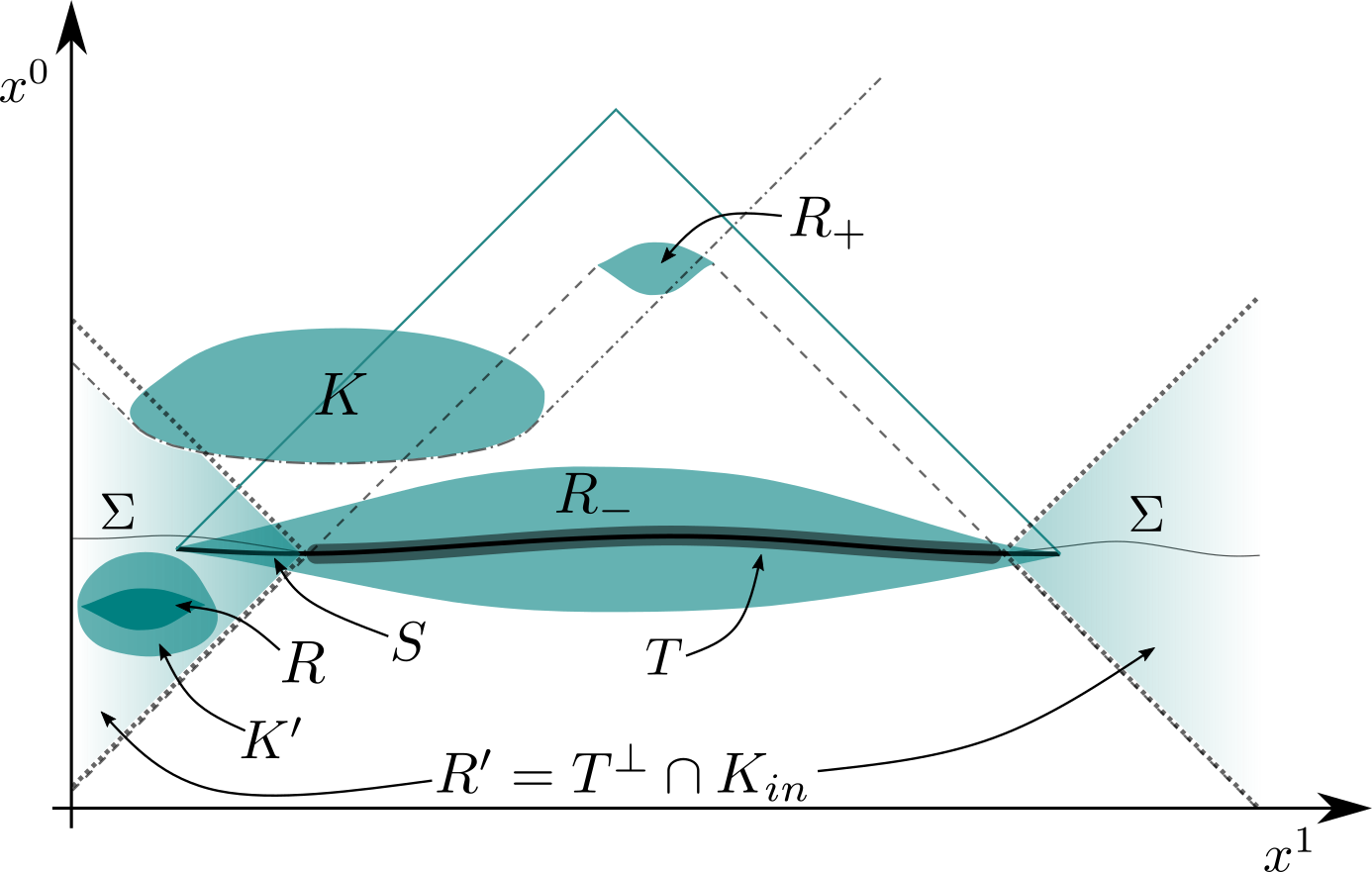}
    \caption{An illustration of the surfaces $\Sigma$ (thin line), $S$ (medium thickness line), and $T$ (thick line lying over the part of $S$ to the past of $\overline{R}_+$). The past-directed lightlike dashed lines from $R_+$ indicate the boundaries of the past set $J^-(\overline{R}_+)$ (note $J^-(\overline{R}_+)$ contains these boundary points). The region $T^{\perp}$ consists of all points to the left and right of the the dotted lightlike lines (not including the dotted lines) emanating from the endpoints of $T$. The region $K_{in}$ lies strictly to the past of the dotted and dashed line. The area shaded with a gradient is the region $R' = T^{\perp}\cap K_{in}$ used in the proof. Note that the position of $K$ in this example means that $R'$ is not the entirety of $T^{\perp}$. This is not always the case, however, and for certain setups $R'=T^{\perp}$. Finally, examples of the subsets $K'\subset R'$ and $R\subset K'$ have been shown to help illustrate the proof.}
    \label{fig:psni_proof}
\end{figure}

Importantly,~\eqref{eq:psni_commutant} is true if we pick $K'$ in a way that matches the setup of the PSNI property. To do this we first pick any pair of regions $R_+\subseteq K_{out}$ and $R_- \subseteq K_{in}$ such that $\overline{R}_+\subset D(R_- )$ (\textit{c.f.} the PSNI condition). Since $R_-$ is a region it is globally hyperbolic in its own right, and hence it has a Cauchy surface $S$. Given that $\overline{R}_+\subset D(R_- )$, we know that all past causal curves from $\overline{R}_+$ pass through $S$, specifically through the surface $T = J^-(\overline{R}_+)\cap S$. Note that $T\subset S\subset R_-$. See Fig.~\ref{fig:psni_proof} for an illustration of these surfaces.

Working toward~\eqref{eq:psni_commutant} we now define the region $R' = T^{\perp}\cap K_{in} \subseteq T^{\perp}$ (see Fig.~\ref{fig:psni_proof}), which we will use shortly to introduce the subsets $K'$ and $R\subset K'$ that appear in~\eqref{eq:psni_commutant}.

Before doing that, however, it will be useful to show that $\mathfrak{A}(R') = \mathfrak{A}(T^{\perp})$, which can be seen as follows. As $R'$ is a region it contains a Cauchy surface $\Sigma$. Importantly, $\Sigma$ is also a Cauchy surface for $T^{\perp}$, since any inextendible timelike curve $\gamma\subset T^{\perp}$ will either be entirely contained in $R'$ (in which case it must intersect $\Sigma$ by virtue of it being Cauchy surface for $R'$) or it passes into $J^+(K)$ (note it cannot lie entirely in $J^+(K)$ as $K$ is compact and $\gamma$ must be past inextendible), in which case the curve $\gamma' = \gamma\setminus J^+(K)$ is a timelike curve in $R'$ (with no endpoints in $R'$, and hence inextendible in $R'$) and hence intersects $\Sigma$, thus implying $\gamma$ also intersects $\Sigma$. One can get an intuition for this result via Fig.~\ref{fig:psni_proof}. By the time-slice property of subalgebras we then have $\mathfrak{A}(R') = \mathfrak{A}(T^{\perp})$ as desired.

We are now ready to apply~\eqref{eq:psni_commutant}. Specifically, if we pick any compact $K'\subset R'$, and any region $R\subset K'$ (see Fig.~\ref{fig:psni_proof}), equation~\eqref{eq:psni_commutant} holds. From the setup we also have that $\overline{R}_+\subseteq K'^{\perp}$, and in particular, $R_+\subseteq K'^{\perp}\cap K_{out}$, since $R_+\subseteq K_{out}$ by construction. Therefore, $\mathfrak{A}(R_+)\subseteq \mathfrak{A}(K'^{\perp}\cap K_{out})$, and hence $\mathcal{E}(\mathfrak{A}( R_+ ))\subseteq\mathcal{E}(\mathfrak{A}(K'^{\perp}\cap K_{out}))\subseteq \mathfrak{A}(R)^{\perp}$ by~\eqref{eq:psni_commutant}. Since this is true for all bounded regions $R\subset R'$ we have $\mathcal{E}(\mathfrak{A}( R_+ ))\subseteq\mathfrak{A}(R')^{\perp} = \mathfrak{A}(T^{\perp})^{\perp}$. Finally, using the Haag property, we know that $\mathfrak{A}(T^{\perp})^{\perp}\subseteq \mathfrak{A}(R_-)$, since $R_-$ is a region containing $T$. Therefore, $\mathcal{E}(\mathfrak{A}( R_+ ))\subseteq\mathfrak{A}(R_-)$, \textit{c.f.} the PSNI condition.

\subsubsection{\emph{PSNI} $\Rightarrow$ \emph{strong causality}}

Above we used strongly causal smeared fields kicks to show that any weakly causal map is PSNI. To complete the argument that weak and strong causality are the same in real scalar QFT, and equivalent to PSNI, we will now show that PSNI implies strong causality. Such an argument renders the three properties equivalent as strong causality already implies weak causality.

First, we pick any compact $K'\subset K_{in}$. For any region $R_+\subset K_{out}$, with compact closure $\overline{R}_+ \subset K'^{\perp}$, we follow Lemma's 3 and 4 of~\cite{Bostelmann_2021} to show that $\overline{R}_+$ is contained in the domain of dependence of the region $K'^{\perp}\cap K_{in}$. Setting $R_- = K'^{\perp}\cap K_{in}$ we then have $\mathcal{E}( \mathfrak{A}(R_+) ) \subseteq \mathfrak{A}(K'^{\perp}\cap K_{in})$ by PSNI. As $K'^{\perp}\cap K_{in}\subseteq K'^{\perp}$ we then have $\mathcal{E}( \mathfrak{A}(R_+) ) \subseteq \mathfrak{A}(K'^{\perp})$, and hence any map $\mathcal{E}'(\cdot )$ local to $K'$ act trivially on $\mathcal{E}( \mathfrak{A}(R_+) )$. This argument holds for any region $R_+\subset K_{out}$ whose compact closure is spacelike to $K'$, and hence it holds on the subalgebra $\mathfrak{A}(K'^{\perp}\cap K_{out})$. That is, any map $\mathcal{E}'(\cdot )$ local to $K'$ acts trivially on $\mathcal{E}(\mathfrak{A}(K'^{\perp}\cap K_{out}) )$. This is precisely the condition of strong causality, and thus PSNI implies strong causality.

Notably, this last argument that PSNI implies strong causality uses only the spacetime causal structure and the basic properties of a physical algebra in AQFT, namely Einstein causality, isotony, and the time-slice property. Thus, the argument straightforwardly generalises to the physical subalgebras of complex scalar and fermionic QFT's. That is, the chain of implications \emph{PSNI} $\Rightarrow$ \emph{strong causality} $\Rightarrow$ \emph{weak causality} holds more generally. The implication \emph{weak causality} $\Rightarrow$ \emph{PSNI}, on the other hand, was only shown above for real scalar fields. We leave the extension of this latter implication to other fields for future work.

Going forward we will consider unitary kicks with other, more complicated operators in real scalar QFT, and then Gaussian measurement maps and other 1-parameter families of Kraus operators in Section~\ref{sec:Measurements}. In every case we will determine if the respective map is causal using the PSNI property. In many cases the update maps will not be PSNI, despite being local. In such cases the update maps must be ruled out as physically impossible to implement in experiments. Conversely, to implement them is to open the door to potential causality violations.

\subsection{Other unitary kicks}\label{sec:Other unitary kicks}

Let us now consider the slightly more complicated case of $A = \phi(f)^2$. Using the BCH formula again, one can verify that $\mathcal{U}_{\phi(f)^2}(\cdot)$ acts on the 1-parameter family of Weyl generators, $e^{i t\phi(g)}$, as
\begin{equation}\label{eq:squared_kick_on_weyl}
\mathcal{U}_{\phi(f)^2}(e^{i t\phi(g)}) = e^{-i t^2\Delta(f , g)^2}e^{-i 2 t\Delta(f ,g) \phi(f)}e^{i t \phi(g)} \; .
\end{equation}
If $f$ is supported in some compact $K\subset M$, and $g$ is supported in some compact $K'\subset K_{out}$, then $\Delta(f,g)$ is not necessarily zero. In such a case $\mathcal{U}_{\phi(f)^2}(e^{i t\phi(g)})$ may not be localisable in the support of $g$, as it now depends on $\phi(f)$, which may not be localisable in $\text{supp}g$ if $\text{supp}f\nsubseteq \text{supp}g$. Crucially, $\text{supp}f$ may contain points that are spacelike to $\text{supp}g$, and hence the past-support can now include the past lightcone of the support of $f$. Thus $\mathcal{U}_{\phi(f)^2}(\cdot)$ is not PSNI, and hence not causal. 

For clarity let us show explicitly how a causality violation can arise. Consider the setup in Fig.~\ref{fig:protocol}: two compact subsets $K,K'$, and some region $R$, such that $K$ is spacelike to the compact closure of $R$, $K'\subset K_{in}$ and $R\subset K_{out}$. Consider three test functions $f,g,h$, where $f$ is supported in $K$, $g$ in $R$, and $h$ in $K'$. Let the initial state of the system be $\rho$.

Now consider three independent agents, Alice, Charlie, and Bob, who perform actions in $K'$, $K$, and $R$ respectively. Alice kicks in $K'$ with the smeared field $\phi(\lambda . h) = \lambda\phi(h)$ (for some kick strength $\lambda\in\mathbb{R}$), and hence the state gets updated as $\rho \mapsto \rho'=\tilde{\mathcal{U}}_{\phi(h)}(\rho)$, where we have used the dual map $\tilde{\mathcal{U}}_{\phi(h)}(\cdot)$ since we are updating the state. In $K$ Charlie enacts the operation under question, $\mathcal{U}_{\phi(f)^2}(\cdot)$, and hence the state is further updated as $\rho' \mapsto \rho'' =  \tilde{\mathcal{U}}_{\phi(f)^2}(\rho' )$. Note that Charlie's update is applied after Alice's, as $K'$ has some points to the past of $K$ (if $K$ is spacelike to $K'$ then the order does not matter as the maps commute). In $R$ Bob measures the expected value of $\phi(g)$, which is given by
\begin{equation}
\text{tr}(\rho''\phi(g) ) = \text{tr}(\rho \mathcal{U}_{\lambda\phi(h)}(\mathcal{U}_{\phi(f)^2}(\phi(g))) ) \; ,
\end{equation}
where we have reverted to the update maps on the operators instead of the state. Focussing on $\mathcal{U}_{\phi(f)^2}(\phi(g))$ we have
\begin{align}
& \mathcal{U}_{\phi(f)^2} (\phi(g)) 
\nonumber
\\
& = -i \partial_t \left( \mathcal{U}_{\phi(f)^2}(e^{i t\phi(g)}) \right)\Big|_{t=0}
\nonumber
\\
& = -i \partial_t \left( e^{-i t^2\Delta(f , g)^2}e^{-i 2 t\Delta(f ,g) \phi(f)}e^{i t \phi(g)} \right)\big|_{t=0}
\nonumber
\\
& = \phi(g) - 2 \Delta(f ,g) \phi(f) \, ,
\end{align}
using~\eqref{eq:squared_kick_on_weyl}. Since $R$ is not spacelike to $K$, $\Delta(f ,g)$ is non-zero in general. To compute Bob's expectation value we then need to act with Alice's kick, $\mathcal{U}_{\lambda\phi(h)}(\cdot)$, where we recall that $[\phi(h),\phi(g)]=0$ (since $K'$ is spacelike to $R$), and that $[\phi(h),\phi(f)]=i \Delta(h,f)$, which is non-zero in general since $K'$ is not spacelike to $K$. We find
\begin{align}
\mathcal{U}_{\lambda\phi(h)} & ( \mathcal{U}_{\phi(f)^2} (\phi(g))) 
\nonumber
\\
& = \mathcal{U}_{\lambda\phi(h)}\big(\phi(g) - 2 \Delta(f ,g) \phi(f) \big)
\nonumber
\\
& =\phi(g)- 2 \Delta(f ,g) \mathcal{U}_{\lambda\phi(h)} (\phi(f)) 
\nonumber
\\
& =\phi(g) +  2 i \Delta(f ,g) \partial_t \left( \mathcal{U}_{\lambda\phi(h)}(e^{i t \phi(f)}) \right)\Big|_{t=0}
\nonumber
\\
& =\phi(g) +  2 i \Delta(f ,g)\partial_t \left( e^{i t \lambda \Delta(f,h)}e^{it\phi(f)} \right)\Big|_{t=0}
\nonumber
\\
& =\phi(g) -  2 \Delta(f ,g) \left( \phi(f) + \lambda \Delta(f,h) \right) \, ,
\end{align}
where we have used~\eqref{eq:smeared_field_kick_on_weyl} and the fact that $[\phi(h),\phi(g)]=0$ to say that Alice's kick acts trivially on $\phi(g)$. To achieve a violation of causality we now need to show that there is some initial state $\rho$ for which Bob's expectation value depends on $\lambda$ --- the strength of Alice's kick. An obvious choice is the usual vacuum state, $\rho = \ket{\Omega}\bra{\Omega}$, for which odd $n$-point functions vanish, and hence $\text{tr}(\rho \phi(f))=0$ for any test function $f$. For Bob's expectation value we then find
\begin{align}
\text{tr} & (\rho''  \phi(g)) 
\nonumber
\\
& = \text{tr}\Big(\rho \big( \phi(g) -  2 \Delta(f ,g) \left( \phi(f) + \lambda \Delta(f,h) \right) \big) \Big)
\nonumber
\\
& = 2 \lambda \Delta(h,f) \Delta(f, g) \, .
\end{align}
The fact that this depends on $\lambda$ (Alice's kick strength) enables Alice to superluminally signal Bob, provided they have an agreed-upon `code', e.g. to send the bit `$\mathtt{0}$' Alice does not kick ($\lambda = 0$), which Bob can discern from the vanishing of his expectation value; to send the bit `$\mathtt{1}$' Alice kicks with some sufficient strength $(\lambda\neq 0$), which Bob can pick up from his non-zero expectation value. Note that this signal is statistical, since Bob picks it up at the level of an expectation value. In each realisation of the experiment this value will fluctuate. To make this protocol more robust to fluctuations many copies of the system can be setup in parallel, such that Bob receives a statistically significant amount of data in $R$ to be able to discern, up to some desired accuracy, whether his expected value vanishes or not.

Since Alice (in $K'$) and Bob (in $R$) are spacelike separated, this transmission of information is faster than light! This violates causality, and hence this protocol must be impossible. We know that Alice's kick is causal, and we assume Bob can measure his expectation value without violating causality (in Section~\ref{sec:Extracting expectation values from measurements of smeared fields} we will offer one way to achieve this). Therefore, the only conclusion we can draw is that Charlie's operation, $\mathcal{U}_{\phi(f)^2}(\cdot)$, is impossible to implement in $K$.

To avoid any causality violations we must therefore rule out the map $\mathcal{U}_{\phi(f)^2}(\cdot)$ as physically unrealisable via experiments. This may seem somewhat surprising, given that we have simply unitarily kicked with an operator that is localisable in some bounded region, i.e. $\phi(f)^2$, and given that analogous unitary kicks are standard in NRQM and lattice systems. We will comment further on why this distinction arises between NRQM and the relativistic setting of QFT in Section~\ref{sec:Discussion}.

One can further investigate unitary kicks with other self-adjoint operators. For many simple cases, where $A$ is not a sum of generators, i.e. not of the form $\phi(f)+ c\mathds{1}$ (for some $c\in\mathbb{R}$), we find that the unitary kicks increase past-support, similarly to $A=\phi(f)^2$. This suggests that \emph{only} kicks with generators (smeared fields and the identity) are permissible with respect to causality. We will not provide a more rigorous argument for this claim here. Instead, we postpone that more rigorous discussion for the next section, wherein we will argue the analogous conclusion that Gaussian measurements of smeared fields are the only permissible Gaussian measurements. The arguments used there can then be readily applied to unitary kicks.

\section{Measurements}\label{sec:Measurements}

\subsection{Preliminaries}\label{sec:Preliminaries}

Consider some self-adjoint operator $C\in \mathfrak{A}(R)$, localisable in the region $R$. Consider some function $G : \mathbb{R} \rightarrow \mathbb{C}$. Provided $G$ is a measurable function the operator $G(C)$ can be defined through functional calculus~\cite{reed1981functional}, even if $C$ is unbounded. Specifically, via the Spectral Theorem~\cite{reed1981functional} we can write $C$ using its projection-valued measure, $P_C$, as
\begin{equation}
C = \int_{\mathbb{R}} \lambda \, dP_C(\lambda) \; ,
\end{equation}
from which any measurable function of $C$ is defined as
\begin{equation}\label{eq:functional_calculus}
G(C) = \int_{\mathbb{R}} G(\lambda ) \, dP_C(\lambda) \; .
\end{equation}
Note that the projection-valued measure satisfies $\int_{\mathbb{R}}dP_C(\lambda) = \mathds{1}$.

For any measurable $G$ that is normalised in $L^2(\mathbb{R})$, that is,
\begin{equation}\label{eq:F_normalisation}
\int_{\mathbb{R}}d\alpha \, G(\alpha )^*G(\alpha ) = 1 \, ,
\end{equation}
we can define the corresponding update map for $C$:
\begin{equation}
\mathcal{E}^G_C (B) = \int_{\mathbb{R}} d\alpha \, G(C- \alpha )^{\dagger}B G(C - \alpha) \, ,
\end{equation}
for any $B\in\mathfrak{A}$. The operators $G(C-\alpha )$, for all $\alpha\in\mathbb{R}$, furnish a 1-parameter family of Kraus operators. One can also consider a discrete family of Kraus operators, e.g. a discrete set of projectors in an ideal measurement, but we will see it is more beneficial to consider a continuum of Kraus operators.

The fact that $G$ is normalised, and that~\eqref{eq:F_normalisation} holds even if we change $G(\alpha)^*G(\alpha)\mapsto G(c-\alpha )^*G(c-\alpha )$ for any $c\in\mathbb{R}$, implies that $\mathcal{E}^G_C(\mathds{1}) = \mathds{1}$. Furthermore, since $G(C - \alpha )$ commutes with everything that $C$ commutes with, we have that $\mathcal{E}^G_C (B) = B$ for all $B$ localisable in some region spacelike to $C$. That is, the map $\mathcal{E}^G_C(\cdot)$ is local for any choice of $G$. 

Note that the dual map that acts on the state $\rho$ is given by
\begin{equation}
\tilde{\mathcal{E}}^G_C (\rho) = \int_{\mathbb{R}} d\alpha \, G(C- \alpha )\rho G(C - \alpha)^{\dagger} \, .
\end{equation}

We will often consider the specific case of \emph{Gaussian measurements} involving the Gaussian Kraus operators
\begin{equation}
G^{\sigma}( C - \alpha ) = \frac{e^{-\frac{(C - \alpha )^2}{4\sigma^2}}}{(2\pi \sigma^2)^{\frac{1}{4}} } \; ,
\end{equation}
where $\sigma > 0$ is interpreted as the measurement accuracy. For convenience we denote the corresponding update map as $\mathcal{E}_C^{\sigma}(\cdot)$. Note that $G^{\sigma}( C - \alpha )^{\dagger}=G^{\sigma}( C - \alpha )$. Such Gaussian update maps are ubiquitous in weak measurements and continuous measurement models~\cite{Brun_2000,Jacobs_2006}.

$\mathcal{E}_C^{\sigma}(\cdot)$ describes a non-selective measurement, where no outcome is conditioned on. If one instead conditions on some measurement outcome, say $\alpha$ landing in some interval $[a,b]\subset\mathbb{R}$, then the associated \emph{selective} update map is
\begin{equation}
\mathcal{E}^{\sigma}_{C,[a,b]} (B) = \frac{1}{P_{[a,b]}}\int_{a}^{b}d\alpha \, 
G^{\sigma}(C-\alpha )\, B \, G^{\sigma}(C-\alpha ) \; ,
\end{equation}
where $P_{[a,b]}$ denotes the probability for $\alpha$ to land in the interval $[a,b]$. The appearance of $P_{[a,b]}$ in the denominator ensures that the updated state is normalised, i.e. $\text{tr}(\tilde{\mathcal{E}}^{\sigma}_{C,[a,b]} (\rho)) = 1$. The  probability $P_{[a,b]}$ is given by
\begin{align}
P_{[a,b]} & = \text{tr}\left(\rho  \int_a^b d\alpha\, {G^{\sigma}(C-\alpha )}^2 \right)
\nonumber
\\
& = \int_a^b d\alpha \, p(\alpha) \; ,
\end{align}
where we have written the last line in terms of the probability density function (pdf) for $\alpha$:
\begin{align}\label{eq:pdf_for_alpha}
p(\alpha) & = \text{tr}\left(\rho {G^{\sigma}(C-\alpha )}^2\right)
\nonumber
\\
& = \frac{1}{\sigma\sqrt{2\pi}} \text{tr}\left(\rho e^{-\frac{(C - \alpha)^2}{2\sigma^2}} \right) \; .
\end{align}
The average outcome is then given by the usual formula in terms of this pdf:
\begin{equation}
\mathbb{E}(\alpha) = \int_{\mathbb{R}}d\alpha \, \alpha \, p(\alpha) \; ,
\end{equation}
and similarly for higher moments.

If $C$ has eigenvectors in the Hilbert space, e.g. if $C$ is compact self-adjoint, then the $\sigma\rightarrow 0$ limit of $\mathcal{E}^{\sigma}_C (\cdot)$ describes an ideal measurement of $C$. That is, $\lim_{\sigma\rightarrow 0}\mathcal{E}^{\sigma}_C (\cdot) = \mathcal{E}^0_C (\cdot)$ (using the notation from~\eqref{eq:ideal_measurement}). To see this, we first spectrally decompose $C$ as
\begin{equation}
C = \sum_n c_n E_n \; ,
\end{equation}
where the sum runs over some countable set labelling the distinct eigenvalues $c_n$ and the associated orthogonal projectors $E_n$. For any function $F : \mathbb{R}\mapsto \mathbb{C}$ we can compute that same function of the operator $C$. Specifically, $F(C)$ is the sum over the projectors $E_n$ multiplied by that same function of the eigenvalues, $F(c_n)$. We can therefore write
\begin{equation}
G^{\sigma}(C-\alpha ) = \sum_n G^{\sigma}(c_n - \alpha ) E_n \; .
\end{equation}
The update map, acting on any operator $B$, then simplifies to
\begin{align}
\mathcal{E}^{\sigma}_C (B) & = \int_{-\infty}^{\infty}d\alpha \, \sum_{n,m} G^{\sigma}(c_n - \alpha )G^{\sigma}(c_m - \alpha ) E_n B E_m 
\nonumber
\\
& = \sum_{n,m} e^{-\frac{(c_n - c_m)^2}{8\sigma^2}} E_n B E_m \; ,
\end{align}
after evaluating the integral over $\alpha$ in the first line. We then note that $e^{-\frac{(c_n - c_m)^2}{8\sigma^2}} \rightarrow \delta_{nm}$ as $\sigma\rightarrow 0$, leaving 
\begin{equation}
\lim_{\sigma\rightarrow 0}  \mathcal{E}^{\sigma}_C (B) =
\sum_n E_n B E_n \; ,
\end{equation}
which matches the form of a non-selective ideal measurement of $C$, i.e. $\mathcal{E}^0_C(\cdot)$. Since $\mathcal{E}^{\sigma}_C (\cdot)$ has this limit, one often thinks of this Gaussian measurement as a less sharp ideal measurement, though if $C$ is not compact (but still self-adjoint) then this limit may not be well-defined.

In what follows it will be useful to determine $\mathcal{U}_{t\phi(g)}(G(C))$ for any smeared field $\phi(g)$ and $t\in\mathbb{R}$. First, it is easy to see that
\begin{align}
\mathcal{U}_{t\phi(g)}(\phi(f)) & = e^{it\phi(g)}\phi(f) e^{-it\phi(g)}
\nonumber
\\
& = \phi(f) + t\Delta(f,g) \, 
\end{align}
for any $\phi(f)$. This can be verified by expanding the exponentials. Similarly, for two smeared fields, $\phi(f_1)$ and $\phi(f_2)$, we have
\begin{align}
\mathcal{U}_{t\phi(g)} & (\phi(f_1)\phi(f_2))
\nonumber
\\
 & = e^{it\phi(g)}\phi(f_1)\phi(f_2) e^{-it\phi(g)}
\nonumber
\\
& = e^{it\phi(g)}\phi(f_1)e^{-it\phi(g)}e^{it\phi(g)}\phi(f_2) e^{-it\phi(g)}
\nonumber
\\
& = (\phi(f_1) + t\Delta(f_1,g))(\phi(f_2) + t\Delta(f_2,g)) \, ,
\end{align}
where we have inserted $\mathds{1} = e^{-it\phi(g)}e^{it\phi(g)}$ in the second last line. It is then clear that, for any polynomial in smeared fields $P(\phi(f_1) ,... ,\phi(f_n ))$,
\begin{align}
\mathcal{U}_{t\phi(g)} & (P(\phi(f_1) ,... ,\phi(f_n ))) 
\nonumber
\\
= & P(\phi(f_1)+ t\Delta(f_1 ,g) ,... ,\phi(f_n )+ t \Delta(f_n , g)) \, .
\end{align}
Below we consider operators $C$ of this form, e.g. $C = P(\phi(f_1) ,... ,\phi(f_n ))$. Therefore, defining $C(tg) = \mathcal{U}_{t\phi(g)}(C)$, we have
\begin{align}\label{eq:unitary_action_on_C}
C(tg) = P(\phi(f_1)+ t\Delta(f_1 ,g) ,... ,\phi(f_n )+ t \Delta(f_n , g))  \, .
\end{align} 
We can similarly show that $\mathcal{U}_{t\phi(g)}(C^2) = C(tg)^2$ by inserting $\mathds{1} = e^{-it\phi(g)}e^{it\phi(g)}$ between the $C$'s. Going further, for any any polynomial $Q$ in a single variable we have
\begin{equation}
\mathcal{U}_{t\phi(g)}(Q(C)) = Q(C(tg)) \, .
\end{equation}
This can be extended to any analytic function, $G : \mathbb{R} \rightarrow \mathbb{C}$, using the Taylor expansion for $G$ and inserting $\mathds{1} = e^{-it\phi(g)}e^{it\phi(g)}$ between any two $C$'s. We now have
\begin{equation}\label{eq:smeared_field_kick_on_function_of_C}
\mathcal{U}_{t\phi(g)}(G(C)) = G(C(tg)) \, .
\end{equation}
Since Hermite functions (the basis for the quantum harmonic oscillator) form an analytic orthonormal basis that is dense in $L^2(\mathbb{R})$, we can extend the above action of $\mathcal{U}_{t\phi(g)}(\cdot)$ on $G(C)$ to all $L^2$ functions $G$, i.e. to all square-integrable functions.

\subsection{Operations with a smeared field}\label{sec:Operations with a smeared field}

\subsubsection{A general 1-parameter family of Kraus operators}

Consider the simplest case of $C = \phi(f)$, for $f$ supported in some compact $K$. Given some $L^2$ function, $G$, and the corresponding update map, $\mathcal{E}^G_{\phi(f)}(\cdot)$, we can determine whether it is causal by acting on Weyl generators $e^{it\phi(g)}$ for some $g$ supported in $K_{out}$. We have
\begin{align}\label{eq:smeared_field_operation_on_weyl}
& \mathcal{E}^G_{\phi(f)} (e^{it\phi(g)}) 
\nonumber
\\
& = \int_{\mathbb{R}}d\alpha \, G(\phi(f) - \alpha )^{\dagger}e^{it\phi(g)}G(\phi(f) - \alpha )
\nonumber
\\
& = \int_{\mathbb{R}}d\alpha \, G(\phi(f) - \alpha )^{\dagger}e^{it\phi(g)}G(\phi(f) - \alpha )e^{-it\phi(g)}e^{it\phi(g)}
\nonumber
\\
& = \int_{\mathbb{R}}d\alpha \, G(\phi(f) - \alpha )^{\dagger}G(\phi(f)+ t\Delta(f,g) - \alpha )e^{it\phi(g)}
\nonumber
\\
& = H(t\Delta(f,g)) \, e^{it\phi(g)} \, .
\end{align}
where we have defined the function $H$ in terms of $G$ as
\begin{equation}
H(t) = \int_{\mathbb{R}} d\beta \, G(\beta )^* G(\beta + t) \; ,
\end{equation}
for any $t\in\mathbb{R}$, and we have used~\eqref{eq:smeared_field_kick_on_function_of_C} to write $e^{it\phi(g)}G(\phi(f) - \alpha )e^{-it\phi(g)} = G(\phi(f)+ t\Delta(f,g) - \alpha )$. To get the last line we used~\eqref{eq:functional_calculus}. Specifically, using ~\eqref{eq:functional_calculus} we can write
\begin{align}\label{eq:integral_over_GG_to_H}
& \int_{\mathbb{R}}d\alpha \, G(\phi(f) - \alpha )^{\dagger}G(\phi(f)+ t\Delta(f,g) - \alpha )
\nonumber
\\
& = \int_{\mathbb{R}} \int_{\mathbb{R}}d\alpha \, G(\lambda - \alpha )^*G(\lambda+ t\Delta(f,g) - \alpha ) d P_{\phi(f)}(\lambda )  \, ,
\end{align}
The interior integral in the last line does not depend on $\lambda$, as for any $\lambda\in\mathbb{R}$ we have
\begin{align}
\int_{\mathbb{R}} & d\alpha \, G(\lambda - \alpha )^*G(\lambda+ t\Delta(f,g) - \alpha )
\nonumber
\\
& =  \int_{\mathbb{R}}d\beta \, G(\beta )^*G(\beta+ t\Delta(f,g) ) 
\nonumber
\\
& = H(t\Delta (f,g) ) \; ,
\end{align}
where we have changed variables from $\alpha$ to $\beta = \lambda - \alpha$. Note that $H(t\Delta (f,g) )$ is finite as $G$ is an $L^2$ function, and hence it has a finite $L^2$ product with any other $L^2$ function (by the Cauchy-Schwarz inequality), including a shifted version of itself.

Now~\eqref{eq:integral_over_GG_to_H} becomes
\begin{align}\label{eq:integral_over_GG_to_H_2}
& \int_{\mathbb{R}}d\alpha \, G(\phi(f) - \alpha )^{\dagger}G(\phi(f)+ t\Delta(f,g) - \alpha )
\nonumber
\\
& = \int_{\mathbb{R}} H(t\Delta(f,g)) d P_{\phi(f)}(\lambda ) 
\nonumber
\\
& = H(t \Delta(f,g)) \int_{\mathbb{R}}d P_{\phi(f)}(\lambda )
\nonumber
\\
& = H(t \Delta(f,g)) \mathds{1} \, ,
\end{align}
where $H(t\Delta(f,g))$ can be moved outside the integral as it does not depend on $\lambda$. The last line of~\eqref{eq:smeared_field_operation_on_weyl} now follows.

From~\eqref{eq:smeared_field_operation_on_weyl} we can see that $\mathcal{E}^G_{\phi(f)} (e^{it\phi(g)} )$ is proportional to the original Weyl generator $e^{it\phi(g)}$, with proportionality constant $H(t\Delta(f,g))$, and hence we know that $\mathcal{E}^G_{\phi(f)} (e^{it\phi(g)} )$ is localisable in the same region as $e^{it\phi(g)}$. Therefore, for any region $R\subseteq K_{out}$, $\mathcal{E}^G_{\phi(f)} ( \mathfrak{A}(R) ) \subseteq \mathfrak{A}(R)$, and thus $\mathcal{E}^G_{\phi(f)} (\cdot )$ is causal.

Notably, this is true for any $L^2$ function $G$. The precise form of $G$ will determine the precise form of $H$, and hence the exact effect of the operation on any future measurements. Importantly, the support is never increased for any choice of $G$, and hence maps of this form are always causal.

We also note that the addition of a real constant to $\phi(f)$ does not change the causal nature of these update maps. This can be seen my repeating~\eqref{eq:integral_over_GG_to_H} with $C= \phi(f)+c\mathds{1}$ for some $c\in\mathbb{R}$:
\begin{align}\label{eq:smeared_field_operation_plus_constant_on_weyl}
& \mathcal{E}^G_{C} (e^{it\phi(g)}) 
\nonumber
\\
& = \int_{\mathbb{R}}d\alpha \, G(\phi(f) + c - \alpha )^{\dagger}e^{it\phi(g)}G(\phi(f) + c - \alpha )
\nonumber
\\
& = \int_{\mathbb{R}}d\alpha' \, G(\phi(f) - \alpha' )^{\dagger}e^{it\phi(g)}G(\phi(f) - \alpha' )
\nonumber
\\
& = H(t\Delta(f,g)) \, e^{it\phi(g)} \, ,
\end{align}
where we changed variables to $\alpha' = \alpha - c$ in line 3, and to get the last line we used~\eqref{eq:integral_over_GG_to_H}. Via the linearity of the smeared fields, any linear combination of smeared fields and the identity, i.e. any generator, is of the form $\phi(f) + c\mathds{1}$. Thus, the maps $\mathcal{E}^G_C (\cdot )$ are causal for any generator $C$.

\subsubsection{Gaussian Measurements}

In the specific case of a Gaussian measurement we get
\begin{equation}\label{eq:smeared_field_measurement_on_weyl}
\mathcal{E}^{\sigma}_{\phi(f)} (e^{it\phi(g)} ) = e^{-t^2\frac{\Delta(f , g)^2}{8\sigma^2}}e^{it\phi(g)} \; .
\end{equation}
By taking derivatives w.r.t. $t$ at $t=0$ we find
\begin{align}
\mathcal{E}^{\sigma}_{\phi(f)}(\phi(g)) & = -i \partial_t (\mathcal{E}^{\sigma}_{\phi(f)}( e^{i t\phi(g)}))\big|_{t=0}
\nonumber
\\
& =-i \partial_t (e^{-t^2\frac{\Delta(f , g)^2}{8\sigma^2}}e^{it\phi(g)})\big|_{t=0}
\nonumber
\\
& = \phi(g) \; ,
\end{align}
and
\begin{align}
\mathcal{E}^{\sigma}_{\phi(f)}(\phi(g)^2) & = - \partial_t^2 (\mathcal{E}^{\sigma}_{\phi(f)}( e^{i t\phi(g)}))\big|_{t=0}
\nonumber
\\
& =-\partial_t^2 (e^{-t^2\frac{\Delta(f , g)^2}{8\sigma^2}}e^{it\phi(g)})\big|_{t=0}
\nonumber
\\
& = \phi(g)^2 + \frac{\Delta(f , g)^2}{4\sigma^2} \; .
\end{align}
Therefore, the update map $\mathcal{E}^{\sigma}_{\phi(f)}(\cdot)$ does not alter a single smeared field $\phi(g)$, but it does alter its square by the addition of a constant. This of course does not change the localisation region of $\phi(g)^2$, since $\phi(g)^2 + \frac{\Delta(f , g)^2}{4\sigma^2}$ still commutes with any smeared field $\phi(h)$ that commutes with $\phi(g)$. 

\subsubsection{Towards ideal measurements}

Given that the above derivation applies to $L^2$ functions $G$, one might hope that we can address the case of an ideal measurement by considering indicator functions $G(\alpha ) = 1_{[a,b]}(\alpha )$, where $1_{[a,b]}(\alpha ) = 1$ if $\alpha\in [a,b]$ and $0$ otherwise. This does not yet amount to an ideal measurement, however, as we are still integrating over $\alpha$, rather than summing over a discrete set $\lbrace \alpha_n \rbrace_n$.

Fortunately, this hints at an obvious generalisation of the above derivation. Specifically, we can replace the constant integration measure, $d\alpha$, by some more general measure $d\mu (\alpha)$, which can depend on $\alpha$, and may even be a discrete point measure. In the latter case the integral over $\alpha$ becomes a sum over some discrete set, $\lbrace \alpha_n \rbrace_n$. We can further replace the function $G(\lambda - \alpha)$ by some function $\tilde{G}(\lambda , \alpha)$ of two variables (which still satisfies the relevant normalisation condition).

In this more general picture ideal measurements correspond to using a discrete measure over some set $\lbrace \alpha_n \rbrace_n$, and the choice $\tilde{G}(\lambda , \alpha_n ) = 1_{A_n}(\lambda)$, where we associate to each $\alpha_n$ a subset $A_n\subset \mathbb{R}$, such that all subsets are mutually disjoint and their union covers $\mathbb{R}$.

Repeating~\eqref{eq:integral_over_GG_to_H_2} with these choices gives
\begin{align}\label{eq:ideal_measurement_of_smeared_field_calculation}
\int_{\mathbb{R}}\int_{\mathbb{R}} & d\mu(\alpha ) \, \tilde{G}(\lambda , \alpha )^*\tilde{G}(\lambda+ t\Delta(f,g),\alpha )\, dP_{\phi(f)}(\lambda )
\nonumber
\\
& = \int_{\mathbb{R}}\sum_n \tilde{G}(\lambda , \alpha_n )^*\tilde{G}(\lambda+ t\Delta(f,g),\alpha_n )dP_{\phi(f)}(\lambda )
\nonumber
\\
& = \int_{\mathbb{R}}\sum_n 1_{A_n}(\lambda)1_{A_n}(\lambda+ t\Delta(f,g))dP_{\phi(f)}(\lambda ) \, .
\end{align}
The final sum does not obviously simplify to some function that is independent of $\lambda$. If it does then we know (following~\eqref{eq:smeared_field_operation_on_weyl} and~\eqref{eq:integral_over_GG_to_H_2}) that the resulting update map is causal. If it does not simplify then $H$ in~\eqref{eq:smeared_field_operation_on_weyl} will, in general, depend on $\phi(f)$. That is, the action of the associated update map on a Weyl generator will be of the form
\begin{equation}
\mathcal{E}_{\phi(f)}(e^{it\phi(g)}) = H(\phi(f) , t\Delta(f,g)) e^{it\phi(g)} \, .
\end{equation}
where $ H(\phi(f) , t\Delta(f,g))$ is some non-trivial function of the operator $\phi(f)$, and hence the localisation region of the Weyl generator may have been increased to include the localisation region of $\phi(f)$. In general, the update map will be acausal in this case, as $\text{supp}f$ may contain points that are spacelike to $\text{supp}g$.

This argument suggests that ideal measurements of smeared fields may not be causal, and hence not possible to realise experimentally. In~\cite{Benincasa_2014} it was argued (though no explicit calculation was given) that ideal measurements of smeared fields are in fact causal. The new insight coming from~\eqref{eq:ideal_measurement_of_smeared_field_calculation}, that indicates the contrary, is that the discrete measure arising in an ideal measurement (essentially the sum in~\eqref{eq:ideal_measurement_of_smeared_field_calculation}) is `incompatible', in a certain sense, with the continuous spectrum of a smeared field~\footnote{The spectrum of a smeared field $\phi(f)$, like the position operator $\hat{x}$ in NRQM, is the whole of $\mathbb{R}$.}. While we postpone a more thorough investigation of this conjecture that ideal measurements of smeared fields are acausal, we note that this calculation further motivates the use of 1-parameter families, rather than discrete sets, of Kraus operators.

\subsection{Operations for other operators}\label{sec:Operations for other operators}

Let us now consider the general case, where $C$ is some algebraic combination of smeared fields (and the identity) localisable in some compact $K$. For any $L^2$ function $G$, the associated map acts on a Weyl generator $e^{it\phi(g)}\in\mathfrak{A}(R)$, for a region $R\subseteq K_{out}$, as
\begin{align}\label{eq:GC_operation_on_weyl}
& \mathcal{E}^G_C (e^{it\phi(g)}) 
\nonumber
\\
& = \int_{\mathbb{R}}d\alpha \, G(C - \alpha )^{\dagger}e^{it\phi(g)}G(C - \alpha )
\nonumber
\\
& = \int_{\mathbb{R}}d\alpha \, G(C - \alpha )^{\dagger}e^{it\phi(g)}G(C - \alpha )e^{-it\phi(g)}e^{it\phi(g)}
\nonumber
\\
& = \int_{\mathbb{R}}d\alpha \, G(C - \alpha )^{\dagger}G(C(tg)- \alpha )e^{it\phi(g)} \, ,
\end{align}
using~\eqref{eq:smeared_field_kick_on_function_of_C}.

At this point, if $C$ is such that $[C,C(tg)]=0$, then the remainder of the calculation is fairly simple. If $[C,C(tg)]\neq 0$, then further computation is more challenging without specifying $C$. In Section~\ref{sec:An example with non-commuting smeared fields} we cover one of the simplest such cases, specifically $C = \phi(f_1)\odot \phi(f_2)$, where $\Delta(f_1 , f_2)\neq 0$. Here $\odot$ denotes the symmetric \emph{Jordan product}: $X\odot Y = \frac{1}{2}(XY + YX)$, and is required to make $C$ self-adjoint in this case. 

For now we continue the calculation in the case where $[C,C(tg)]=0$, and for concreteness we will also focus on the Gaussian case involving $G^{\sigma}$. We have
\begin{align}
G^{\sigma}(C - \alpha ) G^{\sigma} & (C(tg)- \alpha )
\nonumber
\\
& = \frac{1}{\sigma\sqrt{2\pi}}e^{-\frac{(C - \alpha)^2}{4\sigma^2}}e^{-\frac{(C(tg) - \alpha)^2}{4\sigma^2}} 
\nonumber
\\
& = \frac{1}{\sigma\sqrt{2\pi}} e^{-\frac{(C_+(tg) - 2\alpha)^2}{8\sigma^2}}e^{-\frac{{C_-(tg)}^2}{8\sigma^2}} \; ,
\end{align}
where $C_{\pm}(tg) = C(tg)\pm C$. Only $e^{-\frac{(C_+(tg) - 2\alpha)^2}{8\sigma^2}}$ depends on $\alpha$, and hence $e^{-\frac{{C_-(tg)}^2}{8\sigma^2}}$ can be brought outside the integral when computing $\mathcal{E}^{\sigma}_C(e^{it\phi(g)})$. The integral over $\alpha$ then evaluates to $1$ and we have
\begin{equation}\label{eq:measurement_of_general_commuting_C_on_weyl}
\mathcal{E}^{\sigma}_C(e^{it\phi(g)}) = e^{-\frac{{C_-(tg)}^2}{8\sigma^2}}e^{it\phi(g)} \; .
\end{equation}  
For some generic $C$ (for which $[C,C(tg)]=0$) then $C_-(tg)$ is some non-trivial operator localisable in $K$. Therefore, $\mathcal{E}^{\sigma}_C(e^{it\phi(g)})$ is not necessarily contained in the subalgebra $\mathfrak{A}(R)$ to which $\phi(g)$ is localisable, and hence the localisation region of the Weyl generator may have been increased to points spacelike to $R$. In such a case $\mathcal{E}^{\sigma}_C(\cdot)$ is not causal.

In Section~\ref{sec:A reasonably general argument} we will argue that, for a large class of operators, only the generators (smeared fields and the identity) give rise to Gaussian measurements that are causal, and hence they are the only operators that are measurable in this way. Before doing that, however, it will be helpful to go through a specific example which is not causal.

\subsection{A simple acausal example}\label{sec:A simple acausal example}

\begin{figure}
    \centering
    \includegraphics[scale=0.7]{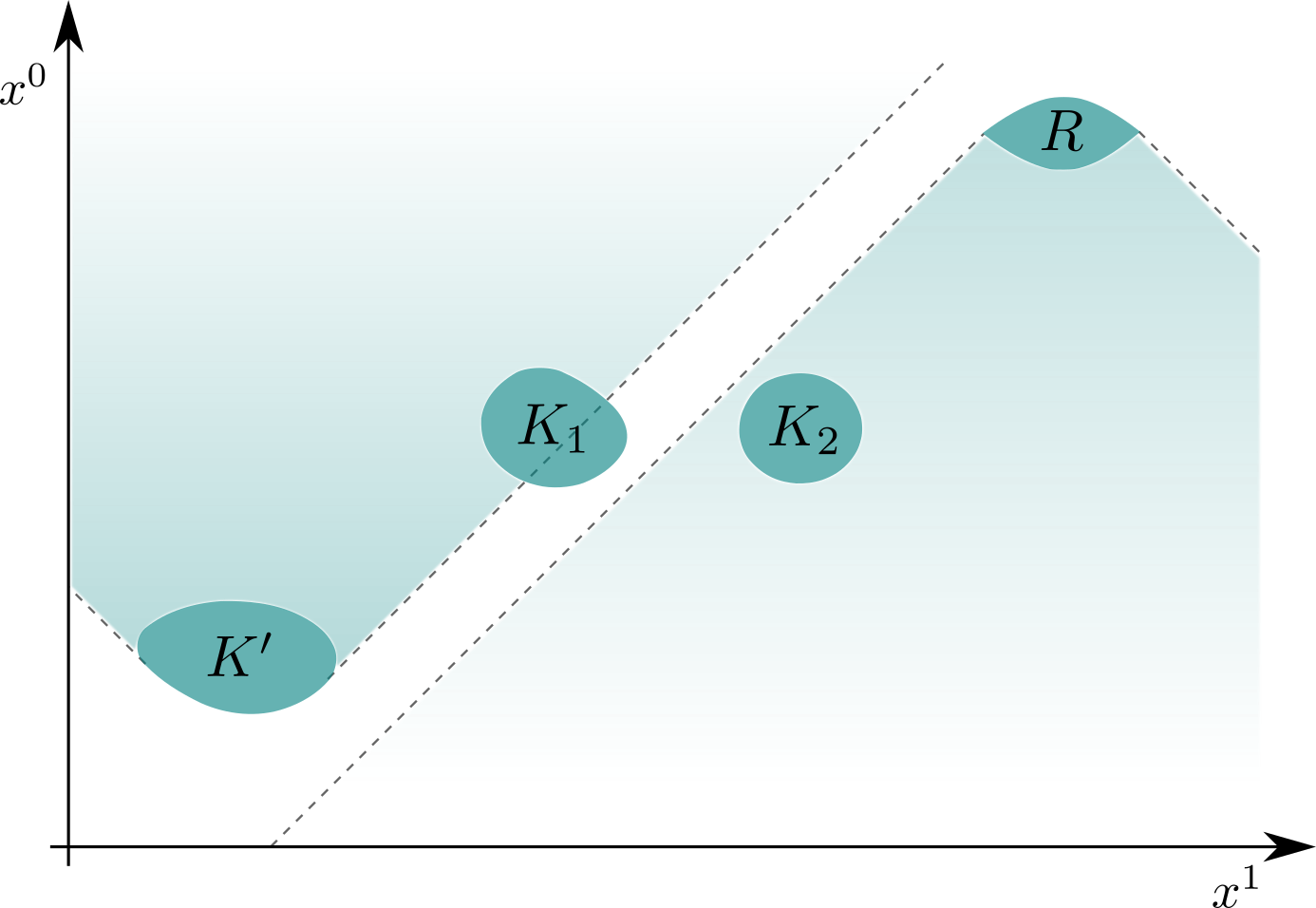}
    \caption{Spacetime diagram of the setup in the simple acausal example. The compact subset $K'$ is spacelike to $K_2$, and $K_1$ is spacelike to the region $R$.}
    \label{fig:commuting_fields_measurement}
\end{figure} 

Consider the operator $C = \phi(f_1) \phi(f_2)$, where $f_1$ and $f_2$ are supported in mutually spacelike compact subsets $K_1$ and $K_2$ (Fig.~\ref{fig:commuting_fields_measurement}). Let $K = K_1 \cup K_2$. Since $[\phi(f_1),\phi(f_2)]=0$, there is no need to invoke the Jordan product $\odot$ to ensure $C$ is self-adjoint. Furthermore, the vanishing of the commutator implies that $[C,C(tg)]=0$ for any test function $g$ supported in some region $R\subseteq K_{out}$. One can also verify that
\begin{align}\label{eq:C_minus_t_for_commuting_product}
C_-(tg) & = t\left( \Delta(f_2 , g) \phi(f_1) + \Delta(f_1 , g) \phi(f_2) \right) 
\nonumber
\\
& + t^2\Delta(f_1 , g)\Delta(f_2 , g) \; .
\end{align}
Note that the appearance of $\phi(f_1)$ and $\phi(f_2)$ (which are localisable in regions that are potentially spacelike to $R$) means that $C_-(tg)\notin \mathfrak{A}(R)$ in general, and hence we are not \textit{a priori} guaranteed that a measurement of $C$ is causal.

Now, consider the action of $\mathcal{E}_C^{\sigma}(\cdot)$ on $\phi(g)$. We have
\begin{align}
\mathcal{E}^{\sigma}_C(\phi(g)) & = -i\partial_t \mathcal{E}^{\sigma}_C(e^{it\phi(g)}) \big|_{t=0} 
\nonumber
\\
& = -i\partial_t \left( e^{-\frac{{C_-(t;g)}^2}{8\sigma^2}}e^{it\phi(g)} \right)\bigg|_{t=0} 
\nonumber
\\
& = \phi(g) \; ,
\end{align}
and hence the map $\mathcal{E}^{\sigma}_C(\cdot)$ does not increase the support of $\phi(g)$. If, however, we act on $\phi(g)^2$, we get
\begin{align}\label{eq:C_product_B_product}
\mathcal{E}^{\sigma}_C & (\phi(g)^2)
\nonumber
\\
& = -\partial_t^2 \mathcal{E}^{\sigma}_C(e^{it\phi(g)}) \big|_{t=0} 
\nonumber
\\
& = \phi(g)^2 + \frac{1}{4\sigma^2}\left( \Delta(f_2 , g) \phi(f_1) + \Delta(f_1 , g) \phi(f_2) \right)^2 \; .
\end{align}
In this case, the support has been increased to include that of $f_1$ and $f_2$. If $\text{supp}f_1$ and/or $\text{supp}f_2$ are outside the past lightcone of $\text{supp}g$, and if $\text{supp}g$ lies partly to the future of $\text{supp}f_1$ and/or $\text{supp}f_2$, then $\mathcal{E}^{\sigma}_C (\cdot)$ increases the support of $\phi(g)^2$ outside its past lightcone.

To highlight the acausal nature of this map we can repeat the protocol with Alice in some compact $K'\subset K_{in}$, Charlie in $K$, and Bob in $R$. Alice unitarily kicks with $\phi(h)$ for $h$ supported in $K'$, Charlie makes the Gaussian measurement under question, $\mathcal{E}^{\sigma}_C(\cdot)$ for $C=\phi(f_1)\phi(f_2)$, and Bob measures the expected value of $\phi(g)^2$ for $g$ supported in $R$.

To simplify the situation we can pick $h$ such that it is supported in some compact $K'$ that is spacelike/timelike to $f_2$/$f_1$, and $g$ supported in some region $R$ that is spacelike/timelike to $f_1$/$f_2$ (Fig.~\ref{fig:commuting_fields_measurement}). We therefore have $\Delta(h,f_2)=\Delta(g,f_1)=0$, but $\Delta(h,f_1)$ and $\Delta(g,f_2)$ non-zero in general.

Working through the example as we did in Section~\ref{sec:Other unitary kicks}, we find (taking the initial state as the vacuum state $\rho = \ket{\Omega}\bra{\Omega}$) that Bob's expected value of $\phi(g)^2$ is given by
\begin{equation}\label{eq:simple_acausal_example_with_expectation_values}
\langle \phi(g)^2 \rangle +  \left(\frac{\Delta(f_2 , g)}{2\sigma}\right)^2 \left( \langle \phi(f_1)^2 \rangle +  \lambda^2 \Delta(f_1 , h)^2 \right) \, ,
\end{equation}
where we have used $\langle X \rangle = \text{tr}(\rho X)$ to denote the vacuum expectation value for brevity. Again, we see it explicitly depends on Alice's kick strength, $\lambda$, and hence Alice and Bob can exploit this to superluminally signal each other (specifically from Alice to Bob). This Gaussian measurement of $C=\phi(f_1)\phi(f_2)$ is therefore not causal, and hence it cannot be physically realisable in any experiment contained in spatial extent and duration in $K$.

It is also not clear in what compact subset, $\tilde{K}$, such a measurement of $C = \phi(f_1) \phi(f_2)$ is physically realisable. Specifically, it does not seem possible to find a compact $\tilde{K}$ such that the map $\mathcal{E}_C^{\sigma}(\cdot)$, when restricted to the subalgebra $\mathfrak{A}(\tilde{K}_{out})$, is PSNI.

\begin{figure}
    \centering
    \includegraphics[scale=0.6]{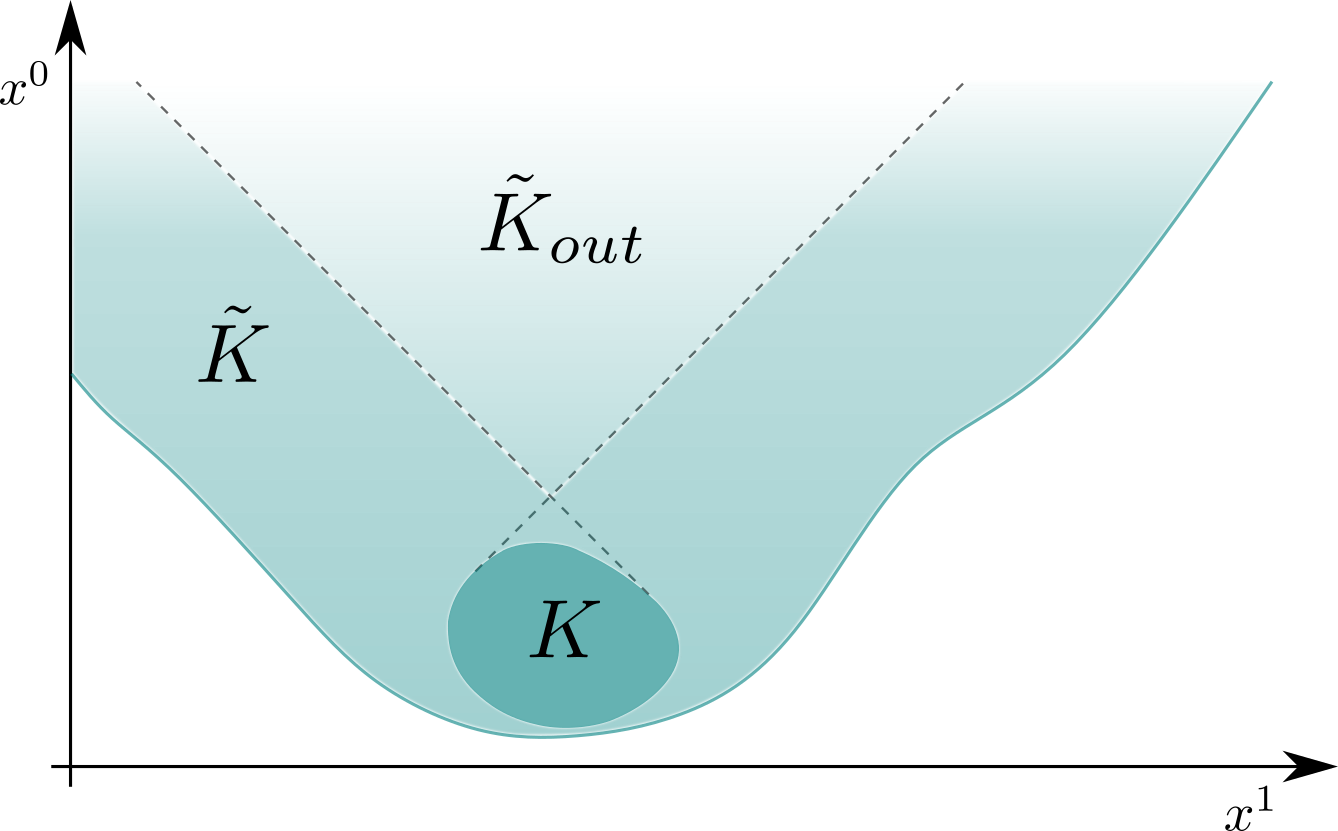}
    \caption{Spacetime diagram of a compact subset $K$ and its total future, denoted here by $\tilde{K}_{out}$. The total future of $K$ is the set of points that contain the \emph{entirety} of $K$ in their pasts. Also shown is an example subset $\tilde{K}$ such that $\tilde{K}_{out} = M\setminus J^-(\tilde{K})$.}
    \label{fig:total_future}
\end{figure}

Alternatively, we can try and find the `largest' out-region, $\tilde{K}_{out}$, such that the update map is causal when acting on any $B\in\mathfrak{A}(\tilde{K}_{out})$, and from that reverse engineer $\tilde{K}$ using the definition $\tilde{K}_{out} = M\setminus J^-(\tilde{K})$. One such candidate for $\tilde{K}_{out}$ is the \emph{total future} of $K$. That is, the set of points $x\in M$ such that $K\subseteq J^-(x)$. This is shown in Fig.~\ref{fig:total_future}. If $\tilde{K}_{out}$ is the total future of $K$ (specifically its interior to ensure we have an open set), then $\tilde{K}$ must, at the very least, be some sort of thickened future lightcone under $\tilde{K}_{out}$ (see Fig.~\ref{fig:total_future}). This ensures that $\tilde{K}_{out} = M\setminus J^-(\tilde{K})$. This choice of $\tilde{K}$ is not unique, and we can even enlarge $\tilde{K}$ to the past in Fig.~\ref{fig:total_future} such that it includes a Cauchy surface for $M$. In fact, if the spacetime $M$ is spatially compact, e.g. the $1+1$ cylinder spacetime $M=\mathbb{R}\times S^1$, then the thickened lightcone will `wrap around', meaning that $\tilde{K}$ will in fact contain a Cauchy surface. In any case, $\tilde{K}$ appears have the property that $\tilde{K}_{in}\cap\tilde{K}_{out}=\emptyset$, i.e. its in- and out-regions are disjoint. Furthermore, its past and future sets cover the entire spacetime. This differs from the case where $K$ is compact and does not contain a Cauchy surface.

Allowing the operation to occur throughout such a $\tilde{K}$ washes out any causality considerations, as every point (outside $\tilde{K}$) is either to the past or future of $\tilde{K}$. The area of spacetime in which the measurement takes place is also not of finite spatial and temporal extent --- a crucial requirement for any locally realisable experiment. In the absence of a less trivial candidate subset $\tilde{K}$, it seems that a Gaussian measurement of $C= \phi(f_1) \phi(f_2)$ is not only impossible in $K$, but is also impossible in any compact region that does not contain a Cauchy surface. 

On the other hand, one intuitively expects a measurement of $C= \phi(f_1) \phi(f_2)$ to be possible in some local sense; surely we do not need to resign to a global operation over all of space just to measure this localisable operator. In this vein it is worth highlighting that we have only ruled out the update map $\mathcal{E}_C^{\sigma}(\cdot)$. Perhaps some modification of $\mathcal{E}_C^{\sigma}(\cdot)$ would make it causal, and would align better with our intuition of a local measurement of $C= \phi(f_1) \phi(f_2)$. The exact modification may depend upon the details of the measurement apparatus, for example, it may involve probe fields as in~\citep{Fewster2020}. Nevertheless, for now we can be sure that $\mathcal{E}_C^{\sigma}(\cdot)$ is not physically realisable.

Following Section~\ref{sec:Operations with a smeared field}, we \emph{can} say however that the measurements $\mathcal{E}^{\sigma}_{\phi(f_1)}(\cdot)$ and $\mathcal{E}^{\sigma}_{\phi(f_2)}(\cdot)$, for the single smeared fields $\phi(f_1)$ and $\phi(f_2)$ respectively, are physically realisable. Given these smeared fields commute, their update maps also commute, i.e. $\mathcal{E}^{\sigma}_{\phi(f_1)}(\mathcal{E}^{\sigma}_{\phi(f_2)}(\cdot)) = \mathcal{E}^{\sigma}_{\phi(f_2)}(\mathcal{E}^{\sigma}_{\phi(f_1)}(\cdot))$, meaning that the measurements can be thought of as occurring in either order. This physically makes sense, as their associated regions are spacelike, and hence there is no causal ordering between them. Therefore, while a Gaussian measurement of $C= \phi(f_1) \phi(f_2)$ appears not to be realisable, independent measurements of $\phi(f_1)$ and $\phi(f_2)$ are.

It is also worth noting that the signal from Alice to Bob gets `weaker' as $\sigma$ increases. Physically this corresponds to a decreasing measurement accuracy. The limit of no measurement accuracy whatsoever, i.e. $\sigma\rightarrow\infty$, is equivalent to no measurement at all. In this limit the past-support increasing term above vanishes, as expected.

At this point one could argue that, if a future measurement of $B = \phi(g)^2$ is somehow limited in its accuracy, then this Gaussian measurement of $C=\phi(f_1)\phi(f_2)$ \emph{is} possible in $K$, provided its accuracy is low enough (or equivalently if $\sigma$ is large enough) to make the second term in the last line of~\eqref{eq:C_product_B_product} smaller than a future experimenter can detect. This connection between the possibility of some measurement and its accuracy was also noted in~\cite{borsten2021impossible}. This resolution is somewhat suspect however, as the allowed accuracy of the Gaussian measurement of $C$ is determined by the accuracies of \emph{all} future measurements. How can someone measuring $C=\phi(f_1)\phi(f_2)$ know the measurement limitations of all future experiments? It makes more sense to turn this restriction around and instead constrain the accuracy of all measurements to the future of $K$, given the accuracy of the measurement of $C=\phi(f_1)\phi(f_2)$. How this would work in practice is not clear. One would have to introduce some mechanism preventing anyone in the future from obtaining some more accurate measurement than is allowed by causality.

\subsection{A reasonably general argument}\label{sec:A reasonably general argument}

\begin{figure*}
    \centering
    \includegraphics[scale=1]{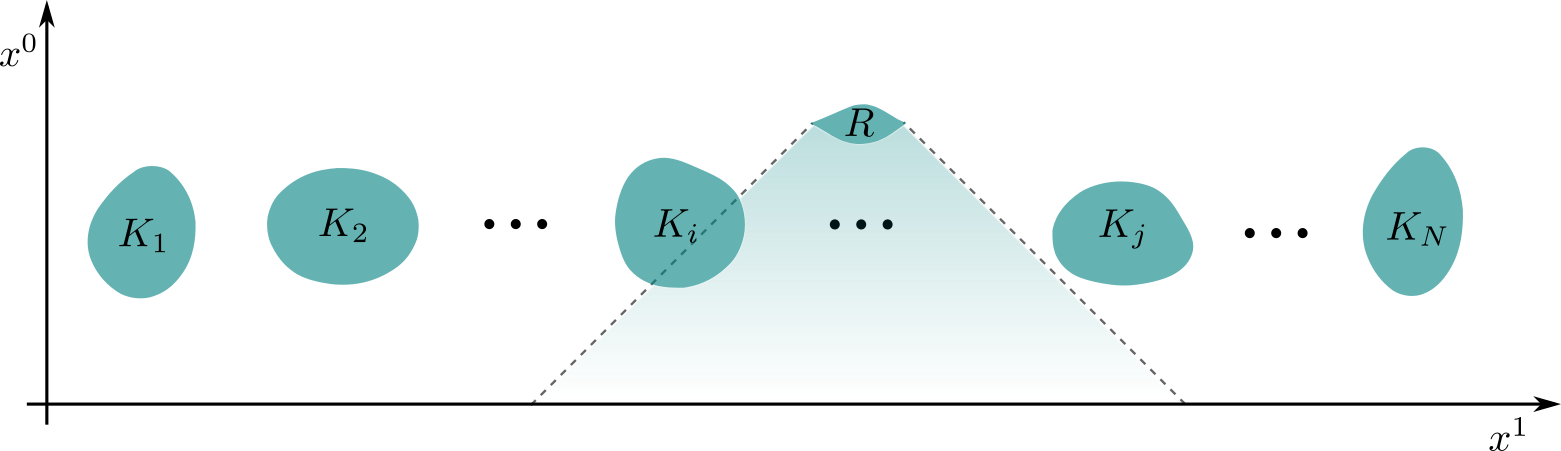}
    \caption{Spacetime diagram of compact subsets $K_1, ... , K_N$ with mutually spacelike closures, within which the smeared fields $\phi(f_1), ... ,\phi(f_N)$ are localisable. In the general argument we choose the region $R$ to be spacelike to all but one $K_i$, as can be seen from the past lightcone of $R$. In case ii) we also pick another $K_j$ to show that the update map has increased the support to points spacelike to $R$.}
    \label{fig:general_argument}
\end{figure*}

We will now argue that if $C$ is \emph{not} a generator then $\mathcal{E}^{\sigma}_C(\cdot)$ is not causal. We will only show this, however, for the following restricted class of operators.

Consider any set of smeared fields that are localisable in compact subsets whose closures are mutually spacelike. This is shown in Fig.~\ref{fig:general_argument}. One can then construct the associated commutative subalgebra, consisting of all algebraic combinations of the identity and the commuting smeared fields. For our general argument we will only consider operators $C$ belonging to some such commutative subalgebra formed from smeared fields in some compact $K$. Furthermore, we restrict to the case in which $C$ only involves a finite number of sums and products of smeared fields.

For example, $\phi(f_1)\phi(f_2)$ satisfies this criteria if $f_1$ and $f_2$ have spacelike supports. On the other hand, $\phi(f_1)\odot \phi(f_2)$, for the supports of $f_1$ and $f_2$ not totally spacelike, does not satisfy this criteria.

Let $K_1 , ... , K_N\subset K$ be $N$ compact subsets whose closures are mutually spacelike, as in Fig.~\ref{fig:general_argument}, and $\phi(f_1), ... ,\phi(f_N)$ the associated smeared fields localisable in the respective subsets. We then set $C = P(\phi(f_1), ... ,\phi(f_N))$, where $P(\cdot)$ is some polynomial in $N$ independent variables. For such a $C$ it is then clear that $[C,C(tg)]=0$, as
\begin{equation}
C(tg) = P( \phi(f_1)+t\Delta(f_1,g), ... ,\phi(f_N)+t\Delta(f_N,g) ) \; ,
\end{equation}
and $\phi(f_i)+t\Delta(f_i,g)$ commutes with any other $\phi(f_j)$.

Given that $[C,C(tg)]=0$, we can apply the derivation in Section~\ref{sec:Operations for other operators} to get
\begin{equation}
\mathcal{E}^{\sigma}_C(e^{i t \phi(g)}) = e^{-\frac{{C_-(tg)}^2}{8\sigma^2}}e^{i t \phi(g)} \; ,
\end{equation}  
where we recall that $C_-(tg) = C(tg) - C$. By expanding in $t$ we get
\begin{equation}
C_-(tg) = \sum_{n=1}^{\infty} \frac{(it)^n}{n!}{\text{Ad}_{\phi(g)}}^n ( C ) \; ,
\end{equation}
where $\text{Ad}_{X}(Y) = [X,Y]$, and ${\text{Ad}_{X}}^{n+1} (\cdot) = \text{Ad}_{X}({\text{Ad}_{X}}^n (\cdot) )$. The sum terminates at some finite $n$ as, at some point, any additional commutators with $\phi(g)$ vanish. For an example see equation~\eqref{eq:C_minus_t_for_commuting_product}. In fact, if $N$ is the degree of the polynomial $P(\cdot )$, then the sum terminates after $N+1$ terms.

Given the regions $K_i$ are all spacelike, and their closures do not touch, we can always pick $g$ supported in some region $R$ that it is spacelike to \emph{all but one} of the regions $K_i$. This is shown in Fig.~\ref{fig:general_argument}. For convenience we let $X \equiv \phi(f_i)$ denote the associated smeared field. $\phi(g)$ then commutes with all the other smeared fields used in the construction of $C$. The action of ${\text{Ad}_{\phi(g)}}^n (\cdot)$ on $C$ then resembles $n$ derivatives of the polynomial $P(\cdot)$ with respect to $X$, up to factors of $i\Delta(f_i , g)$. In this way $C_-(tg)$ looks very similar to a Taylor expansion in $X$ of the polynomial $P(\cdot)$, but with the constant term ($n=0$) thrown away from the sum.

By assumption, $C$ contains some term that is at least quadratic in smeared fields. That is, the degree of the polynomial $P(\cdot)$ is at least 2. Therefore, our choice of $X$ can always be made such that $C$ can be written as
\begin{equation}
C = C_0 + C_1 X + C_m X^m + O(X^{m+1}) \; ,
\end{equation}
where $m\geq 2$, and where either i) $C_m\neq 0$, or ii) $C_m=0$ and all higher order terms vanish, but $C_1$ is some polynomial, $Q(\cdot)$, of degree at least 1 in the other variables $\phi(f_j)\neq X$.

We then have
\begin{align}
C_-(tg) = & t \Delta(f_i ,g) \left( C_1 + m C_m X^{m-1} + O(X^m) \right) 
\nonumber
\\
& + O(t^2) \; ,
\end{align}
and hence
\begin{align}
\mathcal{E}^{\sigma}_C & (\phi(g)^2)
\nonumber
\\
& = -\partial_t^2 \mathcal{E}^{\sigma}_C(e^{it\phi(g)}) \big|_{t=0} 
\nonumber
\\
& = \phi(f_3)^2 
\nonumber
\\
& + \frac{\Delta(f_i,g)^2}{4\sigma^2}\left( C_1 + m C_m X^{m-1}+O(X^m) \right)^2 \; .
\end{align}

For case i) we know that $C_m\neq 0$, and hence the term on the last line is $O(X^{2{m-1}})$, which is at least $O(X^2)$ given that $m\geq 2$. This means that $\mathcal{E}_C^{\sigma}(\phi(g)^2)$ has past support which includes that of $X=\phi(f_i)$, and since we can always pick $g$ such that $\text{supp}f_i$ has points that are spacelike to $\text{supp}g$, this means that $\mathcal{E}_C^{\sigma}(\cdot)$ has increased the past support of $\phi(g)^2$ to outside its past lightcone. Therefore $\mathcal{E}_C^{\sigma}(\cdot)$ is not causal.

For case ii) $C_m=0$ for all $m\geq 2$, but $C_1$ is of degree at least 1 in the other smeared fields $\phi(f_j)$ ($j\neq i$). Therefore, we can always pick some $Y\equiv \phi(f_j)$ ($j\neq i$) such that $C_1$ is at least $O(Y)$, and hence the last line above is at least $O(Y^2)$. From our initial setup the support of $g$ is spacelike to the support of $f_j$, and hence, in this case, $\mathcal{E}_C^{\sigma}(\cdot)$ has increased the past-support of $\phi(g)^2$ to include the past lightcone of $f_j$. Again, the map $\mathcal{E}_C^{\sigma}(\cdot)$ is then not causal.

\subsection{An example with non-commuting smeared fields}\label{sec:An example with non-commuting smeared fields}

\begin{figure}
    \centering
    \includegraphics[scale=0.6]{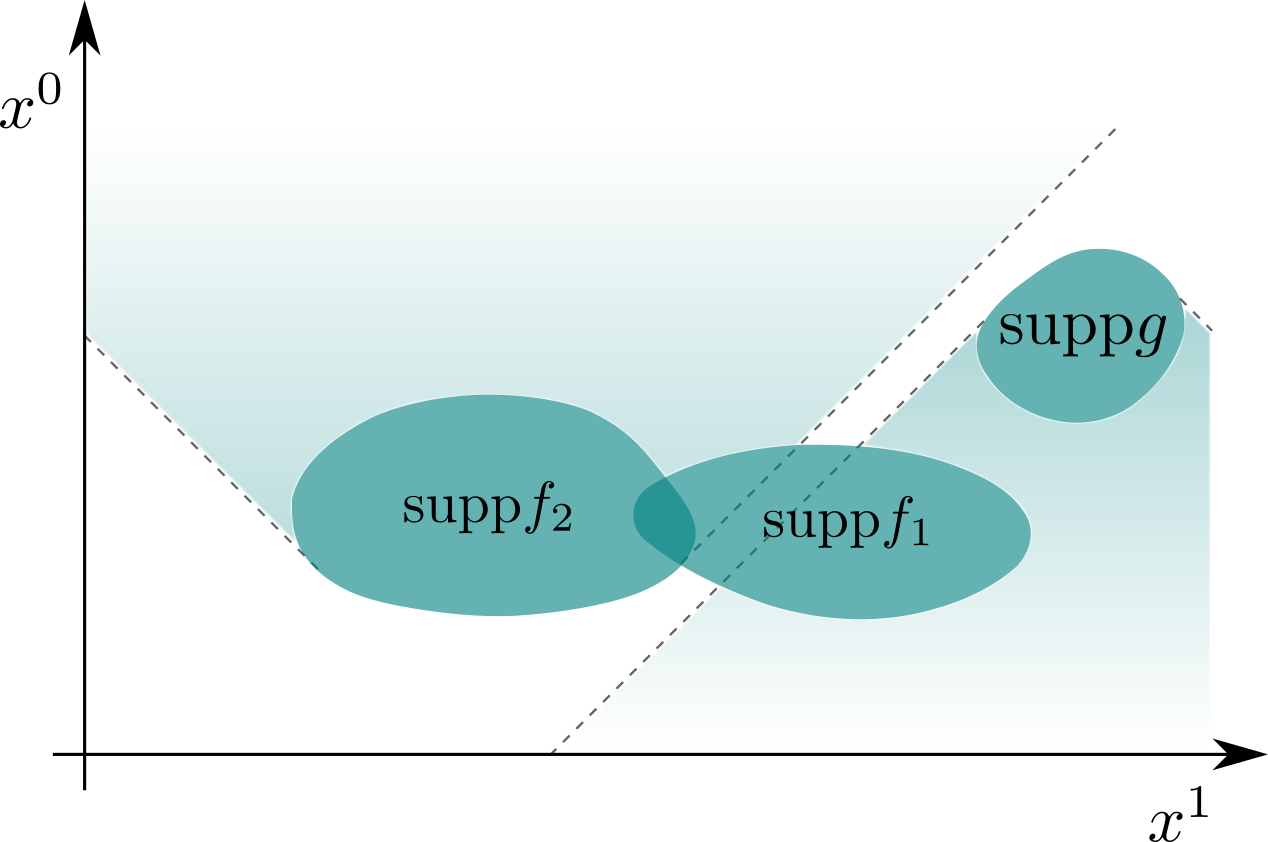}
    \caption{Spacetime diagram of an example pair of smearing functions, $f_1$ and $f_2$, whose supports are not mutually spacelike. Here we have also illustrated that their supports can overlap, though this is not necessary for their supports to not be spacelike. The smeared fields $\phi(f_1)$ and $\phi(f_2)$ do not commute. We have also illustrated the support of the function $g$ used in the calculation. Its support is spacelike to that of $f_2$ but not to $f_1$.}
    \label{fig:overlapping_supports}
\end{figure}

We have just argued for a reasonably wide class of operators that only the generators can be measured in this way. One case we did not consider is when $[C,C(tg)]\neq 0$. In this situation the calculation becomes more complicated, and we do not have a general argument. We can, however, work through one of the simplest examples, namely $C = \phi(f_1)\odot \phi(f_2)$, where the supports of $f_1$ and $f_2$ are not mutually spacelike, as shown in Fig.~\ref{fig:overlapping_supports}. In this case $\phi(f_1)$ and $\phi(f_2)$ do not commute, and hence the Jordan product, $\odot$, has appeared in $C$ to keep it self-adjoint. In the following calculations we will show that this choice of $C = \phi(f_1)\odot\phi(f_2)$ gives rise to an acausal Gaussian measurement, thus adding more evidence to the claim that only generators can be measured in this way.

To show that a Gaussian measurement of $C = \phi(f_1)\odot \phi(f_2)$ is acausal we only need to show that, for some choice of operator, $\mathcal{E}^{\sigma}_C(\cdot)$ increases its support. In this vein we consider $\phi(g)$, where the support of $g$ is spacelike to the support of $f_2$, but not to the support of $f_1$ (see Fig.~\ref{fig:overlapping_supports}). Therefore, $\Delta(f_1,g)\neq 0$ and $\Delta(f_2,g)=0$. We then have
\begin{align}
C(tg) & = \mathcal{U}_{t\phi(g)}(C)
\nonumber
\\
& = \mathcal{U}_{t\phi(g)}(\phi(f_1))\odot \mathcal{U}_{t\phi(g)}(\phi(f_2))
\nonumber
\\
& = (\phi(f_1)+t\Delta (f_1 ,g ) )\odot \phi(f_2)
\nonumber
\\
& = C + t\Delta (f_1 , g)\phi(f_2) \; .
\end{align}
Thus, 
\begin{equation}
e^{it\phi(g)}e^{-\frac{(C - \alpha)^2}{4\sigma^2}} = e^{-\frac{(C - \alpha + t\Delta (f_1 ,g) \phi(f_2))^2}{4\sigma^2}}e^{it\phi(g)} \; ,
\end{equation}
and hence
\begin{equation}\label{eq:measurement_of_non_commuting_product_step_1}
\mathcal{E}^{\sigma}_C(e^{it\phi(g)}) = \frac{1}{\sigma\sqrt{2\pi}}\int_{-\infty}^{\infty}d\alpha \, e^{-A^2}e^{-(A+t B)^2}e^{it\phi(g)} ,
\end{equation}
where we have defined
\begin{align}
A & = \frac{C - \alpha}{2\sigma} \; ,
\\
B & = \frac{\Delta(f_1 , g)\phi(f_2)}{2\sigma} \; ,
\end{align}
for convenience.

As
\begin{equation}
e^{-y^2} = \frac{1}{2\sqrt{\pi}}\int_{-\infty}^{\infty}dx \, e^{-\frac{x^2}{4}} e^{-i x y } \; ,
\end{equation}
for any $y\in\mathbb{R}$, and since $A+tB$ is self-adjoint, we can use the associated projection-valued measure to write 
\begin{equation}\label{eq:measurement_of_non_commuting_product_Fourier_step}
e^{-(A+t B)^2} = \frac{1}{2\sqrt{\pi}}\int_{-\infty}^{\infty}dx \, e^{-\frac{x^2}{4}} e^{-i x (A+t B)} \; .
\end{equation}
Now, since
\begin{equation}
[A , B] = i r B \; ,
\end{equation}
where we have defined the non-zero real number $r=\frac{\Delta(f_1,f_2)}{2\sigma}$, the BCH formula gives
\begin{equation}
e^{-i x (A+t B)} = e^{-i x A}e^{i t \frac{(e^{-x r}-1 )}{r}B} \; .
\end{equation}
By inserting the RHS into~\eqref{eq:measurement_of_non_commuting_product_Fourier_step}, and the result into~\eqref{eq:measurement_of_non_commuting_product_step_1}, we get
\begin{align}
\mathcal{E}^{\sigma}_C( e^{it\phi(g)}) & = \frac{1}{\sigma \sqrt{2}2\pi}
\nonumber
\\
& \times  \int_{\mathbb{R}^2}d\alpha \, dx \, e^{-\frac{x^2}{4}}e^{-A(A+i x A)} e^{i t \frac{(e^{-x r}-1 )}{r}B}
\nonumber
\\
& \times e^{i t \phi(g)} .
\end{align}
Given that the operators in the integrand are bounded, and hence the norm of the integrand is bounded by $e^{-\frac{x^2}{4}}$, we can swap the order of the double integral (by the Fubini-Tonelli Theorem) and evaluate the $\alpha$ integral first. For the parts of the integrand that depend on $\alpha$ this gives
\begin{equation}
\int_{-\infty}^{\infty}d\alpha \, e^{-\frac{x^2}{4}}e^{-A(A+i x A)} = 2\sigma \sqrt{\pi} e^{-\frac{x^2}{4}}\mathds{1} \; ,
\end{equation}
leaving
\begin{equation}\label{eq::measurement_of_non_commuting_product_final_for_weyl}
\mathcal{E}^{\sigma}_C( e^{it\phi(g)}) = \eta\left( t \frac{\Delta(f_1 , g)}{\Delta(f_1,f_2)}\phi(f_2) \right) e^{i t \phi(g)} \; ,
\end{equation}
where we have defined the function
\begin{equation}
\eta(t) = \frac{1}{\sqrt{2\pi}}  \int_{-\infty}^{\infty} dx \, e^{-\frac{x^2}{2}} e^{i (e^{-x r}-1 )t} \; .
\end{equation}
While we do not have a closed form for $\eta (t)$, we note that the integral exists and is bounded for any $t\in\mathbb{R}$. This follows as the absolute value of the integrand is $e^{-\frac{x^2}{2}}$, which integrates to a constant. Therefore, the operator $\eta\left( t \frac{\Delta(f_1 , g)}{\Delta(f_1,f_2)}\phi(f_2) \right)$ is bounded. Furthermore, as $\eta(0) = 1$, we get $\eta\left( 0\times  \frac{\Delta(f_1 , g)}{\Delta(f_1,f_2)}\phi(f_2) \right) = \mathds{1}$

Since the final result for $\mathcal{E}^{\sigma}_C( e^{it\phi(g)})$ in~\eqref{eq::measurement_of_non_commuting_product_final_for_weyl} depends on $\phi(f_2)$ --- a smeared field localisable in a region spacelike to the support of $g$ --- the map $\mathcal{E}^{\sigma}_C( \cdot )$ has increased the past support of $g$ outside the past-lightcone to include that of $f_2$. For completeness we can also take derivatives with respect to $t$ to evaluate $\mathcal{E}^{\sigma}_C( \phi(g))$. Explicitly, since
\begin{align}
\eta'(0) = \frac{i}{\sqrt{2\pi}} \int_{-\infty}^{\infty} dx \, e^{-\frac{x^2}{2}}(e^{-x r}-1) = i( e^{\frac{r^2}{2}}-1 ) \, ,
\end{align}
one can verify that
\begin{align}
\mathcal{E}^{\sigma}_C(\phi(g)) & = \phi(g) 
\nonumber
\\
& \hspace{5mm}+ \left( e^{\frac{\Delta(f_2,f_1)^2}{8\sigma^2}}-1\right)\frac{\Delta(f_1,g)}{\Delta(f_1,f_2)}\phi(f_2) \; ,
\end{align}
which further highlights the increase in support.

We have just shown that a Gaussian measurement of $C = \phi(f_1)\odot \phi(f_2)$ is not causal, and hence is not physically realisable in $K$. One can also verify that the support is increased for higher powers of $\phi(g)$ using further derivatives w.r.t. $t$. It is worth noting that as $\Delta(f_1,f_2)\rightarrow 0$, i.e. in the limit that $\phi(f_1)$ and $\phi(f_2)$ become commuting, we get $\mathcal{E}^{\sigma}_C(\phi(g))=\phi(g)$. This is consistent with our results in Section~\ref{sec:A simple acausal example}.

\subsection{Extracting expectation values from measurements of smeared fields}\label{sec:Extracting expectation values from measurements of smeared fields}

The causality conditions we have imposed on update maps (Section~\ref{sec:Causality conditions on update maps}) all assume that it is possible to measure expectation values of operators that are more complicated than smeared fields (plus the identity), e.g. $\phi(f)^2$ or $\phi(f)\phi(g)$. Given the evidence presented above that we can only kick with and/or enact Gaussian measurements of smeared fields and the identity, one may question whether it is even possible to measure such expectation values. While it may be possible to read out expectation values through some other description of measurement such as in~\cite{Fewster2020,Bostelmann_2021}, here we argue that it can be done using only Gaussian measurements of smeared fields, which were shown to be causal above. 

Consider a Gaussian measurement of some operator $C$ localisable in some compact $K$. As noted in Section~\ref{sec:Preliminaries}, the average outcome is given by
\begin{equation}
\mathbb{E}(\alpha) = \int_{\mathbb{R}}d\alpha \, \alpha \, p(\alpha) \; ,
\end{equation}
in terms of the pdf for $\alpha$, which we recall is given by
\begin{equation}
p(\alpha) = \frac{1}{\sigma\sqrt{2\pi}} \text{tr}\left(\rho e^{-\frac{(C - \alpha)^2}{2\sigma^2}} \right) \; ,
\end{equation}
where $\rho$ is the state. If there are any other measurements occurring in $K_{in}$, then this state should really include the update maps for them. For the present discussion this is irrelevant, however, and hence we will omit any other update maps here for brevity. Substituting this pdf into the expression for $\mathbb{E}(\alpha)$ one can evaluate the integral over $\alpha$ to find
\begin{equation}
\mathbb{E}(\alpha) = \text{tr}(\rho C) \; .
\end{equation}
That is, the expected value of the outcome $\alpha$ matches the expectation value of $C$ coming from quantum theory. One can also verify that
\begin{equation}
\mathbb{E}(\alpha^2) = \text{tr}(\rho C^2) + \sigma^2 \; ,
\end{equation}
and hence
\begin{align}
\text{Var}(\alpha) & = \mathbb{E}(\alpha^2) - \mathbb{E}(\alpha)^2
\nonumber
\\
& = \text{tr}(\rho C^2) + \sigma^2 - \text{tr}(\rho C)^2
\nonumber
\\
& = \Delta C^2 + \sigma^2 \; .
\end{align}
Therefore, the variance, $\text{Var}(\alpha)$, of the Gaussian measurement is always greater than the variance, $\Delta C^2$, computed in the quantum theory. For a perfect measurement, i.e. $\sigma\rightarrow 0$, the two agree.

After repeated Gaussian measurements of the operator $C$ (which can happen simultaneously if multiple copies of the system are set up in parallel) one
can compute the average outcome, or the average of the square of the outcomes etc. In the limit of a large number of experimental realisations this number will approach $\mathbb{E}(\alpha)$, or $\mathbb{E}(\alpha^2 )$ respectively. In this way one can estimate $\text{tr}(\rho C)$ and $\text{tr}(\rho C^2)$ (up to the constant $\sigma^2$), and all higher moments.

This is important for our purposes as it means that one only needs to make Gaussian measurements of the smeared field $\phi(g)$ in order to determine the expectation values needed for the causality conditions in Section~\ref{eq:simple_acausal_example_with_expectation_values}, e.g. $\text{tr}(\rho \phi(g)^2)$. In other words, our above arguments for violations of causality did not require violations of causality to begin with! In fact, someone attempting to measure the expected value of $\phi(g)^2$ may only be able to determine $\mathbb{E}(\alpha^2)$, and hence will only know $\text{tr}(\rho \phi(g)^2)$ up to the (potentially unknown) constant $\sigma^2$. This is not a problem however, as, just like $\text{tr}(\rho \phi(g)^2)$, $\mathbb{E}(\alpha^2)$ must remain unchanged whenever $\rho\mapsto \tilde{\mathcal{E}}'(\rho)$ for some $\tilde{\mathcal{E}}'(\cdot)$ dual to the update map $\mathcal{E}'(\cdot)$ local to $K_{in}$ and spacelike to $\phi(g)$, and hence our causality conditions from Section~\ref{sec:Causality conditions on update maps} still go through. 

This argument can be applied not only to $\text{tr}(\rho \phi(g)^2)$, but also to other more complicated expectation values. For higher powers of $\phi(g)$ one simply computes expectation values of higher powers of $\alpha$ coming from the Gaussian measurement of $\phi(g)$. 

For correlation functions, such as $\text{tr}(\rho \phi(g_1)\phi(g_2))$ (where we restrict $g_1$ and $g_2$ to be supported in mutually spacelike subsets for now), one can do two Gaussian measurements of the smeared fields $\phi(g_1)$ and $\phi(g_2)$. Following these two measurements the state is updated via the composition of the two update maps: $\rho \mapsto \tilde{\mathcal{E}}^{\sigma}_{\phi(g_1)}(\tilde{\mathcal{E}}^{\sigma}_{\phi(g_2)}(\rho))$. The order of these maps does not matter since they commute (this follows from the fact that $[\phi(g_1),\phi(g_2)]=0$). The probability of measuring some value $\alpha\in [a_1,a_2]$ for the measurement of $\phi(g_1)$, \emph{and} some value $\beta\in [b_1,b_2]$ for the measurement of $\phi(g_2)$, is given by
\begin{equation}
P_{[a_1 , a_2]\times [b_1 , b_2]} = \int_{a_1}^{a_2} d\alpha \int_{b_1}^{b_2} d\beta \, p(\alpha , \beta) \; ,
\end{equation}
where the joint pdf is given by
\begin{equation}\label{eq:joint_pdf_commuting_fields}
p(\alpha , \beta) = \frac{1}{2\pi \sigma^2} \text{tr}\left( \rho e^{-\frac{(\phi(g_1) - \alpha)^2}{2\sigma^2}}e^{-\frac{(\phi(g_2) - \beta)^2}{2\sigma^2}} \right) \; .
\end{equation}
Any correlations encoded in the state $\rho$ are revealed here in the sense that the joint pdf $p(\alpha,\beta)$ is not necessarily given by the product of the two marginal pdf's for $\alpha$ and $\beta$. Over many realisations of the two measurements one can compute the average value of the product of the two separate measurement outcomes, i.e. the average of $\alpha\times \beta$. In the limit of a large number of realisations this number will approach
\begin{align}
\mathbb{E}(\alpha\times \beta) & = \int_{\mathbb{R}^2}d\alpha d\beta \,\, \alpha \, \beta \, p(\alpha , \beta) 
\nonumber
\\
& = \text{tr}\left(\rho  \phi(g_1) \phi(g_2) \right) \; .
\end{align}
That is, the correlation function $\text{tr}\left(\rho  \phi(g_1) \phi(g_2) \right)$ can be estimated by making repeated Gaussian measurements of $\phi(g_1)$ and $\phi(g_2)$ and computing the average product of the outcomes. One can also verify that $\mathbb{E}(\alpha + \beta ) = \text{tr}(\rho (\phi(g_1) + \phi(g_2) )= \text{tr}(\rho \phi(g_1 + g_2) )$.

If $g_1$ and $g_2$ are supported in subsets which are not totally spacelike, the recovery of the correlation function $\text{tr}\left(\rho  \phi(g_1) \phi(g_2) \right)$ is slightly more complicated. Since $[\phi(g_1),\phi(g_2)]\neq 0$, the update maps acting on $\rho$ do not necessarily commute in the expression for the probability $P_{[a_1 , a_2]\times [b_1 , b_2]}$. If the supports of $g_1$ and $g_2$ are not totally timelike then there is no canonical order for the update maps. A natural option in any case (which agrees with the totally spacelike case) is the symmetrised Jordan composition, where we average over the two possible orders. 

Computing the joint pdf in this case we find
\begin{equation}
p(\alpha , \beta) = \frac{1}{2}(q_{12}(\alpha , \beta)+q_{21}(\alpha , \beta) ) \, 
\end{equation}
where
\begin{align}
& q_{12} (\alpha,\beta) 
\nonumber
\\
& =  \frac{1}{2\pi \sigma^2} \text{tr}\left( \rho e^{-\frac{(\phi(g_2) - \beta)^2}{4\sigma^2}}e^{-\frac{(\phi(g_1) - \alpha)^2}{2\sigma^2}} e^{-\frac{(\phi(g_2) - \beta)^2}{4\sigma^2}} \right) \; ,
\end{align}
and $q_{21}(\alpha , \beta )$ is the same expression but with the replacements $g_{1/2}\mapsto g_{2/1}$ and $\alpha / \beta \mapsto \beta / \alpha$. Computing the integral of $\alpha\times\beta$ against $q_{12}(\alpha , \beta)$ one finds 
\begin{align}
& \mathbb{E}_{12}(\alpha \times \beta) 
\nonumber
\\
& = \frac{1}{\sigma\sqrt{2\pi}}\int_{\mathbb{R}}d\beta \, \beta \, \text{tr}\left( \rho \,  e^{-\frac{(\phi(g_2) - \beta)^2}{4\sigma^2}}\phi(g_1) e^{-\frac{(\phi(g_2) - \beta)^2}{4\sigma^2}} \right) \; ,
\end{align}
where we have evaluated the integral over $\alpha$. Using~\eqref{eq:comm_g_with_gaussian} we can push $\phi(g_1)$ through the exponential to its right. Following this we can then compute the integral over $\beta$. We find
\begin{equation}
\mathbb{E}_{12}(\alpha \times \beta) = \text{tr}(\rho \phi(g_2) \phi(g_1) ) + \frac{i}{2}\Delta(g_1 , g_2) \; .
\end{equation}
If we add to this the analogous expression, $\mathbb{E}_{21}(\alpha \times \beta)$, computed using the measure $q_{21}(\alpha , \beta)$ associated to the other ordering, we then find (after dividing by 2)
\begin{align}\label{eq:symmetrised_exp_to_correlation_function}
\mathbb{E}(\alpha \times \beta) & = \frac{1}{2}(\mathbb{E}_{12}(\alpha \times \beta)+\mathbb{E}_{21}(\alpha \times \beta) )
\nonumber
\\
& = \text{tr}(\rho \phi(g_1) \odot \phi(g_2) ) \; ,
\end{align}
and thus the symmetrised correlation function is recovered exactly. This last result follows from the fact that $\Delta(\cdot , \cdot)$ is antisymmetric, and hence the $\Delta(\cdot , \cdot)$ terms vanish under the symmetrisation of the Jordan composition.

To recover the correlation function without any symmetrisation, i.e. $\text{tr}(\rho \phi(g_1) \phi(g_2) )$, we can then add to~\eqref{eq:symmetrised_exp_to_correlation_function} the antisymmetrised expression
\begin{equation}
\frac{1}{2}\text{tr}(\rho [\phi(g_1) , \phi(g_2)] ) = \frac{i}{2} \Delta(g_1 , g_2) \; ,
\end{equation}
which can be computed from the classical theory.

It seems then that one can in principle recover any desired expectation value using only Gaussian measurements of smeared fields, even when they do not commute. This is reassuring, as our above arguments suggest that the smeared fields (and the identity) are the only operators which can be measured in this Gaussian manner while still respecting causality. Furthermore, the way in which we tested the causality respecting nature of a given update map, i.e. by using expectation values of products of smeared fields (\eqref{eq:simple_acausal_example_with_expectation_values} for example), can be achieved without any causality violations in and of itself. Thus, it seems, we have an internally consistent and causality respecting model of measurements and unitary kicks in which only generators of the operator algebra can be measured and/or kicked with. Note, this analysis implies that the addition of measurements/kicks for more complicated operators not only introduces causality violations, it is also unnecessary, as any expectation values can already be recovered from the causality respecting smeared field operations.

\subsection{Selective measurements and classical communication}\label{sec:Selective measurements and classical communication}

We have so far been concerned with the causal properties of the non-selective map $\mathcal{E}_C^{\sigma}(\cdot)$. We did not consider the selective map $\mathcal{E}_{C,[a,b]}^{\sigma}(\cdot)$ as the fact that the outcome $\alpha\in[a,b]$ is conditioned on in this case implies some communication between the person measuring $C$ and anyone else, specifically that their outcome landed in $[a,b]$. By assumption we assumed that the person measuring $C$ in $K$ does not communicate with anyone else.

That said, the selective map $\mathcal{E}_{C,[a,b]}^{\sigma}(\cdot)$ can still turn up in protocols where no information is communicated from inside $K$ to other parties, precisely in protocols where information is communicated \emph{within} $K$. For example, the person in $K$ can first make a selective measurement, then, depending on the outcome, choose whether or not to make a second non-selective measurement in $K$. This is just one example of a Local Operations and Classical Communications (LOCC) protocol. Here we will show that such a LOCC protocol, using selective and non-selective Gaussian measurements of smeared fields, amounts to a causal update map. 

\begin{figure}
    \centering
    \includegraphics[scale=0.6]{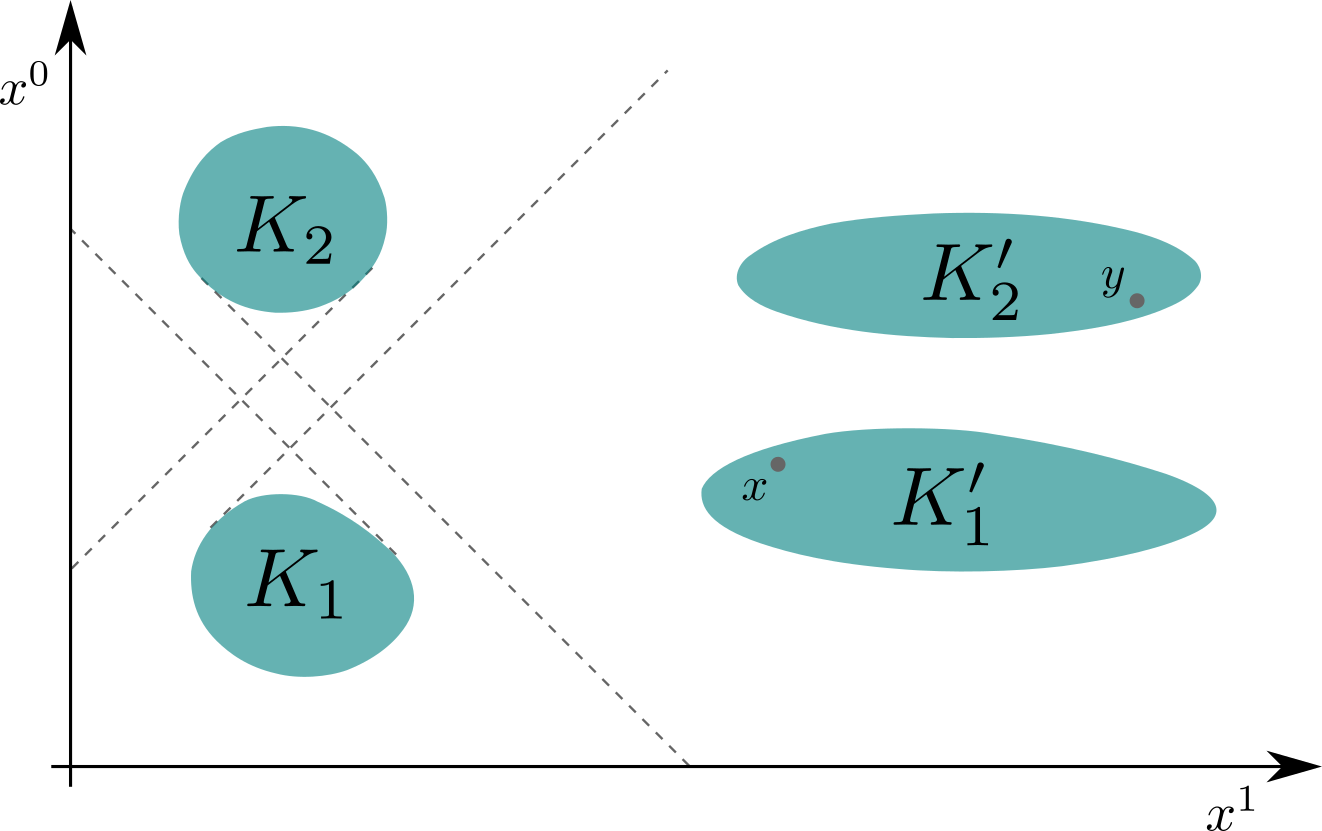}
    \caption{Illustration of \emph{totally timelike} subsets $K_1$ and $K_2$. Every point in $K_1$ is to the past of every point in $K_2$. We have also shown two subsets, $K'_1$ and $K'_2$, that are \emph{not} totally timelike. While $K'_1$ is contained in $J^-(K'_2)$, and $K'_2$ is contained in $J^+(K'_1)$, there are still points from each subset, e.g. $x\in K'_1$ and $y\in K'_2$, that are spacelike.}
    \label{fig:locc_setup}
\end{figure} 

Specifically, we consider two functions $f_1$ and $f_2$ supported in the compact subsets $K_1$ and $K_2$ respectively (Fig.~\ref{fig:locc_setup}). We then selectively measure the smeared field $\phi(f_1)$ in $K_1$, and if the outcome is in $[a,b]$ we non-selectively measure the smeared field $\phi(f_2)$ in $K_2$, otherwise we do nothing in $K_2$. For this to be possible $K_1$ must be \emph{totally timelike} to $K_2$. That is, $K_1/K_2$ is contained in the total past/future of $K_2/K_1$. Such a setup is necessary because we need to collect all the data from our measurement of $\phi(f_1)$ in $K_1$ first, before using the outcome to determine our actions at any point in $K_2$. Put another way, if some point $x_1\in K_1$ is spacelike to some point $x_2\in K_2$, then how would we know what to do at $x_2$? Do we measure or not? The outcome from the measurement at $x_1$ will not have reached $x_2$.

In the case where the outcome of the first measurement of $\phi(f_1)$ in $K_1$ lands in $[a,b]$, which happens with probability $P_{[a,b]}$, the state is updated as $\rho\mapsto \tilde{\mathcal{E}}_{\phi(f_1),[a,b]}^{\sigma}(\rho)$. Following this we make a non-selective measurement of $\phi(f_2)$ in $K_2$, and hence the state is updated as $\tilde{\mathcal{E}}_{C,[a,b]}^{\sigma}(\rho)\mapsto \tilde{\mathcal{E}}_{\phi(f_2)}^{\sigma}(\tilde{\mathcal{E}}_{\phi(f_1),[a,b]}^{\sigma}(\rho))$.

In the other case, where the first outcome lands in $\mathbb{R}\setminus [a,b]$ (with probability $Q_{[a,b]}=1-P_{[a,b]}$), the state is only updated as $\rho\mapsto \tilde{\mathcal{E}}_{\phi(f_1),\mathbb{R}\setminus [a,b]}^{\sigma}(\rho)$, as following this outcome we do nothing in $K_2$.

The final updated state, $\tilde{\mathcal{E}}(\rho)$, after the LOCC protocol has been completed, is given by the sum of these two possibilities, each weighted by its respective probability:
\begin{align}
\tilde{\mathcal{E}}(\rho) = & P_{[a,b]} \tilde{\mathcal{E}}_{\phi(f_2)}^{\sigma}(\tilde{\mathcal{E}}_{\phi(f_1),[a,b]}^{\sigma}(\rho))
\nonumber
\\
& + Q_{[a,b]} \tilde{\mathcal{E}}_{\phi(f_1),\mathbb{R}\setminus [a,b]}^{\sigma}(\rho) \, .
\end{align}
The dual map on some operator $X\in\mathfrak{A}$ is then
\begin{align}
\mathcal{E}(X) = & P_{[a,b]} \mathcal{E}_{\phi(f_1),[a,b]}^{\sigma}(\mathcal{E}_{\phi(f_2)}^{\sigma}(X))
\nonumber
\\
& + Q_{[a,b]} \mathcal{E}_{\phi(f_1),\mathbb{R}\setminus [a,b]}^{\sigma}(X) \, .
\end{align}

Before arguing that $\mathcal{E}(\cdot)$ is causal, we first note that if $X$ is localisable in $(K_2)_{in}$ (\emph{not} in the future of $K_2$) then $\mathcal{E}_{\phi(f_2)}^{\sigma}(X)=X$, and hence the total update map reduces to
\begin{align}
\mathcal{E}(X) & = ( P_{[a,b]} \mathcal{E}_{\phi(f_1),[a,b]}^{\sigma} + Q_{[a,b]} \mathcal{E}_{\phi(f_1),\mathbb{R}\setminus [a,b]}^{\sigma} )(X)
\nonumber
\\
& = \mathcal{E}_{\phi(f_1)}^{\sigma}(X) \; ,
\end{align} 
which we know to be causal. Physically, if $X$ is not in the future of $K_2$, then it does not see the effects of any conditional measurements happening in $K_2$, and hence the update map looks like a non-selective measurement of $\phi(f_1)$ in $K_1$.

Let us now consider the case where $X$ lies partly to the future of $K_2$. To show that $\mathcal{E}(\cdot)$ is causal in this case we can, as above, act with $\mathcal{E}(\cdot)$ on a Weyl generator, $e^{it\phi(g)}$, where $g$ is compactly supported in some region in $(K_2)_{out}$ (and partly to the future of $K_2$ if we want something less trivial). Before attempting this explicit calculation however, we can reason more generally as to why $\mathcal{E}(\cdot)$ is causal in this case.

We first note that $\mathcal{E}_{\phi(f_2)}^{\sigma}(\cdot)$ does not change the localisation region of any operator it acts on. This was shown in Section~\ref{sec:Operations with a smeared field}, specifically~\eqref{eq:smeared_field_operation_on_weyl}. This means that, for any region $R\subseteq (K_2)_{out}$, and any $X\in\mathfrak{A}(R)$, then $Y=\mathcal{E}_{\phi(f_2)}^{\sigma}(X)$ is also localisable in $R$. We then have
\begin{equation}
\mathcal{E}(X) = P_{[a,b]} \mathcal{E}_{\phi(f_1),[a,b]}^{\sigma}(Y) + Q_{[a,b]} \mathcal{E}_{\phi(f_1),\mathbb{R}\setminus [a,b]}^{\sigma}(X) \, ,
\end{equation} 
where $X,Y\in\mathfrak{A}(R)$. Let us denote the two terms on the RHS as $\tilde{Y}$ and $\tilde{X}$ respectively. Both $\tilde{Y},\tilde{X}$ will depend on $\phi(f_1)$ in general. This does not make $\mathcal{E}(\cdot)$ acausal, however, as we now argue.

If $R$ overlaps in any way with the future of $K_2$ then, given that $K_2$ is totally timelike to $K_1$, we know that $K_1$ is entirely contained in the past of $R$. Therefore, any region $R'\subseteq (K_1 \cup K_2 )_{in} = (K_1)_{in}$, and spacelike to $R$, is also spacelike to $K_1$, and hence any $A\in\mathfrak{A}(R')$ commutes, not only with $X$ and $Y$, but also with $\phi(f_1)$, and therefore with $\tilde{X}$ and $\tilde{Y}$. That is, any $A\in\mathfrak{A}(R')$ commutes with $\mathcal{E}(X)$. This means that unitary kicks with $\phi(h)\in\mathfrak{A}(R')$ act trivially on $\mathcal{E}(X)$. From Section~\ref{sec:Kicking causality conditions into shape} we know that this implies the map $\mathcal{E}(\cdot)$ is causal.

For completeness, we will now compute $\mathcal{E}(e^{i t\phi(g)})$ to explicitly show the causality of $\mathcal{E}(\cdot)$. From~\eqref{eq:smeared_field_measurement_on_weyl} we have
\begin{align}
\mathcal{E} ( & e^{it \phi(g)}) 
\nonumber
\\
& = P_{[a,b]} \mathcal{E}_{\phi(f_1),[a,b]}^{\sigma}(\mathcal{E}_{\phi(f_2)}^{\sigma}(e^{it \phi(g)}))
\nonumber
\\
& \hspace{10mm} + Q_{[a,b]} \mathcal{E}_{\phi(f_1),\mathbb{R}\setminus [a,b]}^{\sigma}(e^{it \phi(g)}) 
\nonumber
\\
& = P_{[a,b]} \mathcal{E}_{\phi(f_1),[a,b]}^{\sigma}(
e^{-t^2\frac{\Delta(f_2 , g)^2}{8\sigma^2}}e^{it\phi(g)}
)
\nonumber
\\
& \hspace{10mm}+ Q_{[a,b]} \mathcal{E}_{\phi(f_1),\mathbb{R}\setminus [a,b]}^{\sigma}(e^{it \phi(g)})
\nonumber
\\
& = \frac{1}{\sigma\sqrt{2\pi}} \int_{\mathbb{R}}d\alpha \, e^{-\frac{t^2}{8\sigma^2}(1_{[a,b]}(\alpha)\Delta(f_2,g)^2 + \Delta(f_1,g)^2)}
\nonumber
\\
& \hspace{25mm} \times e^{-\frac{(\tilde{C}(t) - \alpha)^2}{2\sigma^2}}e^{it\phi(g)} \, ,
\end{align}
where $\tilde{C}(t) = \phi(f_1)+(t/2)\Delta(f_1,g)$, and where $1_{[a,b]}(\alpha)$ is an indicator function for $\alpha\in[a,b]$. The Weyl generator, $e^{it\phi(g)}$, can be moved to the right, outside the integral, as it does not depend on $\alpha$. The integral can then be evaluated, resulting in
\begin{align}\label{eq:locc_measurement_on_weyl}
\mathcal{E} & (e^{it\phi(g)} )
\nonumber
\\
= & \left( 1 + \frac{1}{2}\left(1 - e^{-\frac{t^2}{8\sigma^2}\Delta(f_2,g)^2 } \right) D \right)e^{-\frac{t^2}{8\sigma^2}\Delta(f_1,g)^2 }e^{it\phi(g)} ,
\end{align}
where the operator
\begin{equation}
D = \text{erf}\left(\frac{\tilde{C}(t)-b}{\sqrt{2}\sigma} \right)- \text{erf}\left(\frac{\tilde{C}(t)-a}{\sqrt{2}\sigma} \right) \; ,
\end{equation}
is non-trivial in the localisation region for $\phi(f_1)$. Here $\text{erf}(\cdot)$ denotes the standard error function.

It is clear from~\eqref{eq:locc_measurement_on_weyl} that if $g$ is supported in $(K_2)_{in}$ (not in the future of $K_2$), then $\Delta(f_2,g)=0$ and hence $\mathcal{E}(e^{it\phi(g)})$ reduces to the term on the RHS after the brackets, i.e. a Gaussian measurement of $\phi(f_1)$. This agrees with our earlier discussion. As previously stated, the fact that $D$ is non-trivial in $K_1$ does not cause any causality violations. As can be seen from~\eqref{eq:locc_measurement_on_weyl}, the term involving $D$ only appears when $\Delta(f_2,g)\neq 0$, and hence when $g$ is supported partly to the future of $K_2$. In such a case the support of $g$ contains the entirety of $K_1$ in its past (owing to the fact that $K_1$ and $K_2$ are totally timelike). Thus the past-support of $e^{it\phi(g)}$ has not been increased.

\section{Interactions}\label{sec:Interactions}

In~\cite{borsten2021impossible} it was mentioned that the situation for causality violations could be worse in an interacting theory. For example, it may be the case that even smeared fields cannot be measured. Here we sketch an argument as to why this is not the case, at least for interactions that are only turned on in a compact subset $L$. The general idea is to construct a scattering map from smeared fields in the in-algebra $\mathfrak{A}(L_{in})$ to smeared fields in the out-algebra $\mathfrak{A}(L_{out})$, both of which are isomorphic to the entire algebra $\mathfrak{A}$ as $L_{in}$ and $L_{out}$ both contain Cauchy surfaces for $M$. While this mapping is non-linear in the smearing functions, it does not increase the support in an acausal manner, which, as we will see, ensures that measurements/kicks with smeared fields are still causal.

One can either consider self-interactions or interactions with another field. In both cases the argument is very similar, and can be formulated for the most part in the classical theory. In that regard let us briefly review some relevant points about the classical theory and its connection to the quantum theory.

\begin{figure}
    \centering
    \includegraphics[scale=0.6]{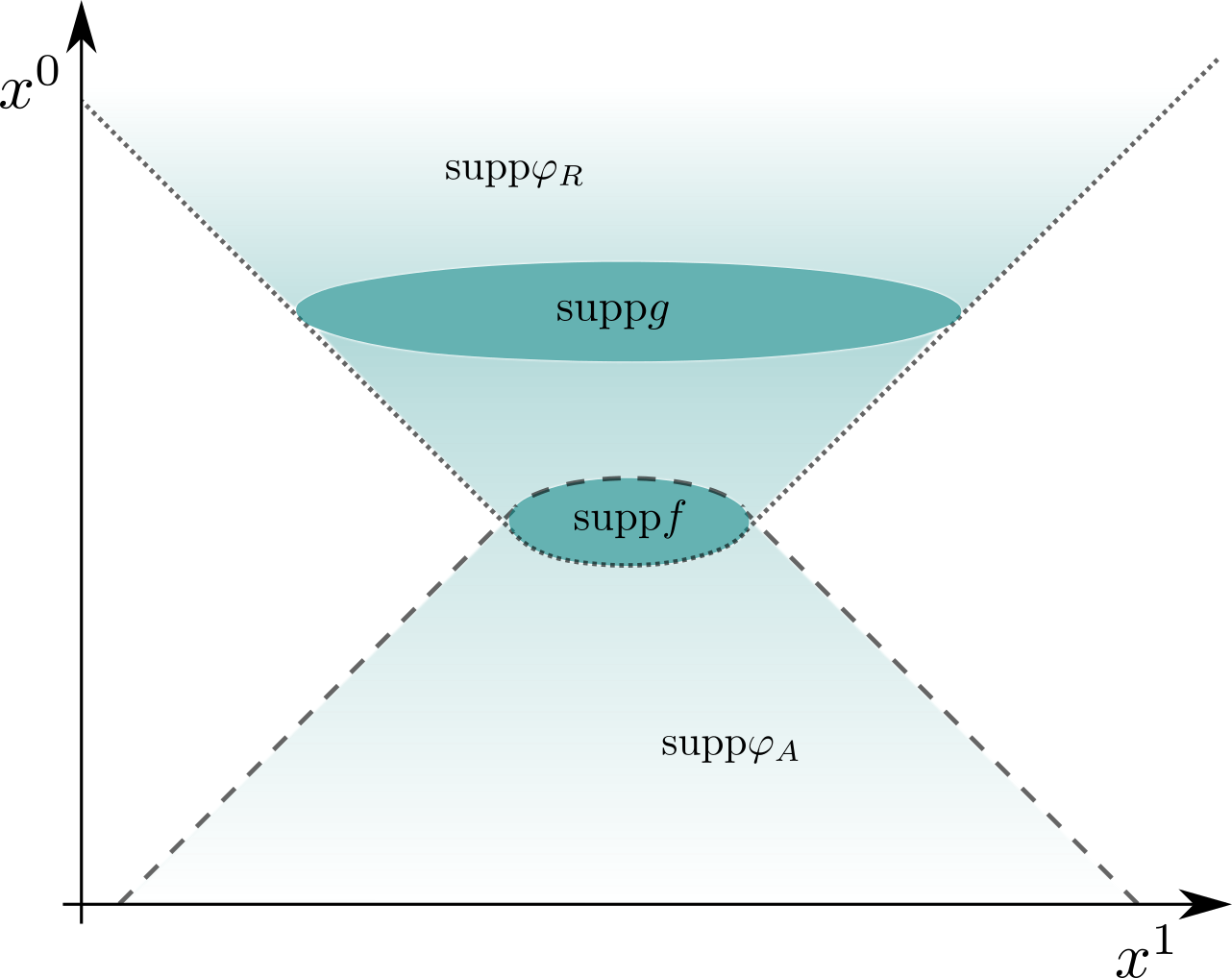}
    \caption{Spacetime diagram of the solution $\varphi$ generated by $f$. The support of $\varphi_{R/A}$ is contained in the future/past of the support of $f$, as shown by the dotted/dashed lines. The support of $\varphi$ is the union $\text{supp}\varphi_A \cup\text{supp}\varphi_R$, and hence $\varphi(x)=0$ at points $x$ that are spacelike to $\text{supp}f$. Also illustrated is an example of another function $g$ that generates the same solution $\varphi$. Note how the wave packet generated by $f$ (or $g$) is of compact spatial support, since it only has a finite width in the $x^1$ direction at any time $x^0$.}
    \label{fig:solution}
\end{figure}

Consider the free equation of motion for the classical field $\varphi$:
\begin{equation}\label{eq:hom_eqn_of_motion}
( \Box + m^2 )\varphi = 0 \; .
\end{equation}
Any spatially compact solution (e.g. a wave packet with finite spatial extent) can be written as
\begin{equation}\label{eq:classical_sol_in_terms_of_delta_and_f}
\varphi(x) = \int_M dy \, \Delta(x,y)f(y) \; ,
\end{equation}
for some smooth and compactly supported test function $f$. We say that $f$ generates the classical solution $\varphi$, and by writing~\eqref{eq:classical_sol_in_terms_of_delta_and_f} as $\varphi = \Delta f$ we can think of $\Delta$ as an operator on test functions $f$. Recall that $x$ and $y$ denote spacetime points, and that the Pauli-Jordan function, $\Delta(x,y)$, is the difference between the retarded and advanced Green functions, $G_{R/A}(x,y)$. We can therefore decompose the solution as $\varphi = \varphi_R - \varphi_A$, where
\begin{equation}\label{eq:classical_adv_ret_sol_in_terms_of_G_and_f}
\varphi_{R/A}(x) = \int_M dy \, G_{R/A}(x,y)f(y) \; .
\end{equation}
is a past/future compact solution to the inhomogeneous equation
\begin{equation}\label{eq:inhom_eqn_of_motion}
( \Box + m^2 )\varphi_{A/R} = f \; .
\end{equation}
We can similarly write $\varphi_{R/A} = G_{R/A}f$. The support of $\varphi_{R/A}$ is contained in the future/past of the support of $f$, as shown in Fig.~\ref{fig:solution}. $f$ can therefore be thought of as the generator of the solution $\varphi = \Delta f$ to the homogeneous equation, \emph{and} as the source of either a past or future compact solution to the inhomogeneous equation.

There is no unique $f$ that generates a spatially compact solution $\varphi$ via $\varphi = \Delta f$. Two different test functions $f$ and $g$ that generate the same solution, i.e. $\Delta f = \Delta g$, can even be supported in disjoint regions, as shown in Fig.~\ref{fig:solution}. This lack of uniqueness appears in the quantum theory too, e.g. $\phi(f)=\phi(g)$ for any two test functions satisfying the classical equation $\Delta f = \Delta g$. In Section~\ref{sec:QFT} this was stated in a different, but equivalent way, as $f-g = (\Box + m^2 ) h$ for some test function $h$. As stated in Section~\ref{sec:QFT}, this is equivalent to imposing the homogeneous equations of motion on the operator-valued distribution $\phi(x)$ in~\eqref{eq:smeared_field_in_terms_of_op_valued_dist}.

Given that the smeared field operators can be the same for different test functions, the region within which a smeared field is localisable is clearly not unique. The two regions in Fig.~\ref{fig:solution} are examples of possible localisation regions. This non-uniqueness is not completely arbitrary; starting from the smearing function $f$ for example, we cannot move the localisation region to a region spacelike to $\text{supp}f$. 

\begin{figure}
    \centering
    \includegraphics[scale=0.6]{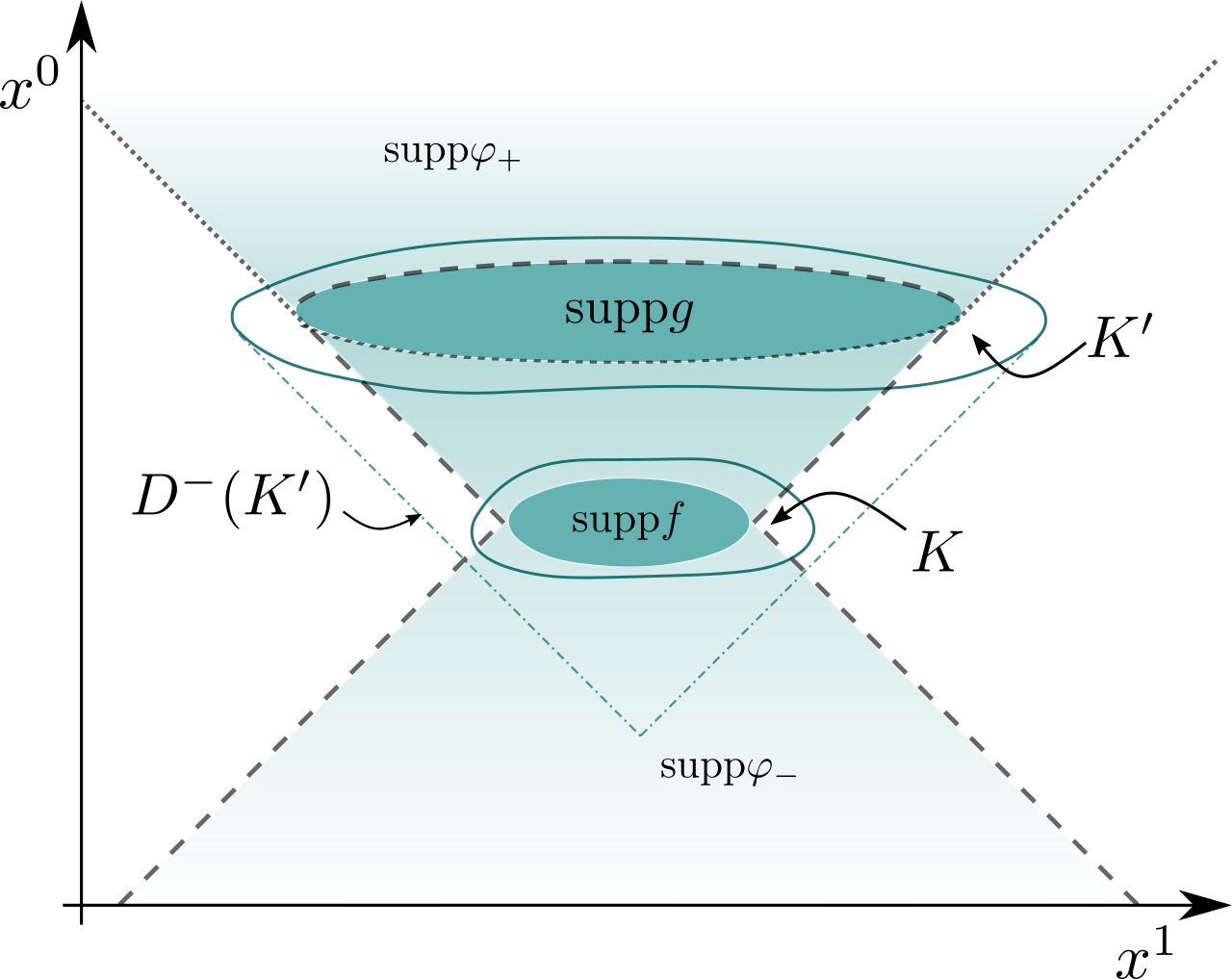}
    \caption{Illustration of the partition $\varphi = \varphi_+ - \varphi_-$. The support of $\varphi_{+/-}$ vanishes to the past/future of $K'$, shown by the dotted/dashed lines. The supports of $\varphi_-$ and $\varphi_+$ only overlap in $K'$. Note that $K\subseteq D^-(K')$, as can be seen by the dotted-and-dashed lines.}
    \label{fig:move_solution}
\end{figure} 

In practice we can `move' the smeared field in the following way. Consider the test function $f$ supported in some compact subset $K$, and the classical solution $\varphi=\Delta f$ that it generates. Now consider some compact $K'$ such that $K\subseteq D(K')$ (Fig.~\ref{fig:move_solution}), in which we want to localise $\phi(f)$. This amounts to the classical problem of finding some $g$ supported in $K'$ that generates $\varphi$ as $\varphi = \Delta g$. To do this we first choose any smooth partition of $\varphi$, i.e. $\varphi = \varphi_+ - \varphi_-$, where the supports of $\varphi_+$ and $\varphi_-$ only intersect in $K'$, and $\varphi_{+/-}$ vanishes to the past/future of $K'$. This is shown in Fig~\ref{fig:move_solution}. Given the supports of $\varphi_{+/-}$ it is clear that, to the future/past of $K'$, we have $\pm\varphi_{+/-}=\varphi$. Since $\varphi$ satisfies the homogeneous equation in~\eqref{eq:hom_eqn_of_motion}, then $\varphi_{+/-}$ also satisfies~\eqref{eq:hom_eqn_of_motion} to the future/past of $K'$. Furthermore, $\varphi_{+/-}$ trivially satisfies~\eqref{eq:hom_eqn_of_motion} to the past/future of $K'$ as it vanishes there. Inside $K'$, however, $\varphi_{+/-}$ may not satisfy~\eqref{eq:hom_eqn_of_motion}. Let $g$ be the function capturing this inability of $\varphi_{+/-}$ to satisfy~\eqref{eq:hom_eqn_of_motion} inside $K'$, i.e. $g = (\Box + m^2)\varphi_{+/-}$. Note that $g$ is the same for both $\varphi_+$ and $\varphi_-$, as the difference $\varphi = \varphi_- - \varphi_+$ satisfies the homogeneous equation everywhere, including inside $K'$. By our previous arguments we also know that $g$ is compactly supported in $K'$. Furthermore, it is by definition a source for the past/future compact solution $\varphi_{+/-}$ of the inhomogeneous equation, and hence we can write $\varphi_{+/-} = G_{R/A} g$. Therefore, we have
\begin{align}
\Delta g & = G_R g - G_A g
\nonumber
\\
& = \varphi_+ - \varphi_- 
\nonumber
\\
& = \varphi \; .
\end{align}
That is, $\varphi = \Delta g$, and so $g$ is equivalent to $f$ in that they generate the same solution to the homogeneous equation. We therefore have the operator equality $\phi(f)=\phi(g)$, and hence this smeared field operator is localisable in $K'$ as desired. The above argument has glossed over some technical details explained more thoroughly in~\cite{Fewster2020}.

Let us now turn on a self-interaction in some compact subset $L$. Specifically, we modify the classical homogeneous equation to
\begin{equation}\label{eq:int_eqn_of_motion}
(\Box + m^2)\varphi = \kappa \chi \varphi^2 \; ,
\end{equation}
where $\kappa\in\mathbb{R}$ is the interaction parameter, and $\chi$ is some smooth function, supported in $L$, which controls the interaction. Here we have picked a $\varphi^2$ interaction (a $\varphi^3$ interaction in the associated action) as an example. The explicit form of the interaction is not so important for our discussion, however, and so one can substitute in some other interaction in what follows.

\begin{figure}
    \centering
    \includegraphics[scale=0.6]{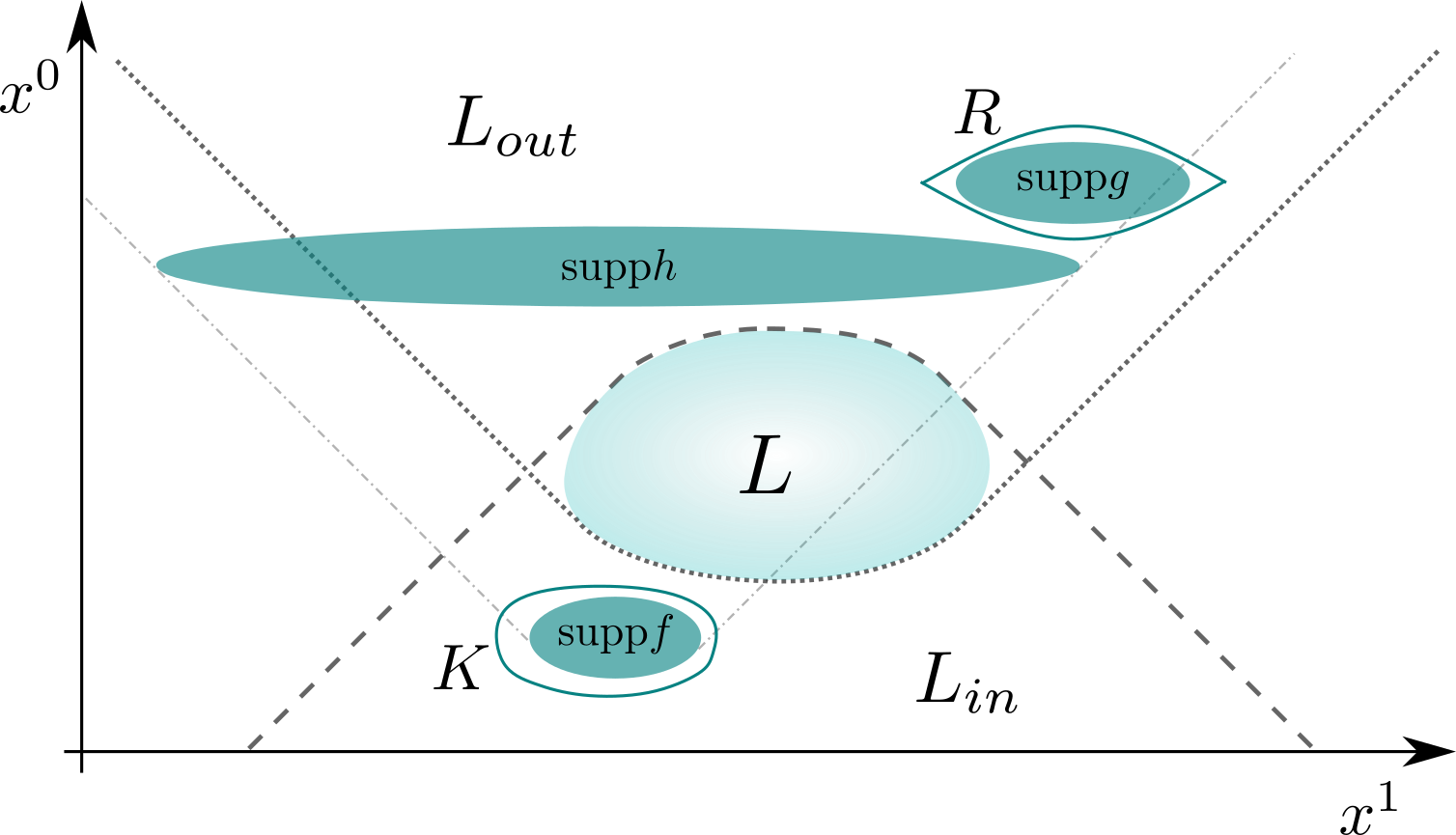}
    \caption{Spacetime diagram of the coupling zone $L$. The \emph{in}/\emph{out}-region, denoted $L_{in}$/$L_{out}$, consists of all the points below/above the dotted/dashed line in the figure, i.e. all the points \emph{not} to the future/past of $L$. Therefore, the in/out-region also contains points spacelike to $L$. Also shown are the supports of the smearing functions $f$ and $g$, whose causal interval intersects the coupling region $L$. Through the argument in the text we can map the smeared field $\phi(f)$ from the \emph{in} algebra to the smeared field $\phi(h)$ in the \emph{out} algebra, after which we can compute the commutator with $\phi(g)$. Note that $\text{supp}h$ lies entirely in the out-region. It also does not need to be disjoint from $\text{supp}g$, although this is the case in the figure.}
    \label{fig:interaction}
\end{figure}

From this compact $L$ we get the associated in- and out-regions $L_{in/out}$. This is illustrated in Fig.~\ref{fig:interaction}. In $L_{in/out}$ the interaction is turned off, and hence the theory matches the free case. $L_{in/out}$ is also globally hyperbolic in its own right, and hence the quantum theory restricted to $L_{in/out}$ goes through as in Section~\ref{sec:QFT}. In particular, for any functions $f$ and $g$ both supported in either $L_{in}$ or $L_{out}$, we have $[\phi(f),\phi(g)]=i \Delta(f,g)$, where $\Delta(f,g)$ is the same as in the free theory. Similarly, for any sequence of measurements/kicks all contained in either $L_{in}$ or $L_{out}$, our previous results go through unchanged.

There will be a change, however, if $f$ is supported in $L_{in}$ say, and $g$ is supported in $L_{out}$, as shown in Fig.~\ref{fig:interaction}. Crucial to our discussion on measurements and kicks is the commutator $[\phi(f),\phi(g)]$. Currently, this commutator is undefined, as $\phi(f)$ belongs to the algebra of operators in the in-region $L_{in}$, and $\phi(g)$ to algebra in the out-region $L_{out}$. To define this commutator we first have to map $\phi(f)$ to some operator in the algebra for the out-region. This can be done as follows.

Given the test function $f$ supported in $L_{in}$ we can generate $\varphi_0 = \Delta f$ which solves the free homogeneous equation throughout the entirety of $M$. Furthermore, $\varphi = \varphi_0$ trivially solves the interacting equation~\eqref{eq:int_eqn_of_motion} in $L_{in}$ as the interaction vanishes there. The only region where we do not yet know the interacting solution $\varphi$ is the future of $L$, $J^+(L)$. There we can perturbatively construct the interacting solution $\varphi = \varphi_0 + \kappa \varphi_1 + O(\kappa^2)$ order by order in $\kappa$. For instance, for $\varphi_1$ we get the equation $(\Box + m^2)\varphi_1 = \chi {\varphi_0}^2$, and hence $\varphi_1 = G_R (\chi {\varphi_0}^2)$. Note that $\varphi_n=0$ for $n>0$ in $L_{in}$.

We now have the perturbative solution $\varphi$ throughout all of $M$, and hence all of $L_{out}$. Since the theory is free in $L_{out}$ we can perturbatively find some $h=h_0 + \kappa h_1 + O(\kappa^2)$, supported in $L_{out}$, that generates the solution as $\varphi = \Delta h$ in $L_{out}$. Order by order we have $\varphi_n = \Delta h_n$, where each $h_n$ can be found using the above prescription, i.e. we partition $\varphi_n = \varphi_{n,+} - \varphi_{n,-}$, where the supports of $\varphi_{n,+}$ and $\varphi_{n,-}$ only intersect in some compact subset of $L_{out}$, and then set $h_n = (\Box +m^2)\varphi_{n,\pm}$. A example of $h$ is shown in Fig.~\ref{fig:interaction}. Note that $h$ is not linearly dependent on $f$. That is, if $f\mapsto \lambda f$, then it is \emph{not} the case that $h\mapsto \lambda h$.

In the quantum theory we then have the map $\phi(f)\leftrightarrow \phi(h)$ between the smeared field operators for the in- and out-regions respectively. We have used the double-ended arrow as we can also map smeared fields from the out- to the in-region following the above procedure. Note this map is non-linear in the test functions. The full \emph{scattering map} is an algebra homomorphism between the algebra of operators for the in- and out-regions, and is defined by extending the map between smeared fields to all sums and products in the obvious way. For example, $\phi(f)^2\leftrightarrow \phi(h)^2$. The commutator $[\phi(f),\phi(g)]$, for $g$ supported in the out region, is then \emph{defined} by first mapping $\phi(f)\mapsto\phi(h)$, and then computing $[\phi(h),\phi(g)] = i \Delta(h,g)$ within $L_{out}$. Essentially, $[\phi(f),\phi(g)]= i\Delta(h,g)$. 

One may worry that the non-linear dependence of $\Delta(h,g)$ on $f$ affects our previous results on causality. In fact, this non-linear dependence is not relevant to the question of causality when measuring/kicking a smeared field. Consider equation~\eqref{eq:smeared_field_operation_on_weyl} where we showed that smeared field operations do not increase the support of some Weyl generator. There $f$ was supported in some compact subset, $K$ say, and $g$ was supported in a region $R\subseteq K_{out}$. Imagine now that the compact subset $L$, where the interaction is turned on, is situated between $K$ and $R$ as in Fig.~\ref{fig:interaction}. Specifically, $K\subset L_{in}$ and $R\subseteq L_{out}$. To take this interaction into account in~\eqref{eq:smeared_field_operation_on_weyl} we simply need to swap $\Delta(f , g)$ for $\Delta(h , g)$, where $h$ is supported in $L_{out}$ and is related to $f$ via the above scattering map. This can be done perturbatively, i.e. order by order in $\kappa$, if convergence is not guaranteed. This changes the precise form of the RHS of~\eqref{eq:smeared_field_operation_on_weyl} as a function of $f$, and hence the precise effect of the operation on any future measurements. Crucially, however, it does not change the fact that the support of the Weyl generator has not increased, and hence the operation is still causal, even in the presence of this self-interaction. We can similarly argue, using~\eqref{eq:smeared_field_kick_on_weyl}, that self-interactions do not make smeared field kicks acausal.

This argument can be readily applied to other self-interactions, and even to interactions with other fields via the scattering map in~\cite{Fewster2020}. In each case the precise form of $\Delta(h , g)$ changes, but importantly the support is never increased. In this sense we can say that interactions do not make the situation for causality violations `worse' than the free case.

\section{Discussion}\label{sec:Discussion}

\subsection{Future directions}

Before discussing the implications of our results, there are a number of future directions to note. While we touched on the case of ideal measurements, the fact that they seem acausal for the simplest operators, i.e. the smeared fields, warrants further study. Many of our results also seem transformable into the probe framework in~\cite{Fewster2020,fewster2019generally,Bostelmann_2021}. It would be useful to construct an explicit dictionary between update maps and specific probe models, and to determine whether this is possible in general. The latter would be analogous to Stinespring's Dilation Theorem~\cite{Stinespring}, but with the added restrictions of locality and causality on the unitary map. On this point we note that our discussion of update maps using only the main field of interest, i.e. with no additional probe fields, is still useful in that it allows us to say which operators of the main (and only) field can be causally measured (in the standard sense of quantum theory) without reference to another probe field, specifically because we can associate to any operator $C$ an update map $\mathcal{E}_C^{\sigma}(\cdot )$. Finally, it would be illuminating to also translate our results from the canonical picture into the path integral framework for quantum theory.

\subsection{Relation to continuous measurement models}

We briefly note the relevance of the above results to continuous measurement models~\cite{Brun_2000,Jacobs_2006}. In these models Gaussian measurements of a chosen operator are enacted repeatedly, in intervals of duration $\Delta t$. The $\Delta t\rightarrow 0$ limit is then taken to make the sequence of measurements effectively continuous. These models are useful in many applications, including \emph{feedback control} (e.g.~\cite{PhysRevLett.96.043003}), where the results of the measurements are used to continuously update the Hamiltonian. For a lattice system one of the simplest cases to study is continuous measurements of operators local to a single site, e.g. local number operators as in~\cite{Cao_2019}. Since such operators are local, in the sense that they commute with operators on different sites, their associated Gaussian measurements do not increase support, and hence are causal, \textit{c.f.} Gaussian measurements of smeared fields. One can also consider models involving Gaussian measurements of operators that couple neighbouring sites. In this case the Gaussian measurements increase support, and are therefore acausal. For a non-relativistic lattice system this means that the Gaussian measurements cannot be implemented faster than the light-travel time between the neighbouring sites. This furnishes a fundamental lower bound on the measurement duration $\Delta t$. That is, `continuous' measurement models such as these can only ever be approximately continuous on timescales much larger than this lower bound. In practice this lower bound may be negligible compared to the timescales present in the lattice model, and hence the assumption of a continuum of measurements is justified.

\subsection{Physical implications}

Returning to QFT, from a philosophical perspective our claim that only the generators can be measured may have important implications for the ontology of the theory. The conventional picture in quantum mechanics is that one can associate to any physical observable a self-adjoint (and gauge invariant) operator. In measuring this observable we usually expect two things from quantum theory: i) a probability distribution over the possible measurement outcomes of the observable, and ii) a map to update the state of the system. The latter is crucial in accurately reflecting the effect of the current measurement on any future measurements.

For a smeared field operator $\phi(f)$ we meet these two requirements: the pdf over possible measurement outcomes is given explicitly in~\eqref{eq:pdf_for_alpha}, and we can update the state via the associated Gaussian update map, $\mathcal{E}^{\sigma}_{\phi(f)}(\cdot)$, since it is causal. 

In the case of more complicated self-adjoint operators requirement ii) is not obviously met, as it seems the associated Gaussian update maps cannot be implemented without violating causality. In this way these operators do not correspond to observables in the usual sense. Requirement i) is still be met however, as expectation values and higher moments of any self-adjoint operators can be recovered from Gaussian measurements of smeared fields \textit{a la} Section~\ref{sec:Extracting expectation values from measurements of smeared fields}. 

While the update map $\mathcal{E}^{\sigma}_C(\cdot)$ for some self-adjoint operator $C$ may not be possible, we can nevertheless meet requirement ii) by instead composing the (causality respecting) update maps $\mathcal{E}^{\sigma}_{\phi(f_i)}(\cdot)$ for the relevant smeared fields $\phi(f_i)$ used in the construction of $C$ (potentially with some symmetrisation). 

In this way we can in fact associate to any self-adjoint operator a causality respecting update map, thus meeting the conventional requirements, i) and ii), of an observable in quantum theory. In doing this, however, we must understand that our use of the update maps $\mathcal{E}^{\sigma}_{\phi(f_i)}(\cdot)$ for the relevant smeared fields implies that, physically speaking, we are really measuring the smeared fields $\phi(f_i)$, and not $C$. The `measurement' of $C$ should be thought of as secondary to the measurement of the smeared fields $\phi(f_i)$, in the sense that any expectation values, or higher moments, of $C$ are actually constructed after the fact, \textit{a la} Section~\ref{sec:Extracting expectation values from measurements of smeared fields}, using the outcomes of the measurements of each $\phi(f_i)$. 

Depending on one's preferred interpretation of quantum mechanics, or one's preferred outlook on QFT, this may amount to a different ontology --- a different picture of what is physically there. Specifically, one way to interpret (and extrapolate from) the above results is that smeared fields (and the identity) are the only physical observables, and that other self-adjoint operators in the algebra simply correspond to different ways to combine the outcomes resulting from measurements of smeared fields. This unconventional viewpoint, where all self-adjoint operators except the smeared fields and the identity are `culled' from the list of QFT observables, necessitates further justification.

To illustrate how one could take this viewpoint we focus on the example in Section~\ref{sec:A simple acausal example}. That is, we have two smeared fields $\phi(f_1)$ and $\phi(f_2)$, localisable in spacelike regions. If we perform two Gaussian measurements, one for $\phi(f_1)$ and one for $\phi(f_2)$, the respective measurement outcomes, $\alpha$ and $\beta$, are distributed according to the joint pdf $p(\alpha , \beta)$ in~\eqref{eq:joint_pdf_commuting_fields}. Now consider the self-adjoint operator $C=\phi(f_1)\phi(f_2)$. We know from Section~\ref{sec:A simple acausal example} that $\mathcal{E}^{\sigma}_C(\cdot)$ is acausal. From the above discussion we can instead take the update map to be the composition of $\mathcal{E}^{\sigma}_{\phi(f_1)}(\cdot)$ and $\mathcal{E}^{\sigma}_{\phi(f_2)}(\cdot)$, in which case the pdf over the possible outcome values for this measurement of $C$, denoted by $\tilde{p}(\gamma)$, is the product pdf over the variable $\gamma = \alpha\beta$ (determined by the joint pdf $p(\alpha ,\beta)$ over the \emph{dependent} random variables $\alpha$ and $\beta$). In this precise sense one could argue that $C$ is not an `independent' observable in its own right. It does not, for example, come with its own pdf of the the form~\eqref{eq:pdf_for_alpha}.

\subsection{Comparison with non-relativistic quantum mechanics}

To highlight how this deviates from our usual intuition, let us examine the analogous situation in non-relativistic quantum mechanics (NRQM). Consider the operator $Z = X Y$, where $X$ and $Y$ are two commuting operators local to separate parts of a bipartite system, e.g. $X = \tilde{X} \otimes \mathds{1}$ and $Y = \mathds{1}\otimes \tilde{Y}$ for some $\tilde{X}$ and $\tilde{Y}$ local to different parts of the system. In this case there is no reason to rule out the map $\mathcal{E}^{\sigma}_Z(\cdot)$ in favour of the composition $\mathcal{E}^{\sigma}_X(\mathcal{E}^{\sigma}_Y(\cdot))=\mathcal{E}^{\sigma}_Y(\mathcal{E}^{\sigma}_X(\cdot))$. In the QFT setting this is precisely what we have done.

To further emphasise the distinction between the QFT and NRQM, recall that the $\sigma\rightarrow 0$ limit of a Gaussian map is an ideal measurement (for compact self-adjoint operators). Therefore, for some sufficiently small $\sigma$, the analogous statement in NRQM is that we cannot make an ideal measurement of $Z=XY$, but we can make two ideal measurements of $X$ and $Y$ (in either order). This is demonstrably not the case; ideal measurements of product operators such as $Z$ are routinely considered in QI.

We have to be careful, though, in making this connection, as we are applying infinite dimensional continuum QFT results to the finite dimensional Hilbert space of the bipartite system. To properly emulate our support increasing Gaussian measurement of $C=\phi(f_1)\phi(f_2)$ in NRQM we should at least consider an operator $Z$ whose ideal measurement is support increasing, or equivalently, is one that enables a (subluminal) signal. Accordingly, we consider the operator $Z = \ket{1}\bra{1}\otimes \sigma^z$ on two qubits $A$ and $B$ (where $\sigma^z = \ket{0}\bra{0}-\ket{1}\bra{1}$ denotes the Pauli-$z$ matrix). Note that $X = \ket{1}\bra{1}\otimes \mathds{1}$ and $Y=\mathds{1}\otimes \sigma^z$ here. In~\cite{borsten2021impossible} it was shown that an ideal measurement of $Z = \ket{1}\bra{1}\otimes \sigma^z$ enables a signal.

Even with this choice of $Z$ it is still the case that we can perform the associated ideal measurement, and we do not need to resort to a composition of ideal measurements of $X = \ket{1}\bra{1}\otimes \mathds{1}$ and $Y=\mathds{1}\otimes \sigma^z$. This non-relativistic example, therefore, \emph{still} differs from the QFT case. In the latter, a Gaussian measurement of $\phi(f_1)\phi(f_2)$ is physically impossible, and one can only do Gaussian measurements of $\phi(f_1)$ and $\phi(f_2)$ separately. Why, then, are these two situations different, and how can we reconcile this? Furthermore, such a reconciliation seems necessary if NRQM is to arise as an effective description of QFT.

To answer these questions we must focus on how an ideal measurement of $Z$ is realised experimentally. As mentioned in~\cite{borsten2021impossible}, one can use the following 2 step LOCC (Local Operations and Classical Communication) protocol: 1) The experimenter first measures the $z$-spin on qubit $A$. 2) If it is down then they do nothing on qubit $B$, and if it is up then they measure the $z$-spin of qubit $B$. One can verify that the associated sequence of update maps amounts to the update map for an ideal measurement of $Z$.

Notably, this realisation requires information about the measurement outcome on qubit $A$ to be sent to qubit $B$ before qubit $B$ is (potentially) measured. The spacetime regions in which the measurements of each qubit take place are therefore \emph{timelike} related, and \emph{not spacelike}. This is the crucial distinction to the QFT case. There the operators $\phi(f_1)$ and $\phi(f_2)$ are localisable in spacelike regions, and even if we `move' the smeared fields around to $\phi(f'_1)$ and $\phi(f'_2)$ say, using the procedure in Section~\ref{sec:Interactions}, we can never make them \emph{totally} timelike related, i.e. such that there are no pairs of points, $x\in\text{supp}f'_1$ and $y\in\text{supp}f'_2$, that are spacelike. This is essentially why the two situations are different; why an ideal measurement of $Z$ is possible in NRQM but a Gaussian measurement of $\phi(f)\phi(g)$ in QFT is not.

Given that the two qubit measurements are totally timelike, the NRQM example is then more comparable to the Gaussian LOCC protocol in Section~\ref{sec:Selective measurements and classical communication}, which we found to be causal. This should provide some reassurance as to why our QFT results are not contradictory with standard NRQM experiments. From a QFT perspective, the update map for an ideal measurement of $Z$ is physically realisable because it is simply an effective description of some underlying causal update map in the QFT setting.

This is not obvious from the form of the operator, $Z = \ket{1}\bra{1}\otimes \sigma^z$, however. Given that $\ket{1}\bra{1}$ and $\sigma^z$ are local to separate parts of the tensor product we get the impression that they are analogous to spacelike observables in QFT. This led us to incorrectly compare $Z$ with $\phi(f_1)\phi(f_2)$, where $f_1$ and $f_2$ are spacelike. `Hidden' in the ideal measurement of $Z$ is knowledge that the measurements of the two qubits happen in timelike regions. To make the situation more comparable to $\phi(f_1)\phi(f_2)$ we can instead ask if the measurement of $Z$ can be performed using spacelike qubit measurements. In other words, can we perform an ideal measurement of $Z$ faster than the light-travel time between the qubits? This, like a Gaussian measurement of $\phi(f_1)\phi(f_2)$, is impossible. As shown in~\cite{borsten2021impossible}, if we were to make such a measurement of $Z$ we would enable a superluminal signal, and hence it is impossible.

In this way, the conclusion that a Gaussian measurement of $\phi(f_1)\phi(f_2)$ is impossible, and hence why $\phi(f_1)\phi(f_2)$ (and many other operators) fail requirement ii) for a typical observable in quantum theory, is more reasonable. Ruling out $\phi(f_1)\phi(f_2)$ as unobservable is analogous to ruling out $Z$ as unobservable on time-scales shorter than its light-travel time -- the latter being perfectly reasonable to those in QI.

%

\section{Conclusion}\label{sec:Conclusion}

Above we precisely formulated Sorkin's additional causality condition that any state update in QFT should obey to respect causality. Through the use of unitary kicks with smeared fields we showed that causal state updates in real scalar QFT are precisely those that are \emph{past-support non-increasing} (PSNI). Moreover, we argued that PSNI state updates are causal more generally, specifically for the physical subalgebras of complex scalar and fermionic QFT. We then went on to consider a variety of update maps in real scalar QFT with a focus on Gaussian measurements. Our calculations suggest that only Gaussian measurements/unitary kicks with the generators (the smeared fields and the identity) are causal, while measurements/kicks with other more complicated operators are acausal. Additionally, ideal measurements of smeared fields appear to be acausal, though a more thorough analysis needs to be done. Using Gaussian measurements of smeared fields alone we then sketched how one could recover expectation values of products of smeared fields, and following this we discussed the addition of a compactly supported interaction.

In the last section we discussed some future directions and relations to continuous measurement models. We then went on to discuss the physical implications of our findings, arguing that the generators of the algebra seem to be the only physical observables, at least in the usual sense of quantum physics. Despite our above reasoning, a shift in ontology as radical as culling all self-adjoint operators, bar smeared fields and the identity, from the list of QFT observables certainly requires further scrutiny before it should be taken seriously. In particular, we have only focussed on a few particular classes of update maps, and even within the set of these maps there are more LOCC protocols that can be investigated.

Lastly, while this potential shift in ontology would not obviously be of any practical importance, it may be relevant in the construction of more fundamental theories. Specifically, a clearer understanding of what is physical in curved spacetime QFT will, most likely, better inform our decisions as to which physical principles to retain in quantum gravity.

{\it Acknowledgements:}
The author would like to acknowledge L. Borsten, G. Kells, L. Coopmans, A. Conlon, M. Ruep, C. Fewster, and N. Curran for helpful discussions, as well as the referee for many helpful suggestions. IJ is supported by a Schr\"{o}dinger Scholarship.

\bibliography{refs.bib}

\begin{thebibliography}{37}%
\makeatletter
\providecommand \@ifxundefined [1]{%
 \@ifx{#1\undefined}
}%
\providecommand \@ifnum [1]{%
 \ifnum #1\expandafter \@firstoftwo
 \else \expandafter \@secondoftwo
 \fi
}%
\providecommand \@ifx [1]{%
 \ifx #1\expandafter \@firstoftwo
 \else \expandafter \@secondoftwo
 \fi
}%
\providecommand \natexlab [1]{#1}%
\providecommand \enquote  [1]{``#1''}%
\providecommand \bibnamefont  [1]{#1}%
\providecommand \bibfnamefont [1]{#1}%
\providecommand \citenamefont [1]{#1}%
\providecommand \href@noop [0]{\@secondoftwo}%
\providecommand \href [0]{\begingroup \@sanitize@url \@href}%
\providecommand \@href[1]{\@@startlink{#1}\@@href}%
\providecommand \@@href[1]{\endgroup#1\@@endlink}%
\providecommand \@sanitize@url [0]{\catcode `\\12\catcode `\$12\catcode
  `\&12\catcode `\#12\catcode `\^12\catcode `\_12\catcode `\%12\relax}%
\providecommand \@@startlink[1]{}%
\providecommand \@@endlink[0]{}%
\providecommand \url  [0]{\begingroup\@sanitize@url \@url }%
\providecommand \@url [1]{\endgroup\@href {#1}{\urlprefix }}%
\providecommand \urlprefix  [0]{URL }%
\providecommand \Eprint [0]{\href }%
\providecommand \doibase [0]{https://doi.org/}%
\providecommand \selectlanguage [0]{\@gobble}%
\providecommand \bibinfo  [0]{\@secondoftwo}%
\providecommand \bibfield  [0]{\@secondoftwo}%
\providecommand \translation [1]{[#1]}%
\providecommand \BibitemOpen [0]{}%
\providecommand \bibitemStop [0]{}%
\providecommand \bibitemNoStop [0]{.\EOS\space}%
\providecommand \EOS [0]{\spacefactor3000\relax}%
\providecommand \BibitemShut  [1]{\csname bibitem#1\endcsname}%
\let\auto@bib@innerbib\@empty
\bibitem [{Note1()}]{Note1}%
  \BibitemOpen
  \bibinfo {note} {Here we work in the Heisenberg picture where operators carry
  the dynamics, and hence it makes sense to talk about operators at points, or
  more accurately in regions, of spacetime.}\BibitemShut {Stop}%
\bibitem [{\citenamefont {Hellwig}\ and\ \citenamefont
  {Kraus}(1970)}]{Hellwig_Kraus}%
  \BibitemOpen
  \bibfield  {author} {\bibinfo {author} {\bibfnamefont {K.~E.}\ \bibnamefont
  {Hellwig}}\ and\ \bibinfo {author} {\bibfnamefont {K.}~\bibnamefont
  {Kraus}},\ }\bibfield  {title} {\bibinfo {title} {Formal description of
  measurements in local quantum field theory},\ }\href
  {https://doi.org/10.1103/PhysRevD.1.566} {\bibfield  {journal} {\bibinfo
  {journal} {Phys. Rev. D}\ }\textbf {\bibinfo {volume} {1}},\ \bibinfo {pages}
  {566} (\bibinfo {year} {1970})}\BibitemShut {NoStop}%
\bibitem [{\citenamefont {Sorkin}(1993)}]{Sorkin:1993gg}%
  \BibitemOpen
  \bibfield  {author} {\bibinfo {author} {\bibfnamefont {R.~D.}\ \bibnamefont
  {Sorkin}},\ }\bibfield  {title} {\bibinfo {title} {{Impossible measurements
  on quantum fields}},\ }in\ \href@noop {} {\emph {\bibinfo {booktitle}
  {{Directions in General Relativity: An International Symposium in Honor of
  the 60th Birthdays of Dieter Brill and Charles Misner}}}}\ (\bibinfo {year}
  {1993})\ \Eprint {https://arxiv.org/abs/gr-qc/9302018} {arXiv:gr-qc/9302018}
  \BibitemShut {NoStop}%
\bibitem [{\citenamefont {Benincasa}\ \emph {et~al.}(2014)\citenamefont
  {Benincasa}, \citenamefont {Borsten}, \citenamefont {Buck},\ and\
  \citenamefont {Dowker}}]{Benincasa_2014}%
  \BibitemOpen
  \bibfield  {author} {\bibinfo {author} {\bibfnamefont {D.~M.~T.}\
  \bibnamefont {Benincasa}}, \bibinfo {author} {\bibfnamefont {L.}~\bibnamefont
  {Borsten}}, \bibinfo {author} {\bibfnamefont {M.}~\bibnamefont {Buck}},\ and\
  \bibinfo {author} {\bibfnamefont {F.}~\bibnamefont {Dowker}},\ }\bibfield
  {title} {\bibinfo {title} {Quantum information processing and relativistic
  quantum fields},\ }\href {https://doi.org/10.1088/0264-9381/31/7/075007}
  {\bibfield  {journal} {\bibinfo  {journal} {Classical and Quantum Gravity}\
  }\textbf {\bibinfo {volume} {31}},\ \bibinfo {pages} {075007} (\bibinfo
  {year} {2014})}\BibitemShut {NoStop}%
\bibitem [{\citenamefont {Borsten}\ \emph {et~al.}(2021)\citenamefont
  {Borsten}, \citenamefont {Jubb},\ and\ \citenamefont
  {Kells}}]{borsten2021impossible}%
  \BibitemOpen
  \bibfield  {author} {\bibinfo {author} {\bibfnamefont {L.}~\bibnamefont
  {Borsten}}, \bibinfo {author} {\bibfnamefont {I.}~\bibnamefont {Jubb}},\ and\
  \bibinfo {author} {\bibfnamefont {G.}~\bibnamefont {Kells}},\ }\href@noop {}
  {\bibinfo {title} {Impossible measurements revisited}} (\bibinfo {year}
  {2021}),\ \Eprint {https://arxiv.org/abs/1912.06141} {arXiv:1912.06141
  [quant-ph]} \BibitemShut {NoStop}%
\bibitem [{\citenamefont {Beckman}\ \emph {et~al.}(2002)\citenamefont
  {Beckman}, \citenamefont {Gottesman}, \citenamefont {Kitaev},\ and\
  \citenamefont {Preskill}}]{Wilson_loops}%
  \BibitemOpen
  \bibfield  {author} {\bibinfo {author} {\bibfnamefont {D.}~\bibnamefont
  {Beckman}}, \bibinfo {author} {\bibfnamefont {D.}~\bibnamefont {Gottesman}},
  \bibinfo {author} {\bibfnamefont {A.}~\bibnamefont {Kitaev}},\ and\ \bibinfo
  {author} {\bibfnamefont {J.}~\bibnamefont {Preskill}},\ }\bibfield  {title}
  {\bibinfo {title} {Measurability of wilson loop operators},\ }\href
  {https://doi.org/10.1103/PhysRevD.65.065022} {\bibfield  {journal} {\bibinfo
  {journal} {Phys. Rev. D}\ }\textbf {\bibinfo {volume} {65}},\ \bibinfo
  {pages} {065022} (\bibinfo {year} {2002})}\BibitemShut {NoStop}%
\bibitem [{Note2()}]{Note2}%
  \BibitemOpen
  \bibinfo {note} {Note we are not saying that the analogous state updates in
  NRQM are impossible, just that this is the case in relativistic
  QFT.}\BibitemShut {Stop}%
\bibitem [{\citenamefont {Tjoa}\ and\ \citenamefont
  {Martín-Martínez}(2019)}]{Tjoa_2019}%
  \BibitemOpen
  \bibfield  {author} {\bibinfo {author} {\bibfnamefont {E.}~\bibnamefont
  {Tjoa}}\ and\ \bibinfo {author} {\bibfnamefont {E.}~\bibnamefont
  {Martín-Martínez}},\ }\bibfield  {title} {\bibinfo {title} {Zero mode
  suppression of superluminal signals in light-matter interactions},\
  }\bibfield  {journal} {\bibinfo  {journal} {Physical Review D}\ }\textbf
  {\bibinfo {volume} {99}},\ \href {https://doi.org/10.1103/physrevd.99.065005}
  {10.1103/physrevd.99.065005} (\bibinfo {year} {2019})\BibitemShut {NoStop}%
\bibitem [{\citenamefont {Martín-Martínez}\ \emph {et~al.}(2021)\citenamefont
  {Martín-Martínez}, \citenamefont {Perche},\ and\ \citenamefont
  {Torres}}]{Mart_n_Mart_nez_2021}%
  \BibitemOpen
  \bibfield  {author} {\bibinfo {author} {\bibfnamefont {E.}~\bibnamefont
  {Martín-Martínez}}, \bibinfo {author} {\bibfnamefont {T.~R.}\ \bibnamefont
  {Perche}},\ and\ \bibinfo {author} {\bibfnamefont {B.~d.~S.}\ \bibnamefont
  {Torres}},\ }\bibfield  {title} {\bibinfo {title} {Broken covariance of
  particle detector models in relativistic quantum information},\ }\bibfield
  {journal} {\bibinfo  {journal} {Physical Review D}\ }\textbf {\bibinfo
  {volume} {103}},\ \href {https://doi.org/10.1103/physrevd.103.025007}
  {10.1103/physrevd.103.025007} (\bibinfo {year} {2021})\BibitemShut {NoStop}%
\bibitem [{\citenamefont {de~Ramón}\ \emph {et~al.}(2021)\citenamefont
  {de~Ramón}, \citenamefont {Papageorgiou},\ and\ \citenamefont
  {Martín-Martínez}}]{de_Ram_n_2021}%
  \BibitemOpen
  \bibfield  {author} {\bibinfo {author} {\bibfnamefont {J.}~\bibnamefont
  {de~Ramón}}, \bibinfo {author} {\bibfnamefont {M.}~\bibnamefont
  {Papageorgiou}},\ and\ \bibinfo {author} {\bibfnamefont {E.}~\bibnamefont
  {Martín-Martínez}},\ }\bibfield  {title} {\bibinfo {title} {Relativistic
  causality in particle detector models: Faster-than-light signaling and
  impossible measurements},\ }\bibfield  {journal} {\bibinfo  {journal}
  {Physical Review D}\ }\textbf {\bibinfo {volume} {103}},\ \href
  {https://doi.org/10.1103/physrevd.103.085002} {10.1103/physrevd.103.085002}
  (\bibinfo {year} {2021})\BibitemShut {NoStop}%
\bibitem [{\citenamefont {Perche}\ and\ \citenamefont
  {Martín-Martínez}(2021)}]{perche2021antiparticle}%
  \BibitemOpen
  \bibfield  {author} {\bibinfo {author} {\bibfnamefont {T.~R.}\ \bibnamefont
  {Perche}}\ and\ \bibinfo {author} {\bibfnamefont {E.}~\bibnamefont
  {Martín-Martínez}},\ }\href@noop {} {\bibinfo {title} {Anti-particle
  detector models in qft}} (\bibinfo {year} {2021}),\ \Eprint
  {https://arxiv.org/abs/2106.03874} {arXiv:2106.03874 [quant-ph]} \BibitemShut
  {NoStop}%
\bibitem [{\citenamefont {Polo-G\'{o}mez}\ \emph {et~al.}(2021)\citenamefont
  {Polo-G\'{o}mez}, \citenamefont {Garay},\ and\ \citenamefont
  {Martín-Martínez}}]{pologomez2021detectorbased}%
  \BibitemOpen
  \bibfield  {author} {\bibinfo {author} {\bibfnamefont {J.}~\bibnamefont
  {Polo-G\'{o}mez}}, \bibinfo {author} {\bibfnamefont {L.~J.}\ \bibnamefont
  {Garay}},\ and\ \bibinfo {author} {\bibfnamefont {E.}~\bibnamefont
  {Martín-Martínez}},\ }\href@noop {} {\bibinfo {title} {A detector-based
  measurement theory for quantum field theory}} (\bibinfo {year} {2021}),\
  \Eprint {https://arxiv.org/abs/2108.02793} {arXiv:2108.02793 [quant-ph]}
  \BibitemShut {NoStop}%
\bibitem [{\citenamefont {Beckman}\ \emph {et~al.}(2001)\citenamefont
  {Beckman}, \citenamefont {Gottesman}, \citenamefont {Nielsen},\ and\
  \citenamefont {Preskill}}]{PhysRevA.64.052309}%
  \BibitemOpen
  \bibfield  {author} {\bibinfo {author} {\bibfnamefont {D.}~\bibnamefont
  {Beckman}}, \bibinfo {author} {\bibfnamefont {D.}~\bibnamefont {Gottesman}},
  \bibinfo {author} {\bibfnamefont {M.~A.}\ \bibnamefont {Nielsen}},\ and\
  \bibinfo {author} {\bibfnamefont {J.}~\bibnamefont {Preskill}},\ }\bibfield
  {title} {\bibinfo {title} {Causal and localizable quantum operations},\
  }\href {https://doi.org/10.1103/PhysRevA.64.052309} {\bibfield  {journal}
  {\bibinfo  {journal} {Phys. Rev. A}\ }\textbf {\bibinfo {volume} {64}},\
  \bibinfo {pages} {052309} (\bibinfo {year} {2001})}\BibitemShut {NoStop}%
\bibitem [{\citenamefont {Popescu}\ and\ \citenamefont
  {Vaidman}(1994)}]{Popescu_1994}%
  \BibitemOpen
  \bibfield  {author} {\bibinfo {author} {\bibfnamefont {S.}~\bibnamefont
  {Popescu}}\ and\ \bibinfo {author} {\bibfnamefont {L.}~\bibnamefont
  {Vaidman}},\ }\bibfield  {title} {\bibinfo {title} {Causality constraints on
  nonlocal quantum measurements},\ }\href
  {https://doi.org/10.1103/physreva.49.4331} {\bibfield  {journal} {\bibinfo
  {journal} {Physical Review A}\ }\textbf {\bibinfo {volume} {49}},\ \bibinfo
  {pages} {4331–4338} (\bibinfo {year} {1994})}\BibitemShut {NoStop}%
\bibitem [{\citenamefont {Martín-Martínez}(2015)}]{Mart_n_Mart_nez_2015}%
  \BibitemOpen
  \bibfield  {author} {\bibinfo {author} {\bibfnamefont {E.}~\bibnamefont
  {Martín-Martínez}},\ }\bibfield  {title} {\bibinfo {title} {Causality
  issues of particle detector models in qft and quantum optics},\ }\bibfield
  {journal} {\bibinfo  {journal} {Physical Review D}\ }\textbf {\bibinfo
  {volume} {92}},\ \href {https://doi.org/10.1103/physrevd.92.104019}
  {10.1103/physrevd.92.104019} (\bibinfo {year} {2015})\BibitemShut {NoStop}%
\bibitem [{\citenamefont {Bostelmann}\ \emph {et~al.}(2021)\citenamefont
  {Bostelmann}, \citenamefont {Fewster},\ and\ \citenamefont
  {Ruep}}]{Bostelmann_2021}%
  \BibitemOpen
  \bibfield  {author} {\bibinfo {author} {\bibfnamefont {H.}~\bibnamefont
  {Bostelmann}}, \bibinfo {author} {\bibfnamefont {C.~J.}\ \bibnamefont
  {Fewster}},\ and\ \bibinfo {author} {\bibfnamefont {M.~H.}\ \bibnamefont
  {Ruep}},\ }\bibfield  {title} {\bibinfo {title} {Impossible measurements
  require impossible apparatus},\ }\bibfield  {journal} {\bibinfo  {journal}
  {Physical Review D}\ }\textbf {\bibinfo {volume} {103}},\ \href
  {https://doi.org/10.1103/physrevd.103.025017} {10.1103/physrevd.103.025017}
  (\bibinfo {year} {2021})\BibitemShut {NoStop}%
\bibitem [{Note3()}]{Note3}%
  \BibitemOpen
  \bibinfo {note} {This follows as $P=\left |\Psi \right \rangle \left \langle
  \Psi \right |$ is a rank 1 operator, and so $P$ cannot be localisable in any
  spacetime region, as all localisable projectors must be of infinite rank ---
  a common feature of type III von Neumann algebras (see~\cite
  {fewster2019algebraic} for example).}\BibitemShut {Stop}%
\bibitem [{\citenamefont {Fewster}\ and\ \citenamefont
  {Verch}(2020)}]{Fewster2020}%
  \BibitemOpen
  \bibfield  {author} {\bibinfo {author} {\bibfnamefont {C.~J.}\ \bibnamefont
  {Fewster}}\ and\ \bibinfo {author} {\bibfnamefont {R.}~\bibnamefont
  {Verch}},\ }\bibfield  {title} {\bibinfo {title} {Quantum fields and local
  measurements},\ }\href {https://doi.org/10.1007/s00220-020-03800-6}
  {\bibfield  {journal} {\bibinfo  {journal} {Communications in Mathematical
  Physics}\ }\textbf {\bibinfo {volume} {378}},\ \bibinfo {pages} {851}
  (\bibinfo {year} {2020})}\BibitemShut {NoStop}%
\bibitem [{\citenamefont {Fewster}(2019)}]{fewster2019generally}%
  \BibitemOpen
  \bibfield  {author} {\bibinfo {author} {\bibfnamefont {C.~J.}\ \bibnamefont
  {Fewster}},\ }\href@noop {} {\bibinfo {title} {A generally covariant
  measurement scheme for quantum field theory in curved spacetimes}} (\bibinfo
  {year} {2019}),\ \Eprint {https://arxiv.org/abs/1904.06944} {arXiv:1904.06944
  [gr-qc]} \BibitemShut {NoStop}%
\bibitem [{\citenamefont {Ruep}(2021)}]{ruep2021weakly}%
  \BibitemOpen
  \bibfield  {author} {\bibinfo {author} {\bibfnamefont {M.~H.}\ \bibnamefont
  {Ruep}},\ }\href@noop {} {\bibinfo {title} {Weakly coupled local particle
  detectors cannot harvest entanglement}} (\bibinfo {year} {2021}),\ \Eprint
  {https://arxiv.org/abs/2103.13400} {arXiv:2103.13400 [quant-ph]} \BibitemShut
  {NoStop}%
\bibitem [{\citenamefont {Brun}(2000)}]{Brun_2000}%
  \BibitemOpen
  \bibfield  {author} {\bibinfo {author} {\bibfnamefont {T.~A.}\ \bibnamefont
  {Brun}},\ }\bibfield  {title} {\bibinfo {title} {Continuous measurements,
  quantum trajectories, and decoherent histories},\ }\bibfield  {journal}
  {\bibinfo  {journal} {Physical Review A}\ }\textbf {\bibinfo {volume} {61}},\
  \href {https://doi.org/10.1103/physreva.61.042107}
  {10.1103/physreva.61.042107} (\bibinfo {year} {2000})\BibitemShut {NoStop}%
\bibitem [{\citenamefont {Jacobs}\ and\ \citenamefont
  {Steck}(2006)}]{Jacobs_2006}%
  \BibitemOpen
  \bibfield  {author} {\bibinfo {author} {\bibfnamefont {K.}~\bibnamefont
  {Jacobs}}\ and\ \bibinfo {author} {\bibfnamefont {D.~A.}\ \bibnamefont
  {Steck}},\ }\bibfield  {title} {\bibinfo {title} {A straightforward
  introduction to continuous quantum measurement},\ }\href
  {https://doi.org/10.1080/00107510601101934} {\bibfield  {journal} {\bibinfo
  {journal} {Contemporary Physics}\ }\textbf {\bibinfo {volume} {47}},\
  \bibinfo {pages} {279–303} (\bibinfo {year} {2006})}\BibitemShut {NoStop}%
\bibitem [{\citenamefont {Fewster}\ and\ \citenamefont
  {Rejzner}(2019)}]{fewster2019algebraic}%
  \BibitemOpen
  \bibfield  {author} {\bibinfo {author} {\bibfnamefont {C.~J.}\ \bibnamefont
  {Fewster}}\ and\ \bibinfo {author} {\bibfnamefont {K.}~\bibnamefont
  {Rejzner}},\ }\href@noop {} {\bibinfo {title} {Algebraic quantum field theory
  -- an introduction}} (\bibinfo {year} {2019}),\ \Eprint
  {https://arxiv.org/abs/1904.04051} {arXiv:1904.04051 [hep-th]} \BibitemShut
  {NoStop}%
\bibitem [{\citenamefont {Wald}(1984)}]{Wald}%
  \BibitemOpen
  \bibfield  {author} {\bibinfo {author} {\bibfnamefont {R.~M.}\ \bibnamefont
  {Wald}},\ }\href {https://cds.cern.ch/record/106274} {\emph {\bibinfo {title}
  {{General relativity}}}}\ (\bibinfo  {publisher} {Chicago Univ. Press},\
  \bibinfo {address} {Chicago, IL},\ \bibinfo {year} {1984})\BibitemShut
  {NoStop}%
\bibitem [{\citenamefont {Haag}(1996)}]{Haag:1992hx}%
  \BibitemOpen
  \bibfield  {author} {\bibinfo {author} {\bibfnamefont {R.}~\bibnamefont
  {Haag}},\ }\href@noop {} {\emph {\bibinfo {title} {{Local quantum physics:
  Fields, particles, algebras}}}}\ (\bibinfo  {publisher} {Springer-Verlag},\
  \bibinfo {address} {Berlin},\ \bibinfo {year} {1996})\BibitemShut {NoStop}%
\bibitem [{\citenamefont {Araki}(1964)}]{araki}%
  \BibitemOpen
  \bibfield  {author} {\bibinfo {author} {\bibfnamefont {H.}~\bibnamefont
  {Araki}},\ }\bibfield  {title} {\bibinfo {title} {Von neumann algebras of
  local observables for free scalar field},\ }\href
  {https://doi.org/10.1063/1.1704063} {\bibfield  {journal} {\bibinfo
  {journal} {Journal of Mathematical Physics}\ }\textbf {\bibinfo {volume}
  {5}},\ \bibinfo {pages} {1} (\bibinfo {year} {1964})},\ \Eprint
  {https://arxiv.org/abs/https://doi.org/10.1063/1.1704063}
  {https://doi.org/10.1063/1.1704063} \BibitemShut {NoStop}%
\bibitem [{Note4()}]{Note4}%
  \BibitemOpen
  \bibinfo {note} {In what follows we will implicitly restrict to states $\rho
  $ for which such expectation values are well defined, specifically quasifree
  states (described in the AQFT framework in~\cite {fewster2019algebraic}) and
  any states that can be constructed from these via the action of elements in
  $\protect \mathfrak {A}$.}\BibitemShut {Stop}%
\bibitem [{Note5()}]{Note5}%
  \BibitemOpen
  \bibinfo {note} {Note that these two approaches are only equivalent given a
  suitable representation of the algebra. We meet this requirement through our
  use of the usual bosonic Fock space and our implicit assumption of an
  appropriate ground state on which the Fock space is built.}\BibitemShut
  {Stop}%
\bibitem [{\citenamefont {Baker}(1905)}]{Baker}%
  \BibitemOpen
  \bibfield  {author} {\bibinfo {author} {\bibfnamefont {H.~F.}\ \bibnamefont
  {Baker}},\ }\bibfield  {title} {\bibinfo {title} {Alternants and continuous
  groups},\ }\href {https://doi.org/https://doi.org/10.1112/plms/s2-3.1.24}
  {\bibfield  {journal} {\bibinfo  {journal} {Proceedings of the London
  Mathematical Society}\ }\textbf {\bibinfo {volume} {s2-3}},\ \bibinfo {pages}
  {24} (\bibinfo {year} {1905})}\BibitemShut {NoStop}%
\bibitem [{\citenamefont {Campbell}(1897)}]{Campbell}%
  \BibitemOpen
  \bibfield  {author} {\bibinfo {author} {\bibfnamefont {J.~E.}\ \bibnamefont
  {Campbell}},\ }\bibfield  {title} {\bibinfo {title} {On a law of combination
  of operators (second paper)*},\ }\href
  {https://doi.org/https://doi.org/10.1112/plms/s1-29.1.14} {\bibfield
  {journal} {\bibinfo  {journal} {Proceedings of the London Mathematical
  Society}\ }\textbf {\bibinfo {volume} {s1-29}},\ \bibinfo {pages} {14}
  (\bibinfo {year} {1897})}\BibitemShut {NoStop}%
\bibitem [{\citenamefont {Hausdorff}(1906)}]{Hausdorff}%
  \BibitemOpen
  \bibfield  {author} {\bibinfo {author} {\bibfnamefont {F.}~\bibnamefont
  {Hausdorff}},\ }\bibfield  {title} {\bibinfo {title} {{Die symbolische
  Exponentialformel in der Gruppentheorie}},\ }\href
  {http://cds.cern.ch/record/435181} {\bibfield  {journal} {\bibinfo  {journal}
  {Ber. Verh. Kgl. SÃ¤chs. Ges. Wiss. Leipzig., Math.-phys. Kl.}\ }\textbf
  {\bibinfo {volume} {58}},\ \bibinfo {pages} {19} (\bibinfo {year}
  {1906})}\BibitemShut {NoStop}%
\bibitem [{\citenamefont {Casas}\ and\ \citenamefont
  {Murua}(2009)}]{Casas_2009}%
  \BibitemOpen
  \bibfield  {author} {\bibinfo {author} {\bibfnamefont {F.}~\bibnamefont
  {Casas}}\ and\ \bibinfo {author} {\bibfnamefont {A.}~\bibnamefont {Murua}},\
  }\bibfield  {title} {\bibinfo {title} {An efficient algorithm for computing
  the baker–campbell–hausdorff series and some of its applications},\
  }\href {https://doi.org/10.1063/1.3078418} {\bibfield  {journal} {\bibinfo
  {journal} {Journal of Mathematical Physics}\ }\textbf {\bibinfo {volume}
  {50}},\ \bibinfo {pages} {033513} (\bibinfo {year} {2009})}\BibitemShut
  {NoStop}%
\bibitem [{\citenamefont {Reed}\ and\ \citenamefont
  {Simon}(1981)}]{reed1981functional}%
  \BibitemOpen
  \bibfield  {author} {\bibinfo {author} {\bibfnamefont {M.}~\bibnamefont
  {Reed}}\ and\ \bibinfo {author} {\bibfnamefont {B.}~\bibnamefont {Simon}},\
  }\href {https://books.google.ie/books?id=rpFTTjxOYpsC} {\emph {\bibinfo
  {title} {I: Functional Analysis}}},\ Methods of Modern Mathematical Physics\
  (\bibinfo  {publisher} {Elsevier Science},\ \bibinfo {year}
  {1981})\BibitemShut {NoStop}%
\bibitem [{Note6()}]{Note6}%
  \BibitemOpen
  \bibinfo {note} {The spectrum of a smeared field $\phi (f)$, like the
  position operator $\protect \cc@accent {"705E}{x}$ in NRQM, is the whole of
  $\protect \mathbb {R}$.}\BibitemShut {Stop}%
\bibitem [{\citenamefont {Stinespring}(1955)}]{Stinespring}%
  \BibitemOpen
  \bibfield  {author} {\bibinfo {author} {\bibfnamefont {W.~F.}\ \bibnamefont
  {Stinespring}},\ }\bibfield  {title} {\bibinfo {title} {Positive functions on
  c*-algebras},\ }\href {http://www.jstor.org/stable/2032342} {\bibfield
  {journal} {\bibinfo  {journal} {Proceedings of the American Mathematical
  Society}\ }\textbf {\bibinfo {volume} {6}},\ \bibinfo {pages} {211} (\bibinfo
  {year} {1955})}\BibitemShut {NoStop}%
\bibitem [{\citenamefont {Bushev}\ \emph {et~al.}(2006)\citenamefont {Bushev},
  \citenamefont {Rotter}, \citenamefont {Wilson}, \citenamefont {Dubin},
  \citenamefont {Becher}, \citenamefont {Eschner}, \citenamefont {Blatt},
  \citenamefont {Steixner}, \citenamefont {Rabl},\ and\ \citenamefont
  {Zoller}}]{PhysRevLett.96.043003}%
  \BibitemOpen
  \bibfield  {author} {\bibinfo {author} {\bibfnamefont {P.}~\bibnamefont
  {Bushev}}, \bibinfo {author} {\bibfnamefont {D.}~\bibnamefont {Rotter}},
  \bibinfo {author} {\bibfnamefont {A.}~\bibnamefont {Wilson}}, \bibinfo
  {author} {\bibfnamefont {F.~m.~c.}\ \bibnamefont {Dubin}}, \bibinfo {author}
  {\bibfnamefont {C.}~\bibnamefont {Becher}}, \bibinfo {author} {\bibfnamefont
  {J.}~\bibnamefont {Eschner}}, \bibinfo {author} {\bibfnamefont
  {R.}~\bibnamefont {Blatt}}, \bibinfo {author} {\bibfnamefont
  {V.}~\bibnamefont {Steixner}}, \bibinfo {author} {\bibfnamefont
  {P.}~\bibnamefont {Rabl}},\ and\ \bibinfo {author} {\bibfnamefont
  {P.}~\bibnamefont {Zoller}},\ }\bibfield  {title} {\bibinfo {title} {Feedback
  cooling of a single trapped ion},\ }\href
  {https://doi.org/10.1103/PhysRevLett.96.043003} {\bibfield  {journal}
  {\bibinfo  {journal} {Phys. Rev. Lett.}\ }\textbf {\bibinfo {volume} {96}},\
  \bibinfo {pages} {043003} (\bibinfo {year} {2006})}\BibitemShut {NoStop}%
\bibitem [{\citenamefont {Cao}\ \emph {et~al.}(2019)\citenamefont {Cao},
  \citenamefont {Tilloy},\ and\ \citenamefont {De~Luca}}]{Cao_2019}%
  \BibitemOpen
  \bibfield  {author} {\bibinfo {author} {\bibfnamefont {X.}~\bibnamefont
  {Cao}}, \bibinfo {author} {\bibfnamefont {A.}~\bibnamefont {Tilloy}},\ and\
  \bibinfo {author} {\bibfnamefont {A.}~\bibnamefont {De~Luca}},\ }\bibfield
  {title} {\bibinfo {title} {Entanglement in a fermion chain under continuous
  monitoring},\ }\bibfield  {journal} {\bibinfo  {journal} {SciPost Physics}\
  }\textbf {\bibinfo {volume} {7}},\ \href
  {https://doi.org/10.21468/scipostphys.7.2.024} {10.21468/scipostphys.7.2.024}
  (\bibinfo {year} {2019})\BibitemShut {NoStop}%
\end{thebibliography}%

\end{document}